\documentclass[11pt,oneside,reqno]{article} 

\usepackage[USenglish]{babel} 
\usepackage[utf8]{inputenc} 

\usepackage[a4paper,top=30mm,bottom=20mm,left=27.5mm,right=27.5mm]{geometry} 
\usepackage{lscape} 

\usepackage{setspace} 
\usepackage{amsmath} 
\usepackage{mathtools} 
\usepackage{array} 
\usepackage{tocloft} 
\usepackage{titlesec} 
    \titleformat*{\section}{\bfseries} 
    \titleformat*{\subsection}{\bfseries} 
    \titleformat*{\subsubsection}{\normalsize\bfseries} 
    \titleformat*{\paragraph}{\small\bfseries} 
    \titleformat{name=\section}{\normalfont\normalsize\bfseries}{\thesection}{0.7em}{\hspace{0em}}{} 
    \titleformat{name=\subsection}{\normalfont\bfseries}{\thesubsection}{0.7em}{\hspace{0em}}{} 
    \titleformat{name=\subsubsection}{\normalfont\normalsize\bfseries}{\thesubsubsection}{0.7em}{\hspace{0em}}{} 
    \titleformat{name=\paragraph}[runin]{\normalfont\small\bfseries}{\theparagraph}{0.7em}{\hspace{0em}}{} 

\usepackage{titletoc}  

\usepackage[hang,flushmargin,symbol]{footmisc} 
    \renewcommand*{\thefootnote}{\fnsymbol{footnote}} 

\usepackage{lipsum} 

\usepackage{amssymb} 
\usepackage{marvosym} 
\usepackage{ifsym} 
\usepackage{wasysym} 
\usepackage{eurosym} 
\usepackage{textcomp} 
\usepackage{MnSymbol} 
\usepackage{makecell}

\usepackage{caption} 
\usepackage{subcaption} 
\usepackage[capposition=top]{floatrow} 
\usepackage{float} 
\usepackage{afterpage} 

\usepackage{xcolor} 
\usepackage{soul}

\usepackage{tabularx} 
    \newcolumntype{L}[1]{>{\raggedright\arraybackslash}p{#1}} 
    \newcolumntype{C}[1]{>{\centering\arraybackslash}p{#1}} 
    \newcolumntype{R}[1]{>{\raggedleft\arraybackslash}p{#1}} 
\usepackage{multirow} 
\usepackage{hhline} 
\usepackage{rotating} 
\usepackage[flushleft]{threeparttable} 
\usepackage{diagbox} 
\usepackage{colortbl} 
\usepackage{arydshln} 
\usepackage{booktabs}

\usepackage{pgfplots} 
    \pgfplotsset{compat=newest}
    \usepgfplotslibrary{statistics} 
    \usepgfplotslibrary{fillbetween} 
    \pgfkeys{/pgf/number format/.cd,1000 sep={}} 
\usepackage{tikz} 
    \usetikzlibrary{decorations.pathreplacing, patterns, intersections} 

\usepackage{enumitem} 

\usepackage{graphicx} 
\usepackage{pdfpages} 

\usepackage{comment} 

\usepackage{bm} 

\usepackage[backend=bibtex,natbib=true,style=authoryear,doi=false,isbn=false,url=false,eprint=false,giveninits=true,uniquename=false,dashed=false,maxnames=3,maxbibnames=30]{biblatex} 

\usepackage{csquotes} 

    \DeclareNameAlias{sortname}{family-given} 
    \DeclareNameAlias{default}{family-given} 

    \DeclareFieldFormat[article,inbook,incollection,inproceedings,patent,thesis,unpublished]{title}{#1\isdot}

    \renewbibmacro{in:}{\ifentrytype{article}{}{\printtext{\bibstring{in}\intitlepunct}}}

    \renewbibmacro*{volume+number+eid}{%
      \printfield{volume}%
      \setunit*{\addnbthinspace}
      \printfield{number}%
      \setunit{\addcomma\space}%
      \printfield{eid}}
    \DeclareFieldFormat[article]{number}{\mkbibparens{#1}}

    \usepackage{xpatch}
    \def\dopatchbibdrivereditorcomma#1{%
      \xpatchbibdriver{#1}
        {\usebibmacro{maintitle+booktitle}%
         \newunit\newblock}
        {\usebibmacro{maintitle+booktitle}%
         \setunit{\addcomma\space}\newblock}
        {}
        {\typeout{failed to patch driver for type #1}}}
    \forcsvlist{\dopatchbibdrivereditorcomma}{inbook,incollection,inproceedings}

    \renewbibmacro*{byeditor+others}{%
      \ifnameundef{editor}
        {}
        {\usebibmacro{byeditor+othersstrg}%
         \setunit{\addspace}%
         \printnames[byeditor]{editor}%
         \clearname{editor}%
         \addcomma}%
      \usebibmacro{byeditorx}%
      \usebibmacro{bytranslator+others}}

    \makeatletter
    \pretocmd{\blx@head@bibintoc}{\phantomsection}{}{\ddt}
    \makeatother

    \AtBeginBibliography{
    }

    \DeclareCiteCommand{\citeyear}
        {}
        {\bibhyperref{\printdate}}
        {\multicitedelim}
        {}
    \DeclareCiteCommand{\citeyearpar}
        {}
        {\mkbibparens{\bibhyperref{\printdate}}}
        {\multicitedelim}
        {}

\usepackage[colorlinks=true,linkcolor=blue,citecolor=blue,urlcolor=black,linkbordercolor=cyan,citebordercolor=cyan,urlbordercolor=cyan,hyperfootnotes=true]{hyperref} 

\pgfplotsset{error bar legend/.style={%
    /pgfplots/legend image code/.prefix code={%
      \pgfkeysgetvalue{/pgfplots/error bars/error mark}{\pgfplotserrorbarsmark}%
      \draw[%
        /pgfplots/every error bar,
        mark=\pgfplotserrorbarsmark,
        /pgfplots/error bars/error mark options,
        sharp plot,
        ##1
      ] plot coordinates {(0.3cm, -0.15cm) (0.3cm, 0.15cm)};%
    }
  }
}

\usepackage{authblk}    

\usepackage{layouts}    

\begin{document}


\title{\textbf{
Scarce Workers, High Wages?}\thanks{Corresponding Author: Erik-Benjamin B\"orschlein, Institute for Employment Research. Address: Regensburger Str. 100, 90478 N\"urnberg, Germany, Email: \href{mailto:erik-benjamin.boerschlein@iab.de}{erik-benjamin.boerschlein@iab.de}.

We particularly thank Nicole G\"urtzgen and Martin Friedrich for helpful discussions and suggestions. Furthermore, we are grateful to Anna Hentschke for excellent research assistance. Earlier versions of this manuscript have been presented at the 2nd BIBB-IAB-ROA Workshop, the RWI Labor Economics Seminar, and the IAB Working Group ``Occupations''.
}
}

\author[1]{\textbf{Erik-Benjamin B\"orschlein}}
\author[2,1,3,4]{\textbf{Mario Bossler}}
\author[1,3,4]{\textbf{Martin Popp}}
\affil[1]{\small Institute for Employment Research (IAB)}
\affil[2]{\small Technical University of Applied Sciences Nuremberg (TH Nuremberg)}
\affil[3]{\small Institute of Labor Economics (IZA)}
\affil[4]{\small Labor and Socio-Economic Research Center (LASER)}

\date{\vspace{-1cm}}

\maketitle

\normalsize


\begin{center}
arXiv Preprint \\
\today

\vspace*{0.5cm}

\begin{abstract}
\onehalfspacing
\noindent Labor market tightness tremendously increased in Germany between 2012 and 2022. We analyze the effect of tightness on wages by combining social security data with unusually rich information on vacancies and job seekers. Instrumental variable regressions reveal positive elasticities between 0.004 and 0.011, implying that higher tightness explains between 7 and 19 percent of the real wage increase. We report greater elasticities for new hires, high-skilled workers, the Eastern German labor market, and the service sector. In particular, tightness raised wages at the bottom of the wage distribution, contributing to the decline in wage inequality over the last decade. [99 words]

\end{abstract}

\end{center}

\vspace*{0.5cm}
\small

\noindent
\textbf{JEL Classification:}\,  
J31, J63, J64


\noindent
\textbf{Keywords:}\, 
labor market tightness, wages, labor shortage, occupations, wage inequality

\renewcommand*{\thefootnote}{\arabic{footnote}}
\setcounter{footnote}{0}

\thispagestyle{empty}
\clearpage
\setcounter{page}{1}




\section{Introduction}
\label{sec:introduction}

While the labor market in reunified Germany was plagued by high unemployment figures until the mid-2000s, the subsequent one-and-a-half decades were -- notwithstanding several crises -- characterized by robust GDP growth and a substantial increase in employment. This turnaround is widely attributed to a sustained period of wage restraint \citep{Dustmann2014, Hoffmann2016} and a set of labor market reforms (called ``Hartz laws'') which aimed at reducing structural unemployment by flexibilizing the labor market \citep{Krause2012, Launov2016, Bradley2019, Hochmuth2021}. As a result, firms were posting an increasing number of vacancies, and -- bolstered by the long-term demographic decline of domestic workers -- the number of unemployed fell. Consequently, the formerly slack labor market in Germany began to tighten. The tightening continued in the following years and accelerated during the mid-2010s. In 2022, labor market tightness in Germany is at an all-time high, implying that the number of vacancies per job seeker has tripled since 2010.

The unprecedented levels of tightness are making recruitment more and more difficult for firms in Germany, while workers are finding it much easier to take up new jobs. Through the lens of different economic models, such as the standard supply-demand, monopsony, search-and-matching, or bargaining model, an increase in labor market tightness is supposed to translate into higher wage levels -- in terms of either a new labor market equilibrium under (im-)perfect competition or an improved bargaining position of workers relative to firms. This so-called ``wage-(setting-)curve relationship'' raises the question of how much wages have risen due to the tightening of the German labor market. Unfortunately, estimating the relationship between tightness and wages proves difficult in two respects. Importantly, detailed information on the relevant numbers of vacancies and job seekers for regional labor markets by occupation is hardly available. Moreover, failure to trace out exogenous variation in labor market tightness to the individual worker will result in biased estimates.

In this paper, we analyze the effects of the tremendous increase in labor market tightness on wages in Germany from 2012 to 2022. Our analysis is based on longitudinal information on workers from the German social security records. We enrich this dataset with exceptionally rich information on vacancies and job seekers that allows us to directly calculate tightness for labor markets in terms of fine-grained combinations of occupations and regions. When estimating the wage-(setting-)curve relationship, we perform Mincer-type wage regressions with multidimensional fixed effects to absorb unobserved systematic differences across years, workers, labor markets, and firms. To address endogeneity from reverse causality and local productivity shocks, we additionally build on the popular leave-one-out instrumental variable strategy from \citet{Azar2022} and, in doing so, we are the first to transfer their design from concentration indices to ratios of vacancies to job seekers. Specifically, we instrument log labor market tightness in a particular occupation, region, and year by the average log tightness in all other regions for the very same occupation and year. By virtue of this leave-one-out design, our instrument harnesses only variation in tightness from national or non-local forces and is therefore immune against feedback effects or shocks from the same labor market.

Our OLS regressions with varying sets of control variables and fixed effects deliver statistically significant elasticities of labor market tightness on wages between 0.007 and 0.009. After successfully passing first-stage diagnostics, our IV regressions show the same positive sign but turn out somewhat higher, seemingly addressing downward bias from reverse causality. Our baseline IV regression with the full set of fixed effects and controls shows a significantly positive elasticity of 0.011, implying that an increase in labor market tightness by 100 log points raises the daily gross wages of regular full-time workers on average by 1.1 percent.

We acknowledge that our baseline IV estimate is upward-biased in the presence of occupation-specific productivity shocks at the national level, which our leave-one-out instrument does not protect against. Therefore, we use several proxies to control for these shocks with varying rigor. Under the most rigorous proxy (i.e., when conditioning on the number of vacancies in an occupation), the elasticity turns out lower but remains significantly positive with a value of 0.004, which plausibly forms a lower bound of the causal effect of log labor market tightness on log wages. Although the upper bound of 0.011 exceeds the lower bound by factor 2.6, both values imply positive but limited wage gains from the rising tightness. In light of our effect interval, the increase in tightness can explain between 7.4 and 19.1 percent of the rise in real wages in the German labor market between 2012 and 2022.

We perform a large variety of robustness checks to test the sensitivity of our baseline IV effect -- namely in terms of narrower or broader labor market definitions, a flow-adjusted version of labor market tightness that takes additionally into account vacancies and job seekers from neighboring occupations, the handling of unregistered vacancies, trimming extreme values, and quadratic effects. In all checks, the effect of tightness on wages turns out to be in close proximity to our upper-bound elasticity of 0.011.

Further analyses of heterogeneous effects for subgroups show that the elasticities are comparatively higher for newly hired workers, specialists, experts, high-skilled workers, and workers in Eastern Germany and in the service sector. Moreover, an additional analysis by wage deciles highlights that workers in the low-wage segment particularly benefit from rising labor market tightness. In light of this pattern, we show that the rising tightness has contributed to the observed decline in wage inequality over the last decade. Moreover, final analyses of firm-level wage-setting indicate that most of our effect stems from low-paying firms raising their overall wage level in response to higher tightness across the set of employed occupations. Put differently, only a small share of the effect can be attributed to firms raising wages differently depending on the tightness increases in the respective occupations.

Our study contributes to a better understanding of wage-setting policies, tight labor markets, and their mutual interdependencies. First, we add to the growing literature on the general consequences of scarcity of labor input. In recent years, the literature has gathered new momentum as labor has become increasingly scarce in U.S.\ and European markets \citep{Abraham2020, Groiss2024}.
While search-and-matching theory advocates the use of tightness (i.e., the ratio of vacancies to job seekers or unemployed), many studies are resorting to alternative scarcity measures like employment rates, job-to-job transition rates, vacancy durations, or self-reported labor shortages. An increasing number of studies quantifies how scarce labor has manifested in lower (growth of) employment in the U.K. \citep{Stevens2007}, the U.S. \citep{Beaudry2018}, Germany \citep{Bossler2024}, and France \citep{LeBarbanchon2024}. While insufficient labor supply may negatively affect firms' productivity \citep{Haskel1993}, bargaining power \citep{Hirsch2018, Pezold2023}, and profits, \citet{Reder1955} argues that firms may counteract hiring difficulties by either raising wages to retain and attract workers or by lowering hiring standards. Otherwise, firms might compensate for the lack of manpower by increasing capital usage \citep{Dacunto2020, Lipowski2024} or adjusting labor contracts of incumbent workers \citep{Houseman2003, Fang2009, Healy2015}.

Causal evidence on the effects of labor market tightness (or proxies thereof) on wages is small but rapidly growing. For the post-Covid U.S.\ economy, \citet{Autor2023} find that rising tightness (as a composite of job-to-job transitions and unemployment) primarily raised wages at the bottom of the wage distribution, which counteracted rising wage inequality. Using Danish register data at the firm level, \citet{Hoeck2023} arrives at an elasticity of wages with respect to tightness between 0.01 and 0.02. For Germany, \citet{Linckh2024} build on cross-sectional information on 533 German workers in 2017/18 and trace out a positive but insignificant semi-elasticity of tightness (in levels) on entry wages (in logs).\footnote{For the German labor market, there are two further studies which analyze the effect of tightness on wages for particular groups of workers: For the years 1995-2014, \citet{Brunow2022} show that entry into a tight labor market (proxied by the ratio of unemployed to employed persons) is associated with a steeper wage development for young workers 10 years after entering the labor market. \citet{Koelling2023} finds that firms with self-reported hiring problems pay a wage premium of around 4 percent to nurses during 2008-2018.} Unlike the aforementioned studies, \citet{Bossler2024} use an instrumental variable approach to explicitly address endogeneity from reverse causality and omitted variable bias from productivity shocks. Using a shift-share design at the firm level, they estimate a tightness elasticity of 0.01 for the German labor market in the years 2012-2019. However, their estimation at the firm level may conceal important heterogeneity in terms of socio-demographic worker characteristics.
Against this background, our study is the first to combine rich information on vacancies and job seekers with an instrumental variable approach to determine the wage effects of labor market tightness on the level of workers. In doing so, we are able to differentiate tightness effects along many dimensions of workers while minimizing bias from spurious variation in tightness.

Second, our paper provides microfoundations for calibrating key parameters of search-and-matching models \citep{Mortensen1994, Mortensen1999, Pissarides2000}. In these models, job-finding rates of workers and job-filling rates of firms are directly related to the ratio of vacancies to job seekers. Under Nash bargaining, the standard DMP model propagates a positive relationship of labor market tightness on wages -- commonly known as the ``wage-setting curve''. In line with \citet{Hoeck2023}, our lower- and upper-bound elasticities mirror a positively sloped but relatively flat wage-setting curve. The flat slope contradicts the calibration in \citet{Shimer2005} but resembles the calibration by \citet{Hagedorn2008}, which performs well from an empirical point of view.

Third, our paper indirectly addresses the literature on the so-called ``wage curve'', a related but (compared to the wage-setting curve) differently nuanced concept of wage formation. The wage curve \citep{Blanchflower1995, Card1995, Nijkamp2005} connects wages to regional unemployment levels and can be derived from bargaining models \citep{Nash1950, Nickell1983} or efficiency wage models \citep{Stiglitz1974, Yellen1984}. The wage curve was first described by \citet{Blanchflower1994}, who established a robust negative relationship between regional unemployment and wage levels. Evidence shows that the effect of unemployment on wages is negative but relatively weak in the German labor market \citep{Bellmann2001, Baltagi2009, Baltagi2012, Bauer2020}. Relative to the literature on the wage curve, our tightness measure constitutes a more comprehensive measure for scarcity of labor than the mere number of unemployed due to consideration of the labor-demand side. Despite the different concepts, our results validate the available evidence on German wage curves in the sense that improved outside options for workers have a positive but rather limited effect on wages.

Fourth, we contribute to the literature on the development of the wage structure in Germany \citep{Dustmann2009, Dustmann2014, Card2013}. This literature has documented that wage inequality has been increasing from the 1990s up until the mid-2000s. The rising inequality was driven by declining real wages in the lower half of the wage distribution, while wages were stagnating in the middle and growing at the top \citep{Dustmann2014}. The literature provides various explanations for this phenomenon, namely (i) skill-biased and routine-biased technological change \citep{Dustmann2009, Antonczyk2018}, (ii) decline in collective bargaining along with increased flexibilization of such agreements \citep{Dustmann2009,Dustmann2014}, (iii) rising dispersion in firm-specific wage premia and increasing assortativeness in the matching of workers to plants \citep{Card2013}, and (iv) domestic outsourcing of service personnel \citep{Goldschmidt2017}. \citet{Biewen2019} show that the rise in income inequality stopped during the mid-2000s. Later, it was documented that the inequality even started to decline during the 2010s \citep{Moeller2016, Bossler2023}, and the declining inequality not only referred to labor incomes but also to wages \citep{Bloemer2023, DrechselGrau2022, Fedorets2020}. Part of this decline can be attributed to the 2015 minimum wage introduction \citep{Bossler2023}. However, the decline already started before the minimum wage was introduced, leaving room for other explanatory factors. Against this backdrop, our paper documents that the tremendously rising labor market tightness, from which workers in low-wage firms were particularly benefiting, contributed to the observed decline in wage inequality.

The study is structured as follows: Section~\ref{sec:theory} provides a succinct overview of the models that propagate a wage effect from tightening labor markets. Section~\ref{sec:data} describes the administrative and survey-based data sources. In Section~\ref{sec:descriptives}, we provide descriptive evidence on the nexus between labor market tightness and wages. In Section~\ref{sec:model}, we describe our empirical model and the design of our leave-one-out instrument. Section~\ref{sec:results} presents the results of our empirical analyses and robustness checks and interprets the effect size. In Section~\ref{sec:heterogeneous}, we present an evaluation of heterogeneous effects for different subgroups. Section~\ref{sec:conclusion} concludes.

\section{Theoretical Background} \label{sec:theory}

Economic theory offers various frameworks that shed light on the relationship between the equilibrium outcomes of labor market tightness and wages. In this section, we briefly explain the predictions from these models to facilitate the interpretation of our later empirical results.

\paragraph{The Standard Model of Labor Demand and Labor Supply.}
In a competitive labor market, the equilibrium wage rate, which equals workers' marginal value product of labor, is determined by the intersection of the downward-sloping labor demand and the upward-sloping labor supply curve. A rise in labor market tightness is caused by more vacancies (e.g., from higher product demand) or fewer job-seeking individuals (e.g., due to demographic decline). If, for all wage levels and given employment, the number of vacancies (i.e., unrealized labor demand) increases, the labor demand curve shifts rightward. Vice versa, if the number of job seekers (i.e., unrealized labor supply) decreases, the labor supply curve shifts leftward. Both shifts cause wages to rise in the new market-clearing equilibrium. This wage rise turns out higher, the more wage-inelastic (i.e., the steeper) the demand and supply curves are.

\paragraph{Monopsonistic Labor Markets.}
In contrast to the competitive labor market, monopsonistic labor markets are characterized by only a limited number of employers \citep{Boal1997, Manning2003}.\footnote{In a broader sense, monopsony power can also result from frictions in atomistic labor markets, for example due to low labor mobility allowing firms to exert wage-setting power.} In the textbook monopsony model of a single firm \citep{Robinson1933}, the monopsonist retains wage-setting power when labor supply is less than perfectly elastic. Unlike firms in the competitive model, the monopsonist does not face an exogenously given market wage but chooses employment along an upward-sloping labor supply curve. Since the monopsonist can hire additional workers only by offering higher wages, the monopsonist chooses a profit-maximizing employment level below the competitive level, resulting in a market wage that is marked down relative to workers' marginal productivity.

If tightness increases in the monopsonistic labor market, wages will also rise. However, the first key difference to the competitive model is that the monopsonist raises wages by less when the marginal value product curve of the market (i.e., the labor demand curve in the competitive model) shifts rightward towards a new imperfectly competitive equilibrium. The reason is that, unlike firms in the competitive model, the monopsonist is not a wage taker who is bound to market forces. However, the monopsonist will still pay higher wages along the upward-sloping labor supply curve but will not fully meet the expansion in demand to avoid further wage increases. However, the wage effect does not necessarily fall short of the wage effect in the competitive model. In particular, the second key difference is that higher tightness is supposed to improve worker's bargaining position and thus flattens the labor supply curve to the firm, which comes along with higher wages by reducing markdowns.

\paragraph{Search and Matching.}
The Diamond-Mortensen-Pissarides (DMP) model offers a more nuanced understanding of the labor market by incorporating labor market frictions from search and matching \citep{Mortensen1999, Pissarides2000}. These frictions impede the smooth clearing of the labor market, leading to a simultaneous coexistence of vacancies and unemployed individuals.

Since the search process is costly, a successful match yields a rent that is shared between workers and firms according to Nash bargaining. Under this exogenous bargaining rule, a rise in labor market tightness raises the job-finding rate of workers and, thus, provides workers with a better outside option (i.e., the present value of unemployment increases). Vice versa, higher tightness reduces the job-filling rate of firms, which worsens their outside option (i.e., the present value of an unfilled vacancy decreases). As a result, the model propagates a positive effect of labor market tightness on wages, which is commonly referred to as the ``wage-setting curve'' \citep{Mortensen1994, Mortensen1999, Pissarides2000}. In the standard DMP model, the slope of the wage-setting curve positively depends on the magnitude of vacancy-posting cost and the relative bargaining power of workers, with productivity being an important co-determining variable of the wage.

\paragraph{Bargaining Models.}
In bargaining models, workers (or unions) and firms negotiate about the distribution of rents \citep{Dunlop1944, Nickell1983}. Higher labor market tightness improves the endogenous bargaining power of workers (relative to firms) due to more outside options. As a consequence, monopsony (or wage-setting) power of firms falls and workers will extract a larger fraction of the overall match surplus, resulting in higher wages.

\paragraph{Efficiency Wages.}
In efficiency wage models, firms with incomplete information raise wages above market-clearing levels for four reasons \citep{Yellen1984}: reduced shirking, improved morale, lower turnover, and better applicants. In these models, the wage acts as an incentive and selection device. However, when labor markets tighten, productivity and availability of workers suffer (due to better outside options). Consequently, firms may further increase the wage rate to uphold its functionality.

\section{Data} \label{sec:data}

In this paper, we assemble three different data sources to study the impact of labor market tightness on wages in the German labor market. First, we obtain data on wages and control variables from the Integrated Employment Biographies (IEB). Second, we use official statistics from the Federal Employment Agency to gather detailed information on registered job seekers and registered vacancies. Third, we extrapolate the number of registered vacancies using information on the shares of registered vacancies as of all vacancies from the IAB Job Vacancy Survey (JVS) to determine the total number of vacancies. For further details on our data, we refer the reader to Appendix~\ref{app:data_details}.

\paragraph{Wages and Control Variables.}
The IEB is the administrative dataset of the German labor market, featuring daily information on the near-universe of individual employment spells \citep{Mueller2020}. The data cover workers' longitudinal employment histories, including their daily gross wages (up to a censoring limit), contract type, 5-digit occupation, workplace location, working time (full-time or part-time), gender, age, education, nationality, and a firm identifier.\footnote{Note that the firm identifier refers to establishments which are local units of a company in our data.} Specifically, the IEB includes employment notifications of all workers subject to social security contributions, the reporting of which is mandatory for employers in Germany.

Our analysis focuses on employment spells that span June 30 of the years 2012-2022.\footnote{Although data availability would allow us to construct our measures of labor market tightness from 2010 onward, we begin our analysis in the year 2012 because there was a major structural break in 2011/12 in the occupation variable in the IEB data.} For lack of information on the number of individual working hours in the IEB, we follow standard practice and limit our analysis to full-time workers in regular employment to ensure uniform working hours. Throughout the study, we focus on workers in the non-agricultural private business sector.\footnote{Specifically, we keep the following NACE 2-digit codes: 5-82, 90, and 92-96.} When constructing our wage variable, we explicitly take into account special payments (such as provisions or Christmas bonuses), which we distribute evenly over the observed spell length and, thus, translate into higher daily wages. To address right-censored wages above the upper earnings limit on social security contributions, we employ a two-step imputation technique based on the Tobit regression method as in \citet{Card2013}.\footnote{Initially, we run Mincer-type Tobit regressions (separately by year, gender, and education) on a set of individual controls to fit wages and derive leave-one-out average wages per firm (i.e., the average wage in a firm excluding the observation at hand). Subsequently, we re-run the Tobit regressions using the firm-level leave-one-out average wage as an additional covariate to account for firm-specific wage premia in the wage imputation.} We deflate our imputed wages by the national consumer price index to depict real wages. To reduce the computational burden, we draw a random 5 percent sample of workers to perform our descriptive and causal analyses.

\paragraph{Labor Market Definition.}
We combine detailed information on occupations and regions (in terms of workplace) to define labor markets. Regarding occupations, we leverage the rich occupational information available in the IEB, namely the German Classification of Occupations (KldB) from 2010, which allows us to distinguish between a maximum of 1,300 5-digit occupations. The first four digits classify occupations based on the area of expertise with increasing levels of detail. The fifth digit assigns the skill requirement level based on four categories: helpers, professionals, specialists, and experts.\footnote{Helper occupations require zero or a maximum of one year's training. Professional occupations cover all jobs for which industrial, commercial, or other vocational training is required (excluding master craftsmen and technicians). Specialists hold a bachelor's degree or have completed master craftsman/technician training. Experts hold a master's degree or an equivalent diploma.} In line with standard practice, we employ the leading three digits (occupational groups) along with the fifth digit (skill requirement level). This aggregation leads to $O=431$ different occupations, which we refer to as ``detailed 3-digit occupations'' in the following. In terms of regions, we utilize the graph-theoretical approach proposed by \citet{Kropp2016} to merge 401 administrative districts into commuting zones. Our optimization, which is guided by home-to-work commuting patterns provided by the Federal Employment Agency, results in $R=52$ zones with strong within-zone but little between-zone commuting.\footnote{Further details of the delineation of commuting zones are summarized in Appendix~\ref{app:lm_delineation}.}
Taken together, our baseline definition of labor markets results in 16,980 occupation-by-zone combinations. To convince the reader that our empirical results are not driven by this specific delineation, we later show that our baseline results hardly change when applying broader or narrower definitions of labor markets.

\paragraph{Labor Market Tightness.}
We follow the method from \citet{Bossler2024} and combine process and survey data to construct measures of labor market tightness at an exceptionally fine level. To begin with, we collect process data on posted vacancies that were registered with the Federal Employment Agency. We obtain official statistics on the stock of registered vacancies for a reference date in mid-June between 2012 and 2022, including details on the targeted occupation and commuting zone (in terms of workplace). Note that, in Germany, firms are not required to register vacancies with the Federal Employment Agency, resulting in incomplete vacancy data. To determine the overall stock of registered plus unregistered vacancies for each labor market and year, we divide the number of registered vacancies by the yearly share of registered vacancies from the IAB Job Vacancy Survey.

The IAB Job Vacancy Survey (IAB-JVS) is a representative and large-scale firm survey which, among other questions, asks firms about their numbers of registered and unregistered vacancies \citep{Bossler2020a}. When constructing the yearly shares of registered vacancies among all vacancies, we differentiate between helpers, professionals, and the two groups of specialists and experts combined. The yearly shares of registered vacancies fluctuate slightly over time (see Appendix Table~\ref{tab:notification_share}) and have the following values on average for the years 2012-2022: helpers (45.1 percent), professionals (44.3 percent), and specialists along with experts (28.9 percent).

In contrast to vacancies, individuals must register as unemployed with the Federal Employment Agency to be eligible for benefits from unemployment insurance or social assistance. For each labor market and year, we extract official information on the number of job seekers, comprising registered unemployed individuals plus employed workers who actively seek employment through the Federal Employment Agency.\footnote{Upon registration at the Federal Employment Agency, job seekers must state their targeted occupation.} \citet{Abraham2020} demonstrate that the count of effective job seekers delivers greater explanatory power in the matching function than the mere tally of unemployed individuals. For each labor market (i.e., for each combination of detailed occupational group and commuting zone) and year, we divide the overall stock of registered plus unregistered vacancies by the count of job seekers to arrive at our measure of labor market tightness. \citet{Bossler2024} show that the resulting ratio performs well as a measure for firms' difficulty in recruiting workers -- in the sense that it positively correlates with pre-match hiring cost, search duration, and the number of search channels while it is negatively correlated with the number of applicants.

\section{Descriptive Results} \label{sec:descriptives}

Before estimating the wage effect of labor market tightness in a regression framework, we scrutinize our data on tightness and wages descriptively. The descriptive analysis provides preliminary insights into the plausibility of the data, the aggregate developments, and the raw correlation between both variables. For more detailed descriptive evidence, we refer the reader to Appendix~\ref{app:description}.

\paragraph{The Development of Labor Market Tightness.}
The development of the German labor market has been remarkable during the last decade, moving from what was once described as the ``sick man of Europe'' to an ``economic superstar'' \citep{Dustmann2014}. The German experience is characterized by robust employment growth, which surged from 42.0 million employees in 2012 to 46.8 million employees in 2022. Similarly, the number of unemployed declined from 2.90 million persons in 2012 to 2.42 million persons in 2022, mirroring the growth in employment and reflecting the demographic change of the German labor force. As a result, the expansion in employment was accompanied by a tremendous increase in the proportion of firms reporting increased labor shortages, which came along with a strongly increasing labor market tightness.

\begin{figure}[ht!]
\centering
\begin{center}
\caption{Labor Market Tightness in Germany over Time}
\label{fig:tightness_by_year}
\begin{center}
\includegraphics[scale=0.95]{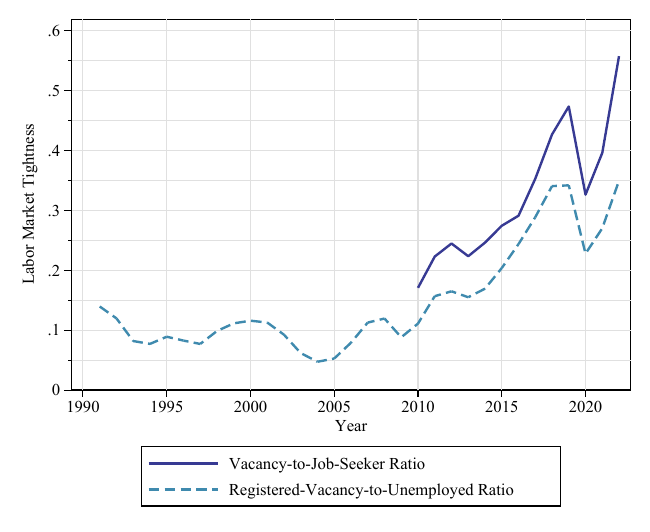}
\end{center}
\begin{tablenotes}
\small \item \textit{Notes:} The figure illustrates the development of labor market tightness in Germany over time. The solid line refers to the economy-wide ratio of vacancies to job seekers. The dashed line shows the ratio of registered vacancies to unemployed persons and, thus, builds on aggregate information that is available already from 1991 onwards. \textit{Sources:} Official Statistics of the German Federal Employment Agency $\plus$ IAB Job Vacancy Survey, 1991-2022.
\end{tablenotes}
\end{center}
\end{figure}

Figure~\ref{fig:tightness_by_year} presents the development of labor market tightness in Germany. Our detailed measure of tightness relates the total number of vacancies to the number of job seekers in each labor market and is available from 2010 onward. The solid line shows the development of this measure when dividing the number of vacancies and job seekers at the national level. The dashed line illustrates the development of an alternative tightness measure relating the nationwide number of registered vacancies to the nationwide number of unemployed, each of which is available in aggregate data of the Federal Employment Agency since 1991. The development of our baseline and the alternative tightness measure is remarkably similar. Despite taking into account employed job seekers in the denominator, our more comprehensive measure turns out to be higher since it features the overall number of vacancies in the numerator, which vastly exceeds the number of registered vacancies. The descriptive development demonstrates that our sample period covers an interesting period of a significant rise in labor market tightness, which has previously remained -- by and large -- constant. In 2010, our detailed measure of tightness was still about 0.17 at the national level, implying, on average, six job seekers for every vacant job. In 2012 (i.e., at the beginning of our regression sample), tightness rose to 0.24, averaging four job seekers per vacancy. Apart from a temporary slump during the Covid-19 pandemic, tightness rose tremendously in the following years to a value of 0.56 by 2022. At the current edge, there are fewer than two job seekers for every vacancy, representing more than a tripling in tightness since 2010. In precise terms, tightness rose by 229.4 percent during 2010-2022 and by 133.3 percent during 2012-2022.

However, we observe remarkable differences in the level and development of tightness by the underlying requirement level (see Appendix Figure~\ref{fig:tightness_by_level_year}). Among the four groups, tightness for helper occupations turns out by far the lowest, while the markets for professionals, specialists, and experts are generally tighter. During our period of analysis, all groups of requirement levels experienced a tremendous increase in tightness. Between 2010 and 2022, tightness for helpers and experts doubled while it tripled and quadrupled for specialists and professionals, respectively. For professionals, specialists, and experts, the number of vacancies even exceeds the number of job seekers at the current edge in 2022. When abstracting from the requirement level, we observe that all labor markets in terms of 2-digit occupations (see Appendix Figure~\ref{fig:tightness_by_kldb10_2}) or commuting zones (see Appendix Figure~\ref{fig:tightness_by_zone}) tightened. Importantly, however, there are substantial differences in the tightness increase across occupations and regions over time, which is the variation we are leveraging in our causal analysis.

\begin{figure}[ht!]
\centering
\begin{center}
\caption{Real Wages in Germany over Time}
\label{fig:wages_by_year}
\begin{center}
\includegraphics[scale=0.95]{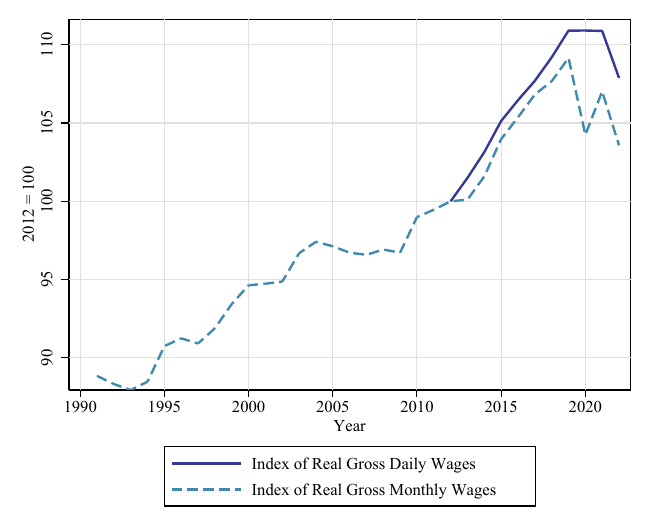}
\end{center}
\begin{tablenotes}
\small \item \textit{Notes:} The figure illustrates the development of wages in Germany over time. The solid line refers to real gross daily wages (including special payments) for regular full-time workers in the non-agricultural private business sector. The dashed line shows the index of real gross monthly wages for full-time workers excluding special payments. The latter time series refers to the manufacturing sector until 2006 and to manufacturing and services from 2007 onwards.
\textit{Sources:} Integrated Employment Biographies + Official Data of the German Federal Statistical Office, 1991-2022.
\end{tablenotes}
\end{center}
\end{figure}

\paragraph{The Development of Wages.} Against the backdrop of the wage(-setting) curve, the main question of interest is to what extent the tremendous increase in tightness has led to wage increases. Figure~\ref{fig:wages_by_year} shows the development of average real wages since the beginning of the 1990s based on data from the German Federal Statistical Office and in the period of our analysis sample based on the IEB data. The resulting time series shows relatively flat growth in real wages during the 2000s, which has been argued to have contributed to the favorable development of the German labor market by increasing the country's competitiveness through lower labor costs \citep{Dustmann2014}. Most importantly for our analysis, however, wage growth accelerated during the 2010s, suggesting that some of the wage growth could be driven by rising labor market tightness. In sum, we observe that average real gross daily wages rose from 106.25 Euro to 114.62 Euro per day, that is, by 7.9 percent between 2012 and 2022.\footnote{The qualitative assessment of the wage development remains unchanged (albeit steeper) when looking at nominal wages, as depicted in Appendix Figure~\ref{fig:nominal_wages_by_year}.}

\paragraph{Raw Correlation of Labor Market Tightness and Wages.}

Before estimating the effect of labor market tightness on wages, we examine the raw correlation between the two variables.

\begin{figure}[ht!]
\centering
\caption{Raw Correlation of Labor Market Tightness and Wages}
\label{fig:binscatter_w_vu}
\begin{center}
\includegraphics[scale=0.95]{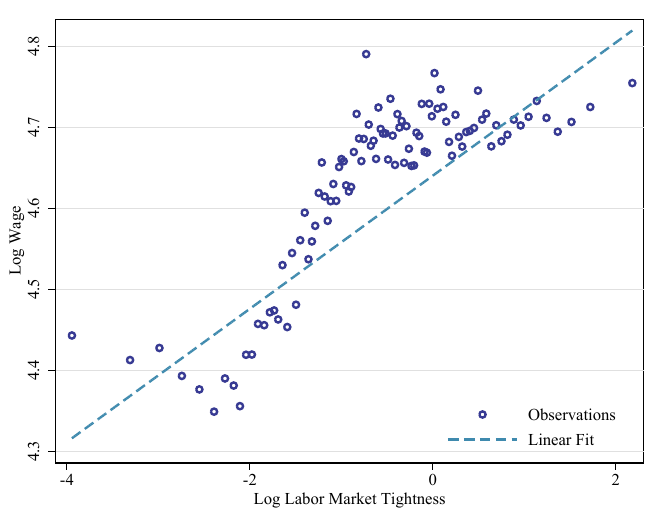}
\end{center}
\vspace{-0.5cm}
\begin{tablenotes}
\small \item \textit{Notes:} The figure shows a binned scatterplot with 100 markers to illustrate the raw correlation between log labor market tightness and log real daily wages. \textit{Sources:} Integrated Employment Biographies $\plus$ Official Statistics of Federal Employment Agency + IAB Job Vacancy Survey, 2012-2022.
\end{tablenotes}
\end{figure}

Figure~\ref{fig:binscatter_w_vu} depicts the raw correlation between both variables, displaying the raw average log wage for each percentile bin of the log tightness distribution together with a linear fit based on the respective bivariate linear regression. By visual inspection, the correlation between both variables is clearly positive, demonstrating that high tightness is associated with high wages. This aligns with our theoretical expectations that wages respond positively to a rising tightness. However, the picture does not necessarily imply that there is a causal effect of tightness on wages since the raw correlation may be confounded by uncontrolled variables (e.g., requirement level) or reverse causality.

Finally, it is worth noting that the relationship between tightness and wages is far from perfectly linear. While wages seem to rise in response to tightness, the wage curve becomes flatter beyond a certain level of tightness. This particular pattern suggests that it might be worth allowing for a non-linear relationship of both variables, which we will address when estimating heterogeneous effects in Section~\ref{sec:heterogeneous}.

\section{Empirical Model} \label{sec:model}

Our descriptive analysis suggests a positive relationship between labor market tightness and wages. To estimate the causal effect of higher labor market tightness on wages, we employ high-dimensional fixed-effects regression models using the following log-linear Mincer-type wage equation:
\begin{equation}\label{eq:hdfe_baseline}
ln\,W_{it} \,=\, \alpha \cdot ln\,\theta_{ort} \,+\, X_{it}\beta \,+\, \gamma_{i} \,+\, \delta_{t} \,+\, \phi_{or} \,+\, \psi_{j} \,+\, \varepsilon_{it}
\end{equation}

We regress daily wages $W$ of worker \textit{i} in year \textit{t} on labor market tightness $\theta$ (all in logs) while allowing for an idiosyncratic error term $\epsilon$. Our measure of labor market tightness is defined as the ratio of vacancies to job seekers for each combination of occupation $o$ and region $r$: $\theta_{ort}=\frac{V_{ort}}{U_{ort}}$. The occupation identifier $o$ corresponds to the occupation that worker $i$ has in year $t$ such that $o=O(i,t)$.\footnote{The region identifier $r$ corresponds to the workplace location of worker $i$ in year $t$.} Our main coefficient of interest is $\alpha$, which denotes the elasticity of wages with respect to labor market tightness. In our first and all other specifications, we follow standard practice and control for worker fixed effects $\gamma_{i}$ and year fixed effects $\gamma_{t}$ to absorb unobserved heterogeneity across workers and time. In addition, we control for a labor-market fixed effect $\phi_{or}$ to ensure that our effect is driven by variation in tightness over time rather than by occupational or regional mobility of workers between labor markets. In our second specification, we additionally include a set of time-varying variables $X_{it}$ to control for omitted variable bias, namely education, age, squared age, new hire, and a binary variable indicating whether the firm is located in Western or Eastern Germany. In our third specification, we initially dispense with control variables but add firm fixed effects $\psi_{j}$ to absorb unobserved heterogeneity between employers, thus ruling out that the effect is driven by mobility between employers. The firm identifier $j$ corresponds to the firm that employs worker $i$ in year $t$ such that $j=J(i,t)$. In our fourth and preferred specification, we include both control variables and firm fixed effects. Thus, our effect is identified by comparing changes in workers' wages in labor markets with a changing tightness while netting out composition effects by time, labor market, firm, and socio-demographic characteristics. For inference, we allow for a cluster-robust covariance matrix of the error term, defining clusters at the level of occupation-by-region cells (i.e., at the level of variation of our explanatory variable).

\paragraph{Threats to the Identification.} Guided by the theoretical models in Section \ref{sec:theory}, we explicitly acknowledge that labor market tightness is an equilibrium outcome that is itself influenced by economic forces. The first threat to the identification is reverse causality between wages and tightness, which the inclusion of fixed effects does not protect against. Specifically, higher wages may raise labor supply, thus increasing the number of job seekers in the market. Vice versa, higher wages also reduce labor demand via the job creation curve. As shown by \citet{Bassier2023}, increased wages reduce vacancy durations and, consequently, the number of vacancies in the market. Both the reverse labor supply and the labor demand channel imply a negative feedback effect on labor market tightness, which would manifest in a downward-biased elasticity. The feedback effect of a single atomistic firm is certainly negligible.\footnote{This assumption is less likely to hold when using (self-reported) measures of labor shortages or hiring difficulties at the firm level as explanatory variable because firms may face these problems simply because they are low-paying. By instead using tightness at the market level to measure the scarcity of the labor input, we already rule out reverse causality from wage policies of single atomistic firms, which cannot reasonably influence market outcomes.} However, when firms are not small in relation to the market or act in concert, their wage-setting policies may have a reverse effect on labor market tightness. To minimize downward bias from reverse causality, we block these feedback mechanisms by instrumenting the explanatory tightness variable in a two-stage least squares regression. We draw on an internal leave-one-out instrumental variable strategy, which became popular through several recent monopsony studies that analyze the effect of higher labor market concentration on wages (e.g., \citealp{Azar2022}; \citealp{Bassanini2024}), and transfer this design to the analysis of labor market tightness. Specifically, we instrument any value of log tightness in a certain region $c$ by the average of the log tightness in all other regions for the same occupation and time period:
\begin{equation}
\label{eq:2}
Z^{1}_{oct} \,\equiv\, \overline{ln\,\theta\,}^{\,-c}_{ot} \,=\, \dfrac{\sum_{r \neq c\vphantom{/}} \, ln\,\tfrac{V_{ort}}{U_{ort}}}{R-1}
\end{equation}
Favorably, the instrument delivers variation in local labor market tightness that is driven by national, non-local forces in the respective occupations. Due to the leave-one-out property (i.e., by excluding vacancies and job seekers from the focal region), the instrument is immune against changes in occupations in the very same region and, thus, protects against a feedback effect of wages on tightness in the very same labor market.

By construction, calculating the leave-one-out average assigns all other regions the same weight. To ensure that larger labor markets receive a correspondingly higher weight, we construct the following alternative leave-one-out instrument for region $c$:
\begin{equation}
\label{eq:3}
Z^{2}_{oct}  \,\equiv\, ln\,\theta^{-c}_{ot} \,=\, ln\,\frac{\,V^{-r}_{ot}}{U^{-r}_{ot}}\,=\, ln\, \bigg( \frac{ \sum_{r \neq c\vphantom{/}} \, V_{ort}}{\sum_{r \neq c\vphantom{/}}U_{ort}} \bigg)
\end{equation}
This ``sum-based'' instrument does not rely on averaging tightness but, for the very same occupation and time period, adds up the number of vacancies and job seekers in all other regions and calculates the ratio of both leave-one-out sums. Thus, larger regions with many vacancies or job seekers contribute more strongly to the instrument than smaller regions.

The second threat to our identification is omitted variable bias from productivity shocks. If productivity increases at the local or national level, the labor demand curve shifts rightward, leading to a simultaneous rise in wages and vacancies. Thus, if productivity shocks are not properly controlled for, the estimated elasticity will be upward-biased. Favorably, our instrumental variable protects against omitted variable bias from local productivity shocks since the leave-one-out average always excludes the focal region. However, our instrument may pick up productivity shocks if these shocks correlate across regions (i.e., if there are national shocks to productivity in the respective occupation).

We acknowledge that our instrument is not fully exogenous in the presence of occupation-wise productivity shocks at the national level. Since we do not observe productivity directly, we perform four different checks with different proxies for national productivity shocks to empirically assess the magnitude of this bias. First, we control for occupation-by-year fixed effects in our OLS specification.\footnote{Note that we cannot control for occupation-by-year fixed effects in our IV specification since our instrument is essentially defined at this level.}
Second, we directly control for log productivity at the firm (rather than occupation) level in our IV specification by combining our IEB data with survey information from the IAB Establishment Panel.
Third, we control for industry-by-year fixed effects, which assumes that productivity shocks are industry-specific, in our IV specification.
Fourth, and most rigorously, we control for the leave-one-out sum of vacancies in the respective occupation-by-year cell in our IV specification. By controlling for these vacancies, we safeguard that the variation in tightness stems exclusively from supply-side changes in the number of job seekers which are plausibly unrelated to productivity shocks. Taken together, these checks will allow us to establish a lower bound for the elasticity of wages with respect to tightness.

\paragraph{Heterogeneous Effects.} When estimating heterogeneous effects of the relationship between tightness and wages, we reformulate Equation~(\ref{eq:hdfe_baseline}) and run the following interacted model:

\begin{equation}\label{eq:hdfe_interacted}
ln\,W_{it} \,=\, \sum_{s=1}^{S} \, \alpha_{s} \cdot D^{s}_{it} \cdot ln\,\theta_{ort} \,+\, X_{it}\beta \,+\, \gamma_{i} \,+\, \delta_{t} \,+\, \phi_{or} \,+\, \psi_{j} \,+\, \varepsilon_{it}
\end{equation}
where $D^{s}$ represents a dummy variable that is 1 when the underlying categorical variable has value $s$ (zero otherwise). In this formulation, $\alpha_{s}$ directly captures the tightness effects of the respective subgroup $s$ without relying on a certain reference group. Note that the group identifier $D$ is always included in the set of control variables or absorbed by the fixed effects. Thus, the base effect for all interacted variables is always controlled for. For our instrumental variable strategy, we interact our baseline instrument $Z_{ort}$ by the respective group identifier $D^{s}$, resulting in $S$ instruments for the $S$ potentially endogenous variables of interest.

\section{Empirical Results}\label{sec:results}

\paragraph{OLS.}
We begin our regression analysis by estimating the wage elasticity with respect to labor market tightness from OLS-based fixed effects regression, as specified in Equation~(\ref{eq:hdfe_baseline}). Table~\ref{tab:baseline_ols} presents the respective results for our four different specifications.\footnote{Full tables of the regression estimates are presented in Appendix~\ref{app:full_regression}.} Column (1) presents the regression with year, worker, and labor-market fixed effects. The elasticity turns out to be positive, implying that a doubling in labor market tightness (i.e., an increase by 100 percent) raises the daily gross wages of full-time workers by 0.8 percent. The effect size remains fairly constant when we additionally control for socio-demographic characteristics in Column (2) or firm fixed effects in Column (3). In Column (4), we present our most comprehensive OLS specification with socio-demographic controls and year, worker, labor-market, and firm fixed effects. Again, the effect remains fairly robust at an elasticity of 0.074. In sum, the OLS regressions point towards an elasticity of wages with respect to tightness in the range of 0.007-0.009, which hints towards a positive but rather small effect. Finally, note that all tightness effects are statistically significant at 1 percent levels.

\begin{table}[ht!]
\centering
\begin{center}
\caption{OLS Effects of Labor Market Tightness on Wages} \label{tab:baseline_ols}
\begin{center}
\begin{tabular}{lcccc}
\toprule
 & (1) & (2) & (3) & (4) \\
 & Log Wage & Log Wage & Log Wage & Log Wage \\
\midrule
\multirow{2}{*}{Log Tightness} & \hphantom{00}0.0079***\hphantom{0} & \hphantom{00}0.0088***\hphantom{0} & \hphantom{00}0.0065***\hphantom{0} & \hphantom{00}0.0074***\hphantom{0} \\
 & \hphantom{0}(0.0008)\hphantom{000} & \hphantom{0}(0.0006)\hphantom{000} & \hphantom{0}(0.0006)\hphantom{000} & \hphantom{0}(0.0005)\hphantom{000} \\
Year FE & yes & yes & yes & yes \\
Worker FE & yes & yes & yes & yes \\
Labor Market FE & yes & yes & yes & yes \\
Firm FE &  &  & yes & yes \\
Controls &  & yes &  & yes \\
\midrule
Observations \phantom{XXXXXX} &  8,584,726 &  8,584,726 &  8,454,953 &  8,454,953 \\
Clusters &     13,806 &     13,806 &     13,716 &     13,716 \\
\bottomrule
\end{tabular}
\end{center}
\begin{tablenotes}[para]
\small \item \textit{Notes:} The table displays OLS regressions of log daily wages of regular full-time workers on log labor market tightness. Control variables include binary variables for new hires, workplace location in Eastern Germany, three levels of professional education and continuous variables for age and squared age. Labor markets are combinations of detailed 3-digit occupations and commuting zones. Standard errors (in parentheses) are clustered at the labor market level: * = p$<$0.10. ** = p$<$0.05. *** = p$<$0.01. Including firm fixed effects reduces the number of observations and clusters due to singleton groups, which are excluded from the estimation. \textit{Sources:} Integrated Employment Biographies + Official Statistics of the German Federal Employment Agency + IAB Job Vacancy Survey, 2012-2022.
\end{tablenotes}
\end{center}
\end{table}

\paragraph{IV.}
As mentioned in Section~\ref{sec:model}, the OLS-based elasticities may be biased due to reverse causality or uncontrolled productivity shocks. To counteract this bias, we instrument log labor market tightness by the leave-one-out average of log tightness in all other regions but for the very same occupation and time period, as given by Equation~(\ref{eq:2}). Table~\ref{tab:baseline_iv} displays the IV results. Column (1) delivers an elasticity of 0.0124 from the most parsimonious specification with year, worker, and labor-market fixed effects only. The respective first-stage coefficient is significantly positive and large (0.89), corroborating that our leave-one-out instrument adequately predicts changes in labor market tightness. Favorably, the first stage is sufficiently strong, as indicated by an F-Statistic of the excluded instrument of 2,847. In Columns (2) and (3), we separately include socio-demographic controls and firm effects, which lead to slightly smaller coefficients. In our baseline specification in Column (4) with socio-demographic controls as well as year, worker, labor-market, and firm fixed effects, we arrive at a similar elasticity of 0.011, implying that an increase in labor market tightness by 100 percent raises daily wages of workers ceteris paribus by 1.1 percent. In this baseline specification, the first-stage coefficient is 0.88 and the F~Statistic is 2,740, thus lending credence to the plausibility of our IV approach. Finally, note that all IV effects are statistically significant at 1 percent levels.

\begin{table}[ht!]
\begin{center}
\caption{IV Effects of Labor Market Tightness on Wages} \label{tab:baseline_iv}
\begin{center}
\begin{tabular}{lcccc}
\toprule
 & (1) & (2) & (3) & (4) \\
 & Log Wage & Log Wage & Log Wage & Log Wage \\
\midrule
\multirow{2}{*}{Log Tightness} & \hphantom{00}0.0124***\hphantom{0} & \hphantom{00}0.0114***\hphantom{0} & \hphantom{00}0.0114***\hphantom{0} & \hphantom{00}0.0113***\hphantom{0} \\
 & \hphantom{0}(0.0018)\hphantom{000} & \hphantom{0}(0.0014)\hphantom{000} & \hphantom{0}(0.0015)\hphantom{000} & \hphantom{0}(0.0012)\hphantom{000} \\
Year FE & yes & yes & yes & yes \\
Person FE & yes & yes & yes & yes \\
Labor Market FE & yes & yes & yes & yes \\
Firm FE &  &  & yes & yes \\
Controls &  & yes &  & yes \\
\midrule
Observations \phantom{XXXXXX} &  8,584,317 &  8,584,317 &  8,454,543 &  8,454,543 \\
Clusters &     13,783 &     13,783 &     13,692 &     13,692 \\
\midrule
First-Stage Coefficient & \hphantom{00}0.8914***\hphantom{0} & \hphantom{00}0.8915***\hphantom{0} & \hphantom{00}0.8814***\hphantom{0} & \hphantom{00}0.8814***\hphantom{0} \\
\makecell[l]{First-Stage F-Statistic \\of Excluded Instrument} & \makecell[c]{     2,847} & \makecell[c]{     2,848} & \makecell[c]{     2,737} & \makecell[c]{     2,740} \\
\bottomrule
\end{tabular}
\end{center}
\begin{tablenotes}
\small \item \textit{Notes:} The table displays IV regressions of log real daily wages of regular full-time workers on log labor market tightness. The instrumental variable refers to leave-one-out averages of labor market tightness in all other commuting zones but for the same occupation and time period. Control variables include binary variables for new hires, workplace location in Eastern Germany, three levels of professional education, and continuous variables for age and squared age. Labor markets are combinations of detailed 3-digit occupations and commuting zones. Standard errors (in parentheses) are clustered at the labor market level: * = p$<$0.10. ** = p$<$0.05. *** = p$<$0.01. Including firm fixed effects reduces the number of observations and clusters due to singleton groups, which are excluded from the estimation. \textit{Sources:} Integrated Employment Biographies + Official Statistics of the German Federal Employment Agency + IAB Job Vacancy Survey, 2012-2022.
\end{tablenotes}
\end{center}
\end{table}

In sum, the elasticities of the IV estimation turn out to be somewhat larger than their OLS counterparts. The larger IV elasticities suggest that the leave-one-out instrument successfully addresses reverse causality, which manifests in downward-biased OLS estimates.

\paragraph{Productivity Shocks at the National Level.}

While our instrument counteracts reverse causality and omitted variable bias from local productivity shocks, it is not immune against occupational productivity shocks at the national level. To assess the magnitude of this potential upward bias, we perform four robustness checks with differently rigorous proxies for national productivity shocks in Table~\ref{tab:productivity_shocks}.\footnote{Full regression tables including the coefficients of the control variables are presented in Appendix Table ~\ref{tab:productivity_shocks_full}.}

In Column (1), we address productivity shocks by occupations at the national level by additionally controlling for occupation-by-year fixed effects in our preferred OLS specification.\footnote{Note that occupation-by-year fixed effects may not only absorb confounding productivity shocks but also absorb desired mediating variation in tightness from occupation-wise shocks to vacancies and job seekers at the national level.} By virtue of the occupation-by-year fixed effects, our OLS estimate slightly decreases from 0.0074 to 0.0060, indicating only little omitted variable bias from national productivity shocks.
In Column (2), we directly control for productivity, albeit not at the occupational but at the firm level. To do so, we build on the sub-sample of IEB workers whose firms were interviewed in the IAB Establishment Panel. The combination of both datasets forms the so-called ``Linked-Employer-Employee Dataset of the IAB'' (LIAB), which allows us to calculate firm productivity by dividing annual revenues by headcount employment. When controlling for log firm productivity at the worker level, the IV elasticity slightly decreases from 0.011 to 0.010.\footnote{As surveyed firms in the IAB Establishment Panel may constitute a non-random sample of firms, we employ survey weights when estimating the effect of labor market tightness on wages using the LIAB sample. The baseline effects in the LIAB sample are 0.0064 (OLS) and 0.0122 (IV), that is, in close proximity to the baseline effects of the 5 percent IEB sample.}
In Column (3), we alternatively include industry-by-year fixed effects in our IV specification. As a consequence, our baseline IV elasticity shrinks from 0.011 to 0.006.
In Column (4), we perform the most rigorous approach to address productivity shocks in our IV setting by controlling for the sum of vacancies in all other regions but for the very same occupation and time period. In doing so, we ensure that the identifying variation purely stems from variation in the supply side of the labor market and arrive at a significantly positive elasticity of 0.004.

\begin{table}[ht!]
\centering
\begin{center}
\caption{Addressing Productivity Shocks at the National Level} \label{tab:productivity_shocks}
\begin{center}
\begin{tabular}{lcccc}
\toprule
 & (1) & (2) & (3) & (4) \\
 & Log Wage & Log Wage & Log Wage & Log Wage \\
\midrule
\multirow{2}{*}{Log Tightness}                                         & \hphantom{00}0.0060***\hphantom{0} & \hphantom{00}0.0101***\hphantom{0} & \hphantom{00}0.0057***\hphantom{0} & \hphantom{00}0.0044***\hphantom{0} \\
                                                                                                       & \hphantom{0}(0.0004)\hphantom{000} & \hphantom{0}(0.0028)\hphantom{000} & \hphantom{0}(0.0011)\hphantom{000} & \hphantom{0}(0.0017)\hphantom{000} \\
\multirow{2}{*}{Log Firm Productivity}                         & - & \hphantom{00}0.0214***\hphantom{0} & - & - \\
                                                                                                       & - & \hphantom{0}(0.0021)\hphantom{000} & - & - \\
\multirow{2}{*}{Log Leave-One-Out Sum V}                       & - & - & - & \hphantom{00}0.0114***\hphantom{0} \\
                                                                                                       & - & - & - & \hphantom{0}(0.0016)\hphantom{000} \\
Year FE & yes & yes & yes & yes \\
Worker FE & yes & yes & yes & yes \\
Labor Market FE & yes & yes & yes & yes \\
Firm FE & yes & yes & yes & yes \\
Occupation-by-Year FE & yes & & & \\
Industry-by-Year FE & & & yes & \\
Controls & yes & yes & yes & yes \\
\midrule
Source & 5\% IEB & LIAB & 5\% IEB & 5\% IEB \\
Survey Weights & & yes & & \\
\midrule
Observations \phantom{XXXXXX} &  8,454,869 &  3,117,429 &  8,454,541 &  8,454,541 \\
Clusters &     13,698 &      8,865 &     13,692 &     13,692 \\
\midrule
First-Stage Coefficient &  & \hphantom{00}0.8569***\hphantom{0} & \hphantom{00}0.8558***\hphantom{0} & \hphantom{00}0.7920***\hphantom{0} \\
\makecell[l]{First-Stage F-Statistic \\of Excluded Instrument} & \makecell[c]{} & \makecell[c]{     2,098} & \makecell[c]{     2,679} & \makecell[c]{     1,629} \\
\bottomrule
\end{tabular}
\end{center}
\begin{tablenotes}[para]
\small \item \textit{Notes:} The table displays OLS and IV regressions of log real daily wages of regular full-time workers on log labor market tightness. The instrumental variable refers to leave-one-out averages of labor market tightness in all other commuting zones but for the same occupation and time period. Control variables include binary variables for new hires, workplace location in Eastern Germany, three levels of professional education, and continuous variables for age and squared age. Labor markets are combinations of detailed 3-digit occupations and commuting zones. Column (1) shows OLS estimates when additionally controlling for detailed 3-digit-occupation-by-year fixed effects. Column (2) features IV estimates when controlling for log firm productivity in the LIAB worker sample. Column (3) displays IV estimates when controlling for 1-digit industry-by-year fixed effects (based on the NACE 2.0 classification). Column (4) presents IV estimates when conditioning on the log leave-one-out sum of the number of vacancies in all other commuting zones but for the very same occupation and time period. Standard errors (in parentheses) are clustered at the labor market level: * = p$<$0.10. ** = p$<$0.05. *** = p$<$0.01. \textit{Sources:} Integrated Employment Biographies + Official Statistics of the German Federal Employment Agency + IAB Job Vacancy Survey + Linked Employer-Employee Dataset of the IAB, 2012-2022.
\end{tablenotes}
\end{center}
\end{table}

In line with the theoretical expectation of an upward bias from national (occupation-specific) productivity shocks, the inclusion of each of the four productivity proxies results in lower elasticities than our baseline IV elasticity of 0.011, which we view as an upper bound of the causal effect of tightness on wages. We choose a conservative approach and view the elasticity of 0.004 from the most rigorous check, that is, when controlling for the leave-one-out sum of vacancies, as lower bound. In the following, since it is \textit{a priori} unclear which of our three proxies performs best (i.e., productivity is controlled for without eliminating other useful variation in tightness), we proceed with reporting robustness checks and heterogeneity analyses only for our baseline IV specification representing the upper bound.

\paragraph{Further Robustness Checks.}

\begin{figure}[ht!]
\centering
\begin{center}
\caption{Robustness Checks}
\label{fig:robustness}
\begin{center}
\includegraphics[scale=1.0]{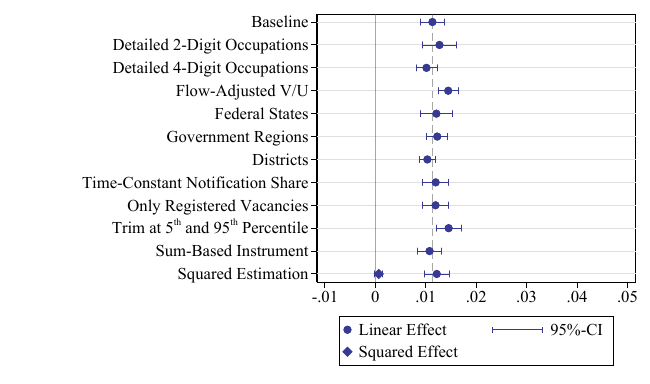}
\end{center}
\begin{tablenotes}
\small \item \textit{Notes:} The figure displays estimated elasticities and 95\% confidence intervals from IV regressions of log real daily wages of regular full-time workers on log labor market tightness for a variety of specifications to test the sensitivity of our baseline IV effect. Each specification includes fixed effects for years, workers, labor markets, and firms as well as control variables. \textit{Sources:} Integrated Employment Biographies + Official Statistics of the German Federal Employment Agency + IAB Job Vacancy Survey, 2012-2022.
\end{tablenotes}
\end{center}
\end{figure}

In addition to addressing the role of productivity shocks, we test the robustness of our baseline instrumental variable effect of tightness on wages in various other dimensions, as illustrated in Figure~\ref{fig:robustness}.

First, we use a broader definition of occupational labor markets using the leading two (rather than three) digits (``occupational main group'') along with the fifth digit (``requirement level'') to form 132 ``detailed 2-digit occupations''. Second, we use a narrower definition of occupational labor markets and exploit all five digits of the occupational classification, resulting in labor markets for 1,300 different occupational types (``detailed 4-digit occupations''), which leads to a total number of 36,596 occupational labor markets. Third, we build a flow-adjusted measure of occupational labor market tightness to account for the fact that individuals can search for a job in a neighboring occupation (see Appendix~\ref{app:flow-adjusted}). Specifically, we additionally include vacancies and job seekers in all other detailed 3-digit occupations with weights that refer to transition probabilities from moving from the focal to the neighboring detailed 3-digit occupation.

Fourth, we rely on an administrative rather than a functional delineation of regions by using 16 federal states, which allows for the possibility that individuals search for a job at a longer regional distance. Fifth, we use another administrative delineation of 38 government regions. Sixth, we employ 400 administrative districts to delineate regions at a very narrow level.

Seventh, we construct the total number of vacancies per labor market by using time-constant instead of time-varying registration shares from the IAB Job Vacancy Survey (see Appendix Table~\ref{tab:notification_share}). While this robustness check does not account for temporal variation in registration shares with the Federal Employment Agency, it may reduce potential measurement errors in the time-varying registration shares over time. Eighth, we solely rely on administratively registered vacancies without calculating the total number of vacancies from registration shares of the IAB Job Vacancy Survey.

Ninth, we trim the dependent variable and the explanatory variable below the 5\textsuperscript{th} and above the 95\textsuperscript{th} percentile to ensure that the result is not driven by outliers. Tenth, we use our sum-based leave-one-out instrument, as given by Equation~(\ref{eq:3}), to ensure that larger labor markets are attributed more importance in the leave-one-out calculation.

\begin{figure}[ht!]
\centering
\caption{Conditional Relationship between Labor Market Tightness and Wages}
\label{fig:binscatter_w_vu_cond}
\begin{center}
\includegraphics[scale=0.95]{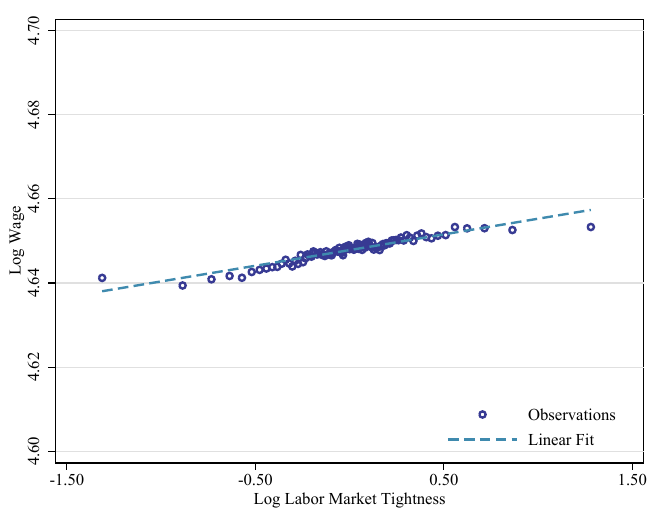}
\end{center}
\vspace{-0.5cm}
\begin{tablenotes}
\small \item \textit{Notes:}
The figure shows a binned scatterplot with 100 markers to illustrate the correlation between log labor market tightness (instrumented by the leave-one-out average) and log real daily wages, after conditioning on control variables. \textit{Sources:} Integrated Employment Biographies $\plus$ Official Statistics of Federal Employment Agency + IAB Job Vacancy Survey, 2012-2022.
\end{tablenotes}
\end{figure}

Eleventh, we estimate an alternative specification with an additional quadratic effect of tightness on wages. While we do not have a theoretical expectation for the functional form of the relationship, this specification is empirically motivated by our descriptive scatter plot in Figure~\ref{fig:binscatter_w_vu} that suggested a concave relationship. We use the same instrumental variable specification as in our baseline regression but add a quadratic term for labor market tightness along with the squared instrumental variable. The results, as illustrated in the last row of Figure~\ref{fig:robustness}, show an unchanged linear effect and a quadratic effect which is virtually zero, rejecting the hypothesis of a quadratic effect pattern. Since the absence of a quadratic relationship contradicts our initial visual inspection, we replicate the graphical illustration after conditioning on observed and unobserved heterogeneity and after accounting for the instrumentation. Figure~\ref{fig:binscatter_w_vu_cond} presents the respective binned scatter plot, which -- in line with the regression -- shows a fairly linear relationship between tightness and wages. In addition, the graphical illustration indicates that the slope is much flatter (compared with the raw correlation) after conditioning for all the confounding forces, which is in line with our baseline interpretation of only a slightly positive wage elasticity.

In summary, the estimated coefficients of our eleven robustness checks turn out to be very similar to our baseline elasticity of 0.011.

\paragraph{Interpretation of Effect Size.} Using OLS and IV regressions, we provide an interval for the causal effect of log labor market tightness on log wages between 0.004 (lower bound) and 0.011 (upper bound). Although the upper bound exceeds the lower bound by factor 2.6, both values imply that there were positive but rather limited wage increases from tremendously rising tightness in Germany. Specifically, the bounded interval implies that the observed increase in aggregate tightness in the German economy between 2012 and 2022 by 133.3 percent (see Figure~\ref{fig:tightness_by_year}) raised, on average, gross wages of regular full-time workers ceteris paribus by 0.6 to 1.5 percent. Against the backdrop that real wages grew by 7.9 percent during the same period (see Figure~\ref{fig:wages_by_year}), the increase in tightness can explain only between 7.4 and 19.1 percent of the wage rise in the German economy.\footnote{In absolute terms, real annual wages grew on average by 3,055 Euro between 2012 and 2022, of which the rise in tightness can explain between 209 and 569 Euro.}

Our results can help to calibrate the parameters in search-and-matching models. The wage-setting curve in the Diamond-Mortensen-Pissarides model propagates a positive linear effect of tightness (in levels) on the wage rate (in levels and normalized by productivity), which is the product of workers' relative bargaining power and vacancy-posting cost. Following standard practice, however, we have estimated elasticities of wages with respect to tightness using log-linear models. To translate log effects into level effects, we weight our upper- and lower-bound elasticities by the 2012 ratio of daily wages and aggregate tightness. After normalizing by average daily gross value added per worker, our bounds correspond to a coefficient for the tightness level between 0.013 and 0.032.\footnote{Using information from the German Statistical Agency from 2012 \citep{Destatis2013}, we calculate the average daily gross value added per worker by dividing the overall gross value added (2,363.9 billion Euro) by the workforce size (41.586 million workers) and 365 days per year.} Specifically, these bounds imply a relatively flat wage-setting curve and are markedly smaller than the calibrated value of 0.153 by \citet{Shimer2005}. This higher value can partly explain why the Shimer model generates more wage volatility than what empirical evidence suggests. By contrast, our bounded interval includes the calibrated value of 0.030 from \citet{Hagedorn2008}, whose model performs better in terms of observed wage volatility.\footnote{Building on Hosios' \citeyearpar{Hosios1990} rule, \citet{Shimer2005} assigns vacancy-posting cost a value of 0.213 (as a fraction of a worker's productivity). At the same time, he sets a comparatively high value of 0.72 for workers' relative bargaining power, thus specifying a relatively steep wage-setting curve. \citet{Hagedorn2008} choose higher vacancy-posting cost of 0.548 but assign workers a much lower relative bargaining power of 0.052, which mirrors a relatively flat wage-setting curve.}

Building on our theoretical considerations of Section \ref{sec:theory}, we want to highlight several explanations as to why the effect of tightness on wages turns out rather small. First, through the lens of the standard LS-LD framework, our limited wage effect can be attributed to relatively elastic (i.e., flat) labor supply or labor demand curves. When labor demand shifts rightward along the upward-sloping labor supply curve, the presence of strong substitution effects from leisure towards labor (which render labor supply relatively elastic) would require only small wage increases to clear the market.\footnote{Alternatively, when higher wages make workers reduce their supplied working hours (i.e., in the presence of strong income effects), firms are disincentivized to raise wages. Empirical evidence, however, suggests that income effects tend to be small \citep{Bargain2014,Bargain2016}.} Similarly, when labor supply shifts leftward along the labor demand curve, the presence of strong substitution or scale effects away from labor (which render labor demand relatively elastic) would lower the necessity for wage increases. Second, the limited wage effect can also be rationalized on the grounds of monopsony or bargaining models. In monopsony models, firms may maximize profits by voluntarily accepting lower employment levels (and forego revenues) to avoid costly wage increases. Relatedly, when firms command substantial bargaining power, the effect of higher tightness will not translate into markedly higher wages in the standard-DMP model.\footnote{However, monopsony power does not necessarily have only a moderating influence on the effect of labor market tightness on wages. Monopsony power also constitutes a mediating force when higher tightness counteracts monopsony power and thereby raises marked-down wages. In the textbook monopsony model, this channel manifests in a flattening labor supply curve to the single firm. In the standard-DMP model, this channel is ruled out by treating workers' and firms' relative bargaining power as exogenous.} Third, along the lines of the standard DMP model, firms with low vacancy posting cost (or search durations) have less incentive to increase wages to timely fill their vacancies.

\section{Heterogeneous Effects}\label{sec:heterogeneous}

\begin{figure}[ht!]
\centering
\begin{center}
\caption{Heterogeneous Effects}
\label{fig:het_effects}
\begin{center}
		\includegraphics[scale=1.0]{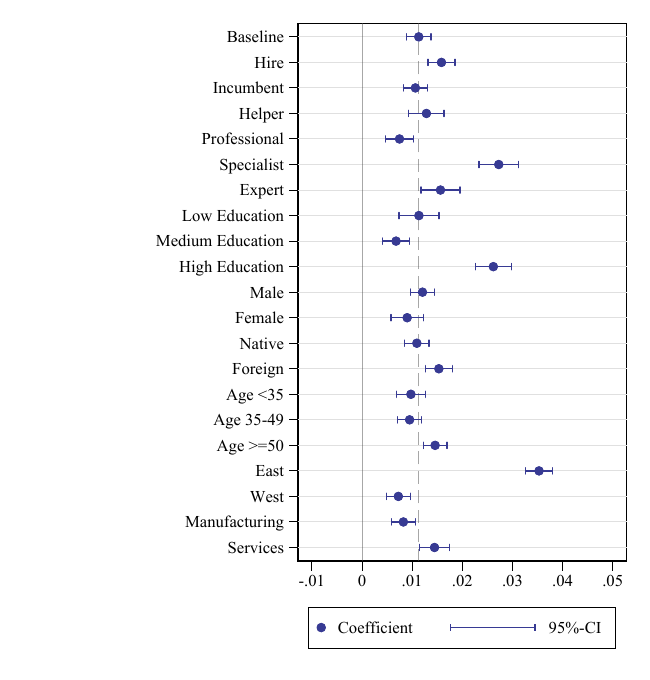}
\end{center}
\begin{tablenotes}
\small \item \textit{Notes:} The figure displays estimated elasticities and 95\% confidence intervals from IV regressions of log real daily wages of regular full-time workers on log labor market tightness to examine the heterogeneity regarding different characteristics (see Equation \ref{eq:hdfe_interacted}). Each specification includes fixed effects for years, workers, labor markets, and firms as well as control variables. \textit{Sources:} Integrated Employment Biographies + Official Statistics of the German Federal Employment Agency + IAB Job Vacancy Survey, 2012-2022.
\end{tablenotes}
\end{center}
\end{figure}

\paragraph{Subgroup Analysis.}
We estimate heterogeneous effects by interacting labor market tightness with the respective group dummies, as described by Equation~(\ref{eq:hdfe_interacted}). Figure~\ref{fig:het_effects} presents the elasticities from these subgroup analyses. The effect turns out markedly larger for new hires than for incumbent workers.\footnote{We define a hire as a worker who is employed in a certain firm on June 30 of year $t$ but not on June 30 in year $t-1$.} This result is in line with a number of studies that show that wages of job changers are significantly more flexible than the wages of incumbent employees \citep{Pissarides2009,Haefke2013, Bassanini2023}.

Next, we differentiate the effects by job requirement level. The effect for helpers roughly corresponds to the baseline effect, while the effect for professionals turns out markedly smaller. The effect for specialists is considerably larger, and the effect for experts is slightly larger than in the baseline. The heterogeneous effects by professional education level closely mirror this pattern: the effect for low-skilled workers is close to the baseline, whereas the effect for medium- and high-skilled workers are smaller and, respectively, larger than the baseline. Through the lens of the LS-LD framework, the larger effect for specialists, experts, and high-skilled can be explained by the lower substitutability of complex tasks (i.e., the labor demand curve for these workers is relatively inelastic/steep), whereas helpers and low-skilled workers perform easier tasks with more scope for substitution.
Labor demand theory also helps to reconcile why professionals and medium-skilled workers feature the smallest effects on wages. Due to the popularity of the dual vocational system in Germany, professionals and medium-skilled workers represent the largest groups in the German workforce. Thus, if wages for these large groups of workers rise, the wage bill increases by a relatively large margin. Consequently, labor demand for professionals and medium-skilled workers becomes relatively elastic due to large scale effects, limiting the scope for wage increases.\footnote{For Germany, \citet{Peichl2022} and \citet{Popp2023} show that demand for medium-skilled workers is more elastic than for low- and high-skilled workers due to relatively large scale effects.}

We observe only marginal differences between men and women, namely that women's wages react slightly less strongly to changes in labor market tightness than men's wages. In terms of nationality and age, the effects are slightly greater for foreigners and for workers over the age of 50.

For workers in Eastern Germany, our effect turns out almost four times as large as for workers in Western Germany. A leading explanation for this pattern is the observation that monopsony power is more pronounced in Eastern Germany \citep{Bachmann2022} and, thus, there is more scope for higher tightness to counteract markdowns. In the service sector, the effect is markedly higher than in the manufacturing sector. This difference seems plausible, as service occupations are relatively labor-intensive and, thus, can be less easily substituted by capital than occupations in the manufacturing sector, manifesting in a relatively inelastic labor demand curve.

\paragraph{Heterogeneous Effects along the Wage Distribution.}
In an additional heterogeneity analysis, we examine differences in the effect magnitude along the wage distribution (see Appendix~\ref{app:ddistribution} for further details). A large empirical literature established an increased wage inequality in Germany during the 1990s and 2000s \citep{Dustmann2009, Dustmann2014,Goldschmidt2017,Antonczyk2009}. Remarkably, wage inequality began declining from 2010 onwards, particularly due to wage increases at the bottom of the wage distribution \citep{Bloemer2023,DrechselGrau2022,Fedorets2020}, which can be partly explained by the introduction of the statutory minimum wage \citep{Bossler2023}. In addition to the effects of the minimum wage, the enormous tightening of the labor market could also have contributed to falling wage inequality. This inequality-reducing effect was first highlighted for the U.S.\ by \citet{Autor2023} with data from after the Covid-19 pandemic.

We build on the work of \citet{Autor2023} and estimate the effect of labor market tightness on wages along the wage distribution. Specifically, we analyze the tightness elasticity along the worker-level wage distribution to identify potentially inequality-reducing effects. Moreover, we also analyze the respective elasticity along the firm-level wage distribution to identify the tightness response from the perspective of firm-level wages. We split the sample into decile groups of the (worker- and firm-level) wage distribution, which we then interact with labor market tightness and our leave-one-out instrument.\footnote{For the analysis of the worker wage distribution, we divide workers into ten equally-sized groups based on their real daily gross wage when they first appear in the IEB as full-time workers during 2012-2022. The use of predetermined wages ensures that workers do not switch decile groups during our period of analysis. For the analysis of the firm wage distribution, we divide workers into ten groups based on their firm's average real daily wage from the firm's first appearance during 2012-2022.}

\begin{figure}[ht!]
\centering
\begin{center}
\caption{Heterogeneous Effects along the Wage Distribution}
\label{fig:het_effects_wage_deciles}
\begin{center}
\includegraphics[scale=1.05]{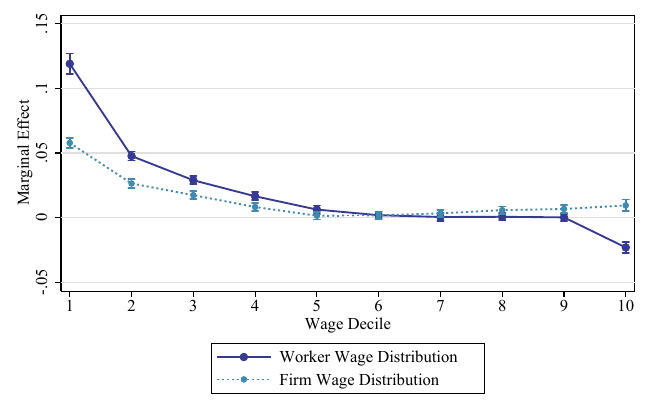}
\end{center}
\begin{tablenotes}
\small \item \textit{Notes:} The figure displays estimated elasticities and 95\% confidence intervals from IV regressions of log real daily wages of regular full-time workers on log labor market tightness for ten decile groups along the worker wage distribution (solid line) and the firm wage distribution (dashed line). Workers are assigned into ten decile groups based on their real daily gross wage when they first appear in the IEB as full-time workers during 2012-2022. Firms are assigned into ten decile groups based on the average real daily wage of their full-time workers when the firm first appears in the IEB sample during 2012-2022. Both curves refer to the baseline specification with fixed effects for years, workers, labor markets, and firms as well as control variables. \textit{Sources:} Integrated Employment Biographies + Official Statistics of the German Federal Employment Agency + IAB Job Vacancy Survey, 2012-2022. \newline
\end{tablenotes}
\end{center}
\end{figure}

Figure~\ref{fig:het_effects_wage_deciles} illustrates the estimated tightness elasticities along the wage distribution.\footnote{The corresponding regression tables and first-stage estimates are shown in Appendix Tables \ref{tab:het_effects_wage_deciles} and \ref{tab:het_effects_wage_deciles_fs}.} The estimates represent the interacted baseline IV specification, including control variables, fixed effects for years, workers, labor markets, and firms. The solid line displays the elasticities for each decile group of the workers' wage distribution. An increase in labor market tightness by 100 percent raises the wages of workers in the lowest decile group by roughly 12 percent. This effect rapidly falls along workers' wage distribution. The effects for the sixth up to the ninth decile group of workers are still slightly positive but no longer statistically different from zero, while the effect in the tenth and highest decile group of workers turns out negative.\footnote{Note that the negative effect at the very top of the distribution could be driven by the top-coding of wages at the social security limit above which we impute wages (see Section~\ref{sec:data}) or mean reversion.}\footnote{Our finding that low-wage workers benefit the most from rising tightness does not contradict the results from our subgroup analysis, which finds lower elasticities for helpers and professionals than for specialists and experts. Appendix Figure~\ref{fig:level_by_wage_decile} shows that helpers and professionals are not only represented at the lower end of the wage distribution. In particular, professionals account for more than fifty percent of workers in each of the bottom eight decile groups while still accounting for more than 40 and 20 percent in the top two decile groups, respectively. Against the backdrop of this pattern, our analysis by workers' decile groups suggests that those workers who earn comparatively low wages within their requirement level benefit most from the increase in labor market tightness. Moreover, when simultaneously differentiating our effects by workers' decile groups and requirement level (see Figure~\ref{fig:level_by_wage_decile_level_interaction}), it turns out that, among the lowest decile groups, specialists and experts benefit more strongly from rising tightness than helpers or professionals.}

The dashed line depicts the tightness elasticities for different firm-level wage decile groups. The pattern largely resembles the pattern from the wage distribution of workers: An increase in labor market tightness by 100 percent raises average wages of firms in the lowest decile group by roughly 6 percent. The effect stays significantly positive in the second, third, and fourth decile group of firms but turns insignificant in the middle of the wage distribution of firms. Unlike for workers, we observe a slightly positive effect in the top three decile groups of firms.

Taken together, the positive wage effects at the bottom of workers' wage distribution point towards an inequality-reducing effect of the rise in labor market tightness. During our period of analysis between 2012 and 2022, real wage dispersion declined by 24.0 percent (when comparing average wages in the highest decile group with average wages in the lowest decile group, see Appendix Figure~\ref{fig:wage_growth_by_decile}). In order to determine what proportion of the decline in wage dispersion can be attributed to the increase in tightness, we calculate the development of wage dispersion at constant tightness. For each decile group, the counterfactual is calculated by subtracting the product of the wage elasticity with respect to tightness and the observed tightness increase. Of these 24.0 percent, about 12.2 percentage points can be ascribed to the increased tightness (see Appendix Figure~\ref{fig:wage_growth_by_decile_counterfactual}). Our finding that primarily low-paying firms raised wages in response to rising tightness shows that the reduction in wage inequality might be driven by wage increases between (rather than within) firms. Overall, our results along the wage distribution lend credence to the notion that higher tightness improved the bargaining power of workers, raising marked-down wages in the subgroup of firms with wage-setting power but less so in the subgroup of firms that pay competitively.

\paragraph{Firm-Level Wage Setting.}

Our finding that primarily low-paying firms raised wages in response to tightness prompts the question of the role of firms in the wage-setting process. Firms explain a large share of the total variation in wages, as they pay a firm-specific premium through rent sharing \citep{VanReenen1996} or a match-specific premium \citep{Abowd1999,Card2013}. In addition, firm-level institutions, such as collective bargaining, are reducing wage dispersion \citep{Freeman1982}, in particular within firms. From our analysis of effects along the wage distribution, we observe pronounced wage increases in low-paying firms, but it remains open whether these firms differentiate between different occupational groups within the same workplace when setting wages. This culminates in the question of whether firms only raise wages in response to an increasing tightness of a certain occupational group or whether they pay all employees a raise, even if the occupation of the person in question has not experienced an increase in tightness but tightness of co-workers.

\begin{table}[ht!]
\begin{center}
\caption{IV Regressions of Wage Setting at the Firm Level} \label{tab:firm_wage_setting}
\begin{center}
\begin{tabular}{L{4.2cm}C{2.6cm}C{2.6cm}}
\toprule
 & (1) & (2) \\
 & Log Wage & Log $\overline{\text{Wage}}^{\text{Firm}}$ \\
\midrule
\multirow{2}{*}{Log Tightness}         & \hphantom{00}0.0024**\hphantom{00} & \hphantom{00}0.0088***\hphantom{0} \\
                                                                       & \hphantom{0}(0.0011)\hphantom{000} & \hphantom{0}(0.0012)\hphantom{000} \\
Year FE & & yes \\
Worker FE & yes & yes \\
Labor Market FE & yes & yes \\
Firm FE & & yes  \\
Firm-by-Year FE & yes & \\
Controls & yes & yes \\
\midrule
Observations \phantom{XXXXXX} &  6,424,250 &  8,454,541 \\
Clusters &     12,676 &     13,692 \\
\midrule
First-Stage Coefficient & \hphantom{00}0.8597***\hphantom{0} & \hphantom{00}0.8814***\hphantom{0} \\
\makecell[l]{First-Stage F-Statistic \\of Excluded Instrument} & \makecell[c]{     3,247} & \makecell[c]{     2,740} \\
\bottomrule
\end{tabular}
\end{center}
\begin{tablenotes}
\small \item \textit{Notes:} The table displays IV regressions of log (average firm-level) real daily wages of regular full-time workers on log labor market tightness. The instrumental variable refers to leave-one-out averages of labor market tightness in all other commuting zones but for the same occupation and time period. Control variables include binary variables for new hires, workplace location in Eastern Germany, three levels of professional education, and continuous variables for age and squared age. Labor markets are combinations of detailed 3-digit occupations and commuting zones. Column (1) shows IV estimates when additionally controlling for firm-by-year fixed effects. Column (2) displays IV estimates when regressing the firm-level average of real daily wages on log labor market tightness. Standard errors (in parentheses) are clustered at the labor market level: * = p$<$0.10. ** = p$<$0.05. *** = p$<$0.01. \textit{Sources:} Integrated Employment Biographies + Official Statistics of the German Federal Employment Agency + IAB Job Vacancy Survey, 2012-2022.
\end{tablenotes}
\end{center}
\end{table}

We take two approaches to address the question of firm-specific wage setting, both displayed in Table~\ref{tab:firm_wage_setting}. First, we re-examine the baseline wage effect (as displayed in Column (4) of Table~\ref{tab:baseline_iv}) and further include firm-by-year fixed effects. These fixed effects eliminate the average wage change that occurs within firms over time. Hence, the resulting effect only captures variation from differential tightness increases of different occupations within a firm. Second, we estimate the effect of tightness on the average wage of all workers in a firm.\footnote{Note that we partial out an individual-level fixed effect before averaging wages at the firm level. This procedure ensures that we control for the same granularity of individual heterogeneity as in the baseline estimation.} In doing so, we examine whether all the co-workers benefit from a tightness increase even if the respective coworkers' tightness does not itself change.

Upon inclusion of firm-by-year fixed effects, our IV effect drastically shrinks in Column (1), indicating that firms hardly differentiate between occupations when the employed occupations experience a differential tightness increase. For example, a manager in a firm receives only a slightly higher pay growth than his or her secretaries if the tightness of managers increases compared to those of secretaries. In line, the use of firm-level averages of wages in Column (2) demonstrates that the average wage of all workers rises almost as much as the individual worker's wage when labor markets tighten. In our example, it implies that the secretary also benefits from the tightness of managers.

\section{Conclusion} \label{sec:conclusion}

Combining exceptionally detailed information on vacancies and job seekers with administrative employment data, we estimate the effect of increased scarcity of the labor input on workers' compensation in Germany for the years 2012 to 2022. Using high-dimensional fixed effects and a new leave-one-out instrument, our empirical analysis delivers a positive upper-bound elasticity of wages with respect to labor market tightness of 0.011. Controlling for potentially confounding (occupation-specific and nationwide) productivity shocks, we arrive at a lower-bound elasticity of 0.004. Taken together, we conclude that labor market tightness has a statistically significant but modest impact on wages. Our estimated effect range implies that the observed increase in labor market tightness by around 130 percent between 2012 and 2022 explains between 7 and 19 percent of the rise in average real wages during the same period.

However, our results point out several interesting heterogeneities which reveal somewhat stronger wage effects. We find particularly large elasticities for newly hired workers, high-skilled workers, the Eastern German labor market, and the service sector. We also observe that wage increases are significantly stronger at the lower end of the wage distribution, which is mainly driven by wage increases from low-paying firms. At the same time, the effects are virtually zero above the fourth decile of wages. Thereby, we document an inequality-reducing effect of higher labor market tightness, thus corroborating evidence from \citet{Autor2023} for the U.S. labor market.

Our results provide some important implications. First, our positive but limited tightness effect mirrors a relatively flat wage-setting curve. Specifically, our upper- and lower-bound log-log elasticities translate into linear coefficients in the range between 0.013 and 0.032, and these relatively low values may guide researchers to their calibrate search-and-matching models. Second, to put our wage effects into perspective, it is worth exploring whether the pay increases were sufficient to overcome firms' hiring frictions that usually arise in tight labor markets \citep{LeBarbanchon2024}. Building on the same data, \citet{Bossler2024} find that, holding all other things equal, the doubling in tightness reduced firms' employment growth on average by 5 percent between 2012 and 2019. When abstracting that our time horizon is three years longer, our estimated wage increases between 0.6 and 1.5 percent (along with potential improvements in non-monetary job amenities) in response to higher tightness were seemingly not strong enough to maintain firms' employment growth. Third, since we do not find any indications of non-linearities in the tightness elasticity, our estimates can provide tentative guidance for what will happen when the German labor market keeps tightening in the upcoming years due to intensifying demographic decline: namely tangible but limited wage increases (despite muted employment growth) that further reduce wage inequality at the bottom end of the wage distribution.


\clearpage
\printbibliography[heading=bibintoc] 

\clearpage

\clearpage
\begin{appendix}
\begin{refsection} 

\renewcommand\thetable{\thesection\arabic{table}} 
\renewcommand\thefigure{\thesection\arabic{figure}} 
\setcounter{figure}{0}
\setcounter{table}{0}

\pdfbookmark[0]{Appendix}{appendix} 

\begin{center}

\vspace*{1cm}

\Large
Appendix \\
\textbf{Scarce Workers, High Wages?}

Erik-Benjamin Börschlein, Mario Bossler, and Martin Popp

\normalsize


\vspace*{1cm}


\startcontents[sections] 
\printcontents[sections]{l}{1}{\setcounter{tocdepth}{2}} 

\end{center}

\clearpage

\counterwithin{equation}{section}

\clearpage
\setcounter{figure}{0}
\setcounter{table}{0}
\renewcommand*{\thefigure}{\thesection\arabic{figure}}
\renewcommand*{\thetable}{\thesection\arabic{table}}
\section{Data on Vacancies: Further Details}
\label{app:data_details}

The official statistics of the German Federal Employment Agency (FEA) collect information on registered vacancies. Registered vacancies refer to those job openings that firms have forwarded to the FEA to facilitate the recruitment of suitable candidates. Upon registration, the firms have to state their targeted 5-digit occupation and the district of the workplace. Vacancies that have not been registered with the FEA, such as those posted exclusively through other channels like newspapers or private online job boards, are not included in the Offical Statistics of the Federal Employment Agency.

To additionally take into account unregistered vacancies, we leverage additional information from the IAB Job Vacancy Survey (IAB-JVS). The IAB Job Vacancy Survey is a quarterly business survey that focuses on labor demand and recruitment practices. In the last quarter of the year, about 15,000 firms disclose the number and structure of their vacancies. Firms are requested to distinguish between their registered and unregistered vacancies and further categorize them based on the requirement level of the associated jobs. Importantly, this information allows us to construct the share of registered vacancies in all vacancies separately by requirement level. As the survey only began differentiating between specialists and experts in 2015, we pool this information and compute notification shares for specialists and experts as a whole.

Table~\ref{tab:notification_share} of this appendix illustrates the evolution of notification shares by requirement level. The notification shares exhibit some temporal variation and generally decline with an increasing level of requirement. While almost half of the vacancies for helper occupations are registered with the FEA, the notification share of vacancies for professionals tends to be slightly lower. For specialists and experts, less than one in three vacancies is registered with the Federal Employment Agency.

To quantify the overall number of registered and unregistered vacancies in a given labor market and year, we make use of the above-mentioned sources and follow the approach from \citep{Bossler2024}. First, we extract the number of registered vacancies for each combination of 5-digit occupation, district, and year from the Federal Employment Agency's Official Statistics. Following standard practice of the FEA, we apply the following filters when drawing these data: we exclude vacancies with an employment duration of less than seven days, subsidized vacancies, vacancies for freelancers, and vacancies from private employment agencies. Second, we divide the resulting number of registered vacancies for each labor market and year by the corresponding notification share. When dividing by the yearly notification share, we use the requirement level (i.e., the fifth digit of the KldB occupation variable) to differentiate between the notification shares of helpers, professionals, and specialists along with experts. As a result, we arrive at the overall number of registered and unregistered vacancies for each combination of 5-digit occupation, district, and year. In the third and final step, we aggregate these yearly numbers for each combination of detailed 3-digit occupation and commuting zone to conform with our baseline definition of a labor market.

\begin{table}[ht!]
\centering
\begin{center}
\caption{Shares of Registered Vacancies in All Vacancies by Requirement Level} \label{tab:notification_share}
\begin{center}
\begin{tabular}{C{2.5cm}C{3.25cm}C{3.25cm}C{3.25cm}} \hline
\multirow{2.4}{*}{Year} & \multirow{2.4}{*}{Helpers} & \multirow{2.4}{*}{Professionals} & \multirow{2.4}{*}{\shortstack{Specialists \\ and Experts}} \\
& & &   \\[0.2cm] \hline
& & &   \\[-0.2cm]
2012 & 36.0 & 45.0 & 33.6  \\[0.1cm]
2013 & 44.2 & 47.0 & 25.7  \\[0.1cm]
2014 & 48.0 & 41.9 & 29.9  \\[0.1cm]
2015 & 48.1 & 46.5 & 29.3  \\[0.1cm]
2016 & 53.4 & 50.5 & 36.6  \\[0.1cm]
2017 & 52.3 & 46.4 & 31.1  \\[0.1cm]
2018 & 44.0 & 46.2 & 32.7  \\[0.1cm]
2019 & 43.5 & 41.9 & 30.4  \\[0.1cm]
2020 & 45.0 & 39.1 & 26.0  \\[0.1cm]
2021 & 46.5 & 43.0 & 27.3  \\[0.1cm]
2022 & 43.1 & 41.8 & 26.9  \\[0.1cm] \hline
& & &   \\[-0.2cm]
2012-2022 & 45.8 & 44.5 & 29.9  \\[0.2cm] \hline
\end{tabular}
\end{center}
\begin{tablenotes}[para]
\small \item \textit{Notes:} The table shows the yearly percentage shares of registered vacancies in all vacancies, separately by requirement levels (i.e., the fifth digit of the KldB-2010 occupation variable). Helper occupations require no training or only a maximum of one year’s training. The group of professionals includes all activities with industrial, commercial or other vocational training (excluding master craftsmen and technicians). Specialist occupations necessitate a bachelor degree or the completion of master craftsman/technician training. Experts hold a master degree or an equivalent diploma. \textit{Source:} IAB Job Vacancy Survey, 2012-2022.
\end{tablenotes}
\end{center}
\end{table}

\clearpage
\setcounter{figure}{0}
\setcounter{table}{0}
\renewcommand*{\thefigure}{\thesection\arabic{figure}}
\renewcommand*{\thetable}{\thesection\arabic{table}}
\section{Delineation of Commuting Zones}
\label{app:lm_delineation}

The definition of a region should mirror the spatial clustering of economic activities as precisely as possible. A ``functional region'' is defined as a cluster of neighboring areas in which a large fraction of economic activities like commuting and trade occurs among resident workers and businesses. Thus, functional regions quite accurately capture the spatial dimension of economic flows. By contrast, administrative delineations, such as districts which are formed by political boundaries, often fall short in capturing economic interactions well. Consequently, a large fraction of economic exchanges occurs across rather than within these administrative borders.

Relying on administrative regions can lead to a mismeasurement of labor market tightness. Assume that there are two cities with many people commuting from city 1 to work for firms in the neighboring city 2 (and vice versa). In such an environment, calculating labor market tightness in city 1 based on vacancies and job seekers only from city 1 mistakenly ignores that vacancies and job seekers from city 2 constitute additional outside options for workers and firms in city 1. When tightness in city 2 is lower (higher) than in city 1, the ratio of vacancies from city 1 to job seekers from city 1 will overestimate (underestimate) the true labor market tightness in city 1 (which must also reflect the vacancies and job seekers from the closely connected city 2).

To address this issue, we create functional labor market regions based on observed home-to-work commuting patterns. These commuting zones are designed to have relatively many connections within zones and relatively few connections between zones to minimize cross-border flows. We employ a graph-theoretical method developed by \citep{Kropp2016} to merge 401 administrative districts into more appropriate commuting zones. This method involves several steps: First, we calculate a matrix of bi-directional commuting flows among administrative regions and identify dominant flows between regions to guide potential mergers. Mergers occur when the dominant flow exceeds a certain threshold, resulting in a consolidated flow matrix. This process is iteratively repeated until further mergers are unnecessary. Second, we repeat the first step with varying threshold values to propose multiple meaningful delineations and select the delineation that delivers the highest modularity value. The concept of modularity is a popular measure to determine the degree of clustering in networks of connections. Specifically, the modularity concept compares the number of ties inside a cluster with the expected number of ties if the network of ties between clusters was random.\footnote{The value of modularity Q equals zero when the delineation does not perform better than a random delineation. Q approaches the maximum value of $Q = 1$ when the network itself is strongly modular (i.e., there are many ties within clusters) and was correctly delineated by the procedure. Values of Q usually range between 0.3 and 0.7.} Third, we ensure that all districts within the newly delineated commuting zone form a coherent area.

We implement this procedure using German FEA data on commuting patterns of the universe of contributory workers in the years 2012-2022. Our starting point refers to 401 administrative districts (i.e., 3-digit NUTS regions), which initially achieve a modularity value of 0.603. After gradually increasing the merger threshold, we find that a threshold of 8 percent yields generates the delineation with the highest modularity (Q = 0.838). After eight iterations of merging dominant flows above this threshold, we arrive at 52 commuting zones with strong internal interactions but limited connections between zones (see Appendix Figure~\ref{fig:commuting_zones}). The use of these commuting zones rather than districts reduces the share of commuters between regions from 39.0 to 10.8 percent.

\begin{figure}[ht!]
\centering
\begin{center}
\caption{Delineation of Commuting Zones}
\label{fig:commuting_zones}
\begin{center}
		\includegraphics[width=0.9\textwidth]{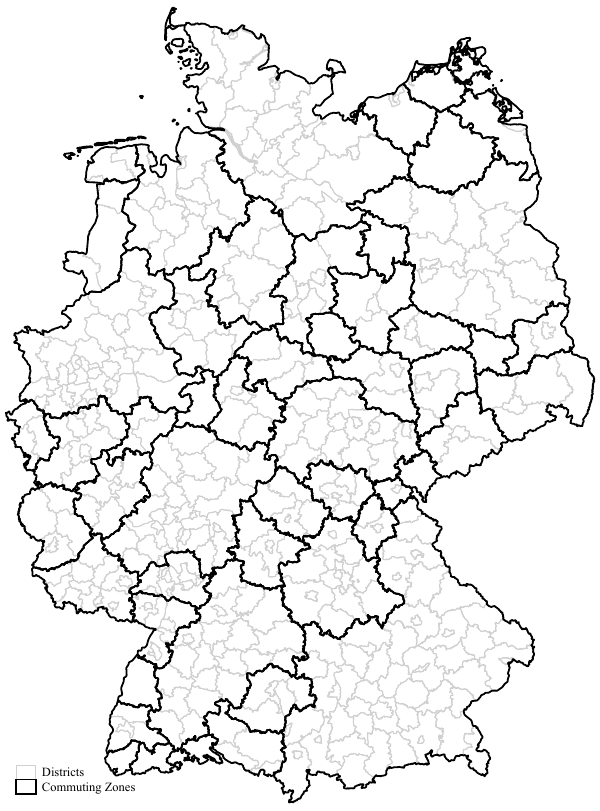}
\end{center}
\begin{tablenotes}
\small \item \textit{Notes:} The figure illustrates the delineation of commuting zones based on 401 German districts (NUTS-3 regions). The districts were merged into 52 commuting zones using the graph-theoretical method \citep{Kropp2016} and register data on German commuting patterns between 2012 and 2022. \textit{Source:} Official Statistics of the German Federal Employment Agency, 2012-2022.
\end{tablenotes}
\end{center}
\end{figure}

\clearpage
\setcounter{figure}{0}
\setcounter{table}{0}
\renewcommand*{\thefigure}{\thesection\arabic{figure}}
\renewcommand*{\thetable}{\thesection\arabic{table}}
\section{The Construction of Flow-Adjusted Labor Market Tightness}
\label{app:flow-adjusted}

When examining the effects of labor market tightness, researchers usually employ regions or, in rare circumstances, occupations to delineate the relevant labor market. However, in doing so, labor markets are divided into mutually exclusive segments. This definition rules out that vacancies and job seekers compete with similar vacancies or job seekers from a different labor market (i.e., a neighboring region or a closely related occupation) in the recruitment and search process, respectively. The mutually exclusive concept of a labor market may result in an overly narrow definition of labor markets but it could, in principle, also be too broad if workers (firms) only search (recruit) within a subsegment of the respective market.

By virtue of our functional delineation of 52 commuting zones (see Appendix~\ref{app:lm_delineation}), we already account for the fact that labor markets are functioning across administratively defined districts. By contrast, the occupational dimension of our baseline labor markets is based on 431 detailed 3-digit occupations, which stem from the German Classification of Occupations 2010 (KldB-2010). Note that the aim of this classification already is to group occupations with similar tasks into the same categories \citep{Dengler2011}. Nevertheless, the mutual exclusiveness of this delineation cannot rule out that workers and firms search and recruit from neighboring occupations.

To overcome this shortcoming, we implement a data-guided approach and take vacancies and job seekers outside the focal occupation as additional outside options into account. Specifically, we closely follow \citet{Bossler2024} and construct a flow-adjusted version of labor market tightness that builds on occupational mobility patterns to determine weights for vacancies and job seekers in neighboring occupations. Their approach is inspired by \citet{Arnold2021} who accounts for jobs in neighboring markets when calculating indices of labor market concentration.

The underlying idea of the flow adjustment is that the relative value of vacancies and job seekers in different occupations can be inferred from labor market flows within and between these occupations. Let $P(h|o)$ denote the probability that a worker in occupation $o$ in year $t$ is employed in occupation $h$ in year $t+1$. When recruiting (searching for a job in) occupation $o$, the firm's (worker's) relative value of a job seeker (vacancy) in occupation $h$ (compared to occupation $o$) then is:
\begin{equation}
\label{eq:G1}
\omega_{oh} = \frac{P(h|o)}{P(o|o)} \cdot \frac{L_{o}}{L_{h}}
\end{equation}
To calculate weights from these flows, it is necessary to consider that flows from one market to another depend on the relative size of the markets. Therefore, we normalize relative transition probabilities by employment in the respective occupations. Note that, by construction, the occupation in question always receives a weight of one, i.e., $\omega_{oo}=1$. When determining the transition probabilities with the administrative IEB data, we pool mobility patterns over labor market regions and the years 2012-2022 to arrive at a stable weighting matrix for mobility between all detailed 3-digit occupations.

Given the weighting matrix, the flow-adjusted number of vacancies in occupation $o$ and region $r$
\begin{equation}
\label{eq:G2}
\tilde{V}_{ort} = \sum_{h=1}^{H} \omega_{oh} V_{ort}
\end{equation}
is calculated as the weighted sum of vacancies in the same occupation and all other occupations in region $r$ and time $t$. By construction, the number of flow-adjusted vacancies always exceeds the number of actual vacancies in a labor market because the flow adjustment takes into account that workers can fill vacancies not only in the same occupation but also in neighboring occupations.

Similarly, the flow-adjusted number of job seekers in occupation $o$ and region $r$ is
\begin{equation}
\label{eq:G3}
\tilde{U}_{ort} = \sum_{h=1}^{H} \omega_{oh} U_{ort}
\end{equation}
which is the sum of job seekers in the same occupation and job seekers of all other occupations in region $r$ and time $t$. Again, the number of flow-adjusted job seekers always exceeds the number of actual job seekers in a labor market because the flow adjustment considers that firms can recruit job seekers not only from the same occupation but also from neighboring occupations.

Note that, when there are no flows between occupations, the weight is zero. In this case, the neighboring occupations do not constitute a relevant outside options and the flow-adjusted measure pins down to the baseline measure, which relies only on the vacancies and job seekers from the focal occupation. When there are random flows between occupations, all occupations receive the same weight. Then, the flow-based number of vacancies and job seekers collapses to the factual number of vacancies and job seekers in the commuting zone.

\clearpage
\setcounter{figure}{0}
\setcounter{table}{0}
\renewcommand*{\thefigure}{\thesection\arabic{figure}}
\renewcommand*{\thetable}{\thesection\arabic{table}}
\section{Descriptive Results: Further Evidence}
\label{app:description}

\begin{figure}[ht!]
\centering
\begin{center}
\caption{Labor Market Tightness by Job Requirement Level over Time}
\label{fig:tightness_by_level_year}
\begin{center}
\includegraphics[scale=1.0]{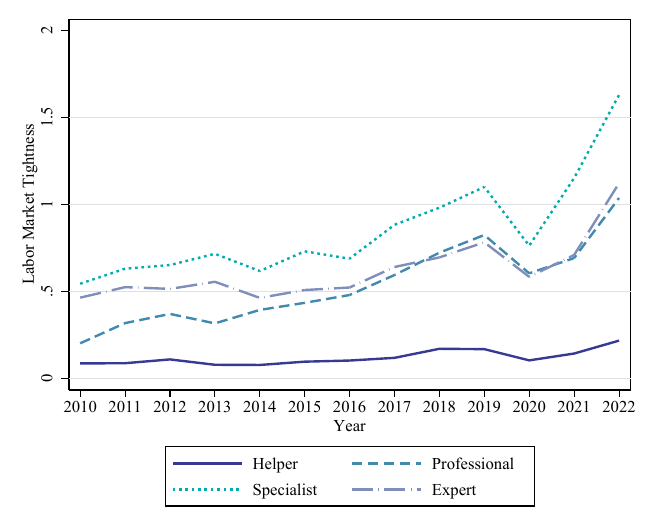}
\end{center}
\begin{tablenotes}
\small \item \textit{Notes:}
The figure illustrates the development of labor market tightness  in Germany over time by requirement levels (i.e., the fifth digit of the KldB-2010 occupation variable). \textit{Sources:} Official Statistics of the German Federal Employment Agency $\plus$ IAB Job Vacancy Survey, 2012-2022.
\end{tablenotes}
\end{center}
\end{figure}

\clearpage

\begin{figure}[ht!]
\centering
\begin{center}
\caption{Labor Market Tightness by 2-Digit Occupation}
\label{fig:tightness_by_kldb10_2}
\begin{center}
\includegraphics[scale=1.0]{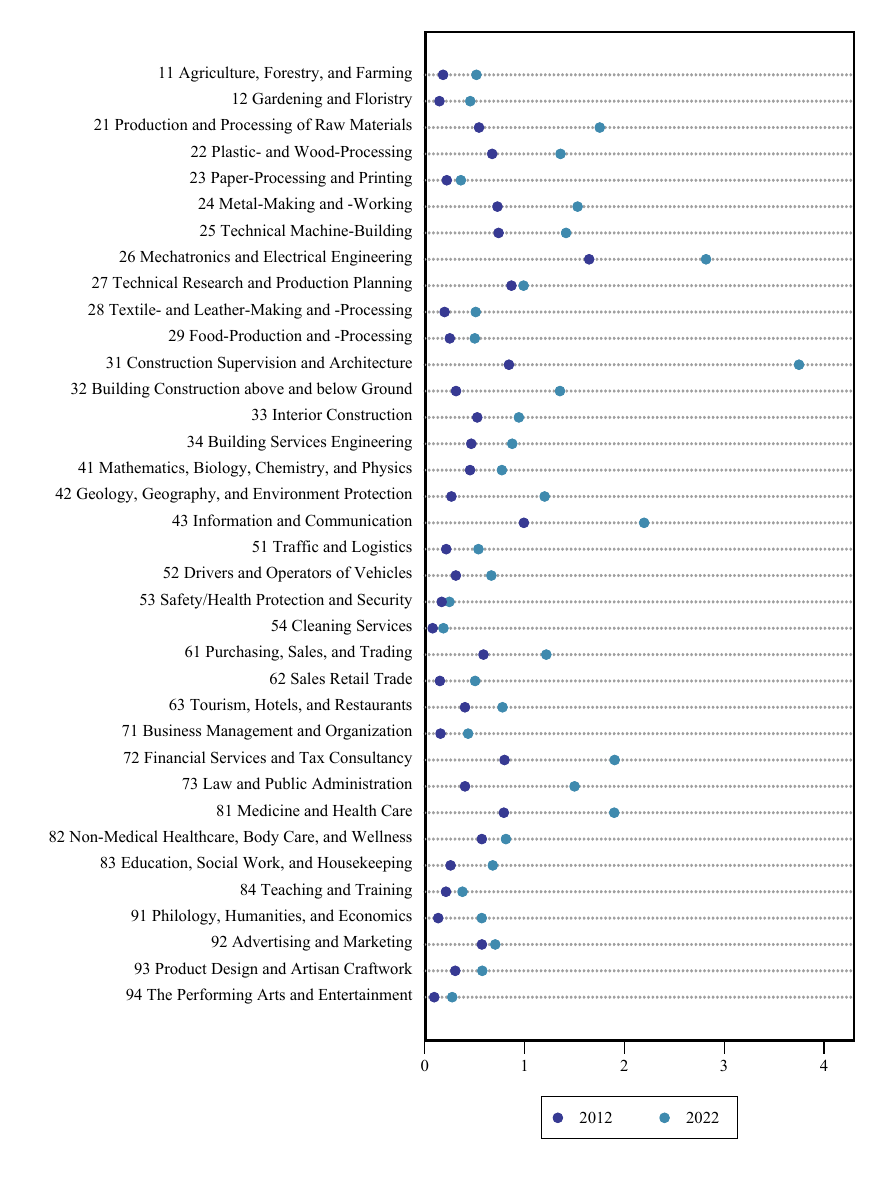}
\end{center}
\begin{tablenotes}
\small \item \textit{Notes:} The figure illustrates the development of labor market tightness  in Germany between 2012 and 2022 by 2-digit occupations. \textit{Sources:} Official Statistics of the German Federal Employment Agency $\plus$ IAB Job Vacancy Survey, 2012-2022.
\end{tablenotes}
\end{center}
\end{figure}

\clearpage

\begin{figure}[ht!]
\centering
\begin{center}
\caption{Labor Market Tightness by Commuting Zone}
\label{fig:tightness_by_zone}
\begin{center}
\includegraphics[scale=0.95]{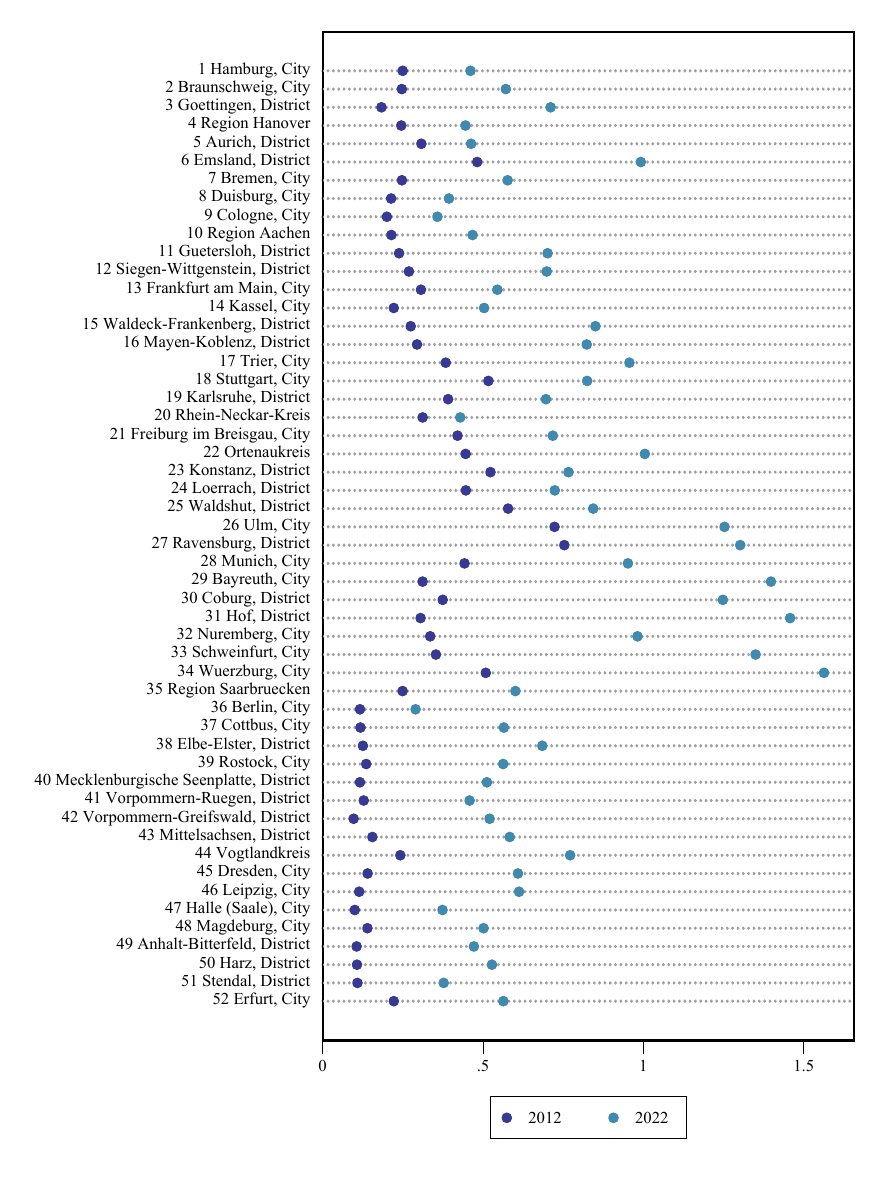}
\end{center}
\begin{tablenotes}
\small \item \textit{Notes:} The figure illustrates the development of labor market tightness  in Germany between 2012 and 2022 by commuting zones. Commuting zones are delineated using the graph-theoretical approach proposed by \citet{Kropp2016}. \textit{Sources:} Official Statistics of the German Federal Employment Agency $\plus$ IAB Job Vacancy Survey, 2012-2022.
\end{tablenotes}
\end{center}
\end{figure}

\clearpage

\begin{figure}[ht!]
\centering
\begin{center}
\caption{Nominal Wages in Germany over Time}
\label{fig:nominal_wages_by_year}
\begin{center}
\includegraphics[scale=1.0]{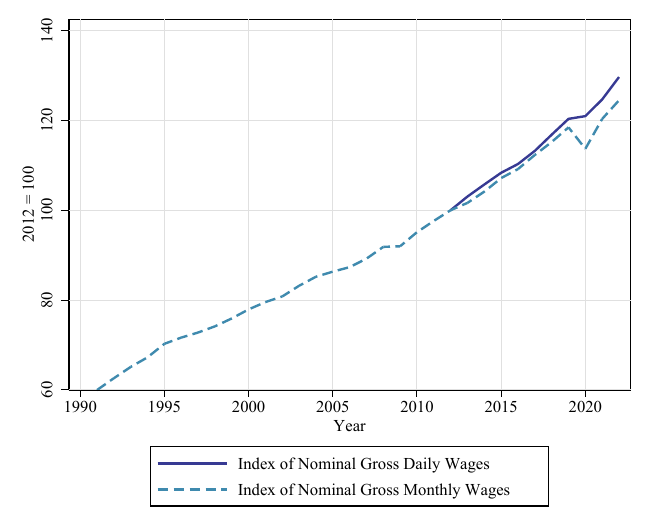}
\end{center}
\begin{tablenotes}
\small \item \textit{Notes:}
The figure illustrates the development of wages in Germany over time. The solid line refers to nominal gross daily wages (including special payments) for regular full-time workers in the non-agricultural private business sector. The dashed line shows the index of nominal gross monthly wages for full-time workers excluding special payments. The latter time series refers to the manufacturing sector until 2006 and to manufacturing and services from 2007 onwards. \textit{Sources:} Integrated Employment Biographies + Official Data of the German Federal Statistical Office, 1991-2022.
\end{tablenotes}
\end{center}
\end{figure}

\clearpage

\begin{table}[ht!]
\begin{center}
\caption{Descriptive Statistics} \label{tab:descriptives}
\begin{center}
{
\def\sym#1{\ifmode^{#1}\else\({#1}\)\fi}
\begin{tabular}{p{5.5cm}cccccc}
\toprule
 & Mean & P25 & P50 & P75 & SD & N \\
\midrule
Labor Market Tightness & 0.97 & 0.27 & 0.56 & 1.12 & 1.41 &  8,885,219  \\[0.1cm]
Gross Daily Wage (in 2012-Euro) & 118.06 & 73.16 & 100.50 & 143.43 & 69.12 &  8,907,310  \\[0.1cm]
Hire & 0.18 & - & - & - & 0.39 &  8,907,310  \\[0.1cm]
Job Requirement Level & & & & & \\[0.1cm]
\quad Helper & 0.12 & - & - & - & 0.33 &  8,907,310  \\[0.1cm]
\quad Professional & 0.58 & - & - & - & 0.49 &  8,907,310  \\[0.1cm]
\quad Specialist & 0.16 & - & - & - & 0.37 &  8,907,310  \\[0.1cm]
\quad Expert & 0.14 & - & - & - & 0.35 &  8,907,310  \\[0.1cm]
Education Level & & & & & \\[0.1cm]
\quad Low & 0.07 & - & - & - & 0.26 &  8,907,310  \\[0.1cm]
\quad Medium & 0.72 & - & - & - & 0.45 &  8,907,310  \\[0.1cm]
\quad High & 0.20 & - & - & - & 0.40 &  8,907,310  \\[0.1cm]
Female & 0.26 & - & - & - & 0.44 &  8,907,310  \\[0.1cm]
Foreign & 0.10 & - & - & - & 0.30 &  8,907,310  \\[0.1cm]
Age & 42.96 & 33.00 & 43.58 & 52.66 & 11.68 &  8,907,310  \\[0.1cm]
East Germany & 0.15 & - & - & - & 0.36 &  8,907,310  \\[0.1cm]
Manufacturing Sector & 0.45 & - & - & - & 0.50 &  8,907,310  \\[0.1cm]
\bottomrule
\end{tabular}
}
\end{center}
\begin{tablenotes}
\small \item \textit{Notes:} The table displays descriptive sample means, selected percentiles, standard deviations, and the numbers of observations of the relevant variables. \textit{Sources:} Integrated Employment Biographies + Official Statistics of the German Federal Employment Agency + IAB Job Vacancy Survey, 2012-2022.
\end{tablenotes}
\end{center}
\end{table}

\begin{table}[ht!]
\centering
\begin{center}
\caption{Descriptive Statistics of Main Variables by Subgroups} \label{tab:descriptive_wages_VU}
\begin{center}
\begin{small}
{
\def\sym#1{\ifmode^{#1}\else\(^{#1}\)\fi}
\begin{tabular}{p{5cm}c*{6}{c}}
\toprule
 & Mean & P50 & SD & Min & Max & N \\
\midrule
Gross Daily Wage (in 2012-Euro) & & & & & \\
\quad Overall & 118.06 & 100.50 & 69.12 & 0.01 & 1845.87 &  8,907,310  \\
\quad Hire & 96.70 & 80.58 & 58.69 & 0.01 & 1403.89 &  1,613,069  \\
\quad Incumbent & 122.79 & 105.01 & 70.35 & 0.01 & 1845.87 &  7,294,241  \\
\quad Helper & 75.49 & 70.63 & 29.03 & 0.01 & 522.24 &  1,084,707  \\
\quad Professional & 99.46 & 91.29 & 43.85 & 0.01 & 1033.36 &  5,131,582  \\
\quad Specialist & 150.46 & 138.03 & 71.50 & 0.01 & 1124.33 &  1,448,068  \\
\quad Expert & 194.30 & 176.67 & 97.76 & 0.33 & 1845.87 &  1,242,953  \\
\quad Low Education & 82.04 & 74.20 & 40.21 & 0.01 & 1586.55 &    661,026  \\
\quad Medium Education & 103.88 & 94.02 & 47.53 & 0.01 & 1121.69 &  6,449,947  \\
\quad High Education & 182.25 & 163.14 & 98.30 & 0.01 & 1845.87 &  1,796,337  \\
\quad Male & 124.73 & 105.55 & 72.03 & 0.01 & 1586.55 &  6,569,545  \\
\quad Female & 99.32 & 86.59 & 56.08 & 0.01 & 1845.87 &  2,337,765  \\
\quad Native & 120.09 & 102.43 & 69.74 & 0.01 & 1845.87 &  8,021,830  \\
\quad Foreign & 99.69 & 84.03 & 60.14 & 0.01 & 1221.98 &    885,480  \\
\quad Age $<$35 & 98.70 & 89.25 & 46.89 & 0.01 & 1104.98 &  2,665,428  \\
\quad Age 35-49 & 126.34 & 106.85 & 74.34 & 0.01 & 1564.56 &  3,297,593  \\
\quad Age $>=$50 & 126.33 & 106.35 & 76.08 & 0.01 & 1845.87 &  2,944,289  \\
\quad East & 88.40 & 74.82 & 47.73 & 0.01 & 1102.33 &  1,337,439  \\
\quad West & 123.31 & 105.34 & 70.96 & 0.01 & 1845.87 &  7,569,871  \\
\quad Manufacturing & 125.56 & 108.60 & 68.08 & 0.01 & 1790.24 &  3,981,248  \\
\quad Services & 112.01 & 92.48 & 69.36 & 0.01 & 1845.87 &  4,926,062  \\
Labor Market Tightness & & & & & \\
\quad Overall & 0.97 & 0.56 & 1.41 & 0.00 & 219.00 &  8,885,219  \\
\quad Hire & 0.87 & 0.50 & 1.30 & 0.00 & 148.00 &  1,610,120  \\
\quad Incumbent & 0.99 & 0.57 & 1.44 & 0.00 & 219.00 &  7,275,099  \\
\quad Helper & 0.29 & 0.17 & 0.46 & 0.00 & 54.00 &  1,084,089  \\
\quad Professional & 1.05 & 0.64 & 1.38 & 0.00 & 148.00 &  5,125,371  \\
\quad Specialist & 1.11 & 0.74 & 1.64 & 0.00 & 219.00 &  1,437,641  \\
\quad Expert & 1.02 & 0.53 & 1.65 & 0.00 & 89.00 &  1,238,118  \\
\quad Low Education & 0.73 & 0.40 & 1.14 & 0.00 & 148.00 &    660,135  \\
\quad Medium Education & 1.00 & 0.58 & 1.43 & 0.00 & 219.00 &  6,435,201  \\
\quad High Education & 0.94 & 0.55 & 1.44 & 0.00 & 148.00 &  1,789,883  \\
\quad Male & 1.07 & 0.62 & 1.53 & 0.00 & 219.00 &  6,552,179  \\
\quad Female & 0.67 & 0.43 & 0.98 & 0.00 & 148.00 &  2,333,040  \\
\quad Native & 0.97 & 0.57 & 1.42 & 0.00 & 219.00 &  8,000,882  \\
\quad Foreign & 0.90 & 0.50 & 1.32 & 0.00 & 148.00 &    884,337  \\
\quad Age $<$35 & 1.02 & 0.60 & 1.45 & 0.00 & 148.00 &  2,659,874  \\
\quad Age 35-49 & 0.94 & 0.55 & 1.38 & 0.00 & 148.00 &  3,289,441  \\
\quad Age $>=$50 & 0.94 & 0.54 & 1.42 & 0.00 & 219.00 &  2,935,904  \\
\quad East & 0.81 & 0.44 & 1.25 & 0.00 & 89.00 &  1,332,228  \\
\quad West & 0.99 & 0.58 & 1.44 & 0.00 & 219.00 &  7,552,991  \\
\quad Manufacturing & 1.20 & 0.70 & 1.63 & 0.00 & 219.00 &  3,969,420  \\
\quad Services & 0.78 & 0.47 & 1.18 & 0.00 & 219.00 &  4,915,799  \\
\bottomrule
\end{tabular}
}
\end{small}
\end{center}
\begin{tablenotes}
\small \item \textit{Notes:} The table displays descriptive sample means, medians, standard deviations, extreme values, and the number of observations of the main variable by subgroups. \textit{Sources:} Integrated Employment Biographies + Official Statistics of the German Federal Employment Agency + IAB Job Vacancy Survey, 2012-2022.
\end{tablenotes}
\end{center}
\end{table}

\clearpage
\setcounter{figure}{0}
\setcounter{table}{0}
\renewcommand*{\thefigure}{\thesection\arabic{figure}}
\renewcommand*{\thetable}{\thesection\arabic{table}}
\section{Full Regression Tables}
\label{app:full_regression}

\begin{table}[ht!]
\centering
\begin{center}
\caption{OLS Effects of Labor Market Tightness on Wages} \label{tab:baseline_ols_full}
\begin{center}
\begin{tabular}{lcccc}
\toprule
 & (1) & (2) & (3) & (4) \\
 & Log Wage & Log Wage & Log Wage & Log Wage \\
\midrule
\multirow{2}{*}{Log Tightness}                         & \hphantom{00}0.0079***\hphantom{0} & \hphantom{00}0.0088***\hphantom{0} & \hphantom{00}0.0065***\hphantom{0} & \hphantom{00}0.0074***\hphantom{0} \\
                                                                                       & \hphantom{0}(0.0008)\hphantom{000} & \hphantom{0}(0.0006)\hphantom{000} & \hphantom{0}(0.0006)\hphantom{000} & \hphantom{0}(0.0005)\hphantom{000} \\
\multirow{2}{*}{Hire}                                          & - & \hphantom{0}-0.0428***\hphantom{0} & - & \hphantom{0}-0.0384***\hphantom{0} \\
                                                                                       & - & \hphantom{0}(0.0007)\hphantom{000} & - & \hphantom{0}(0.0007)\hphantom{000} \\
\multirow{2}{*}{Eastern Germany}                       & - & \hphantom{0}-0.0328***\hphantom{0} & - & - \\
                                                                                       & - & \hphantom{0}(0.0026)\hphantom{000} & - & - \\
\multirow{2}{*}{Age}                                           & - & \hphantom{00}0.0648***\hphantom{0} & - & \hphantom{00}0.0525***\hphantom{0} \\
                                                                                       & - & \hphantom{0}(0.0033)\hphantom{000} & - & \hphantom{0}(0.0047)\hphantom{000} \\
\multirow{2}{*}{Age\textsuperscript{2}}        & - & \hphantom{0}-0.0006***\hphantom{0} & - & \hphantom{0}-0.0005***\hphantom{0} \\
                                                                                       & - & \hphantom{0}(0.0000)\hphantom{000} & - & \hphantom{0}(0.0000)\hphantom{000} \\
\multirow{2}{*}{Low Education}                         & - & base & - & base \\
                                                                                       & - & - & - & - \\
\multirow{2}{*}{Medium Education}                      & - & \hphantom{00}0.1614***\hphantom{0} & - & \hphantom{00}0.1412***\hphantom{0} \\
                                                                                       & - & \hphantom{0}(0.0042)\hphantom{000} & - & \hphantom{0}(0.0061)\hphantom{000} \\
\multirow{2}{*}{High Education}                        & - & \hphantom{00}0.3777***\hphantom{0} & - & \hphantom{00}0.3767***\hphantom{0} \\
                                                                                       & - & \hphantom{0}(0.0082)\hphantom{000} & - & \hphantom{0}(0.0115)\hphantom{000} \\
Year FE & yes & yes & yes & yes \\
Worker FE & yes & yes & yes & yes \\
Labor Market FE & yes & yes & yes & yes \\
Firm FE &  &  & yes & yes \\
Controls &  & yes &  & yes \\
\midrule
Observations \phantom{XXXXXX} &  8,584,726 &  8,584,726 &  8,454,953 &  8,454,953 \\
Clusters &     13,806 &     13,806 &     13,716 &     13,716 \\
\bottomrule
\end{tabular}
\end{center}
\begin{tablenotes}[para]
\small \item \textit{Notes:} The table displays OLS regressions of log daily wages of regular full-time workers on log labor market tightness. Control variables include binary variables for new hires, workplace location in Eastern Germany, three levels of professional education and continuous variables for age and squared age. Labor markets are combinations of detailed 3-digit occupations and commuting zones. Standard errors (in parentheses) are clustered at the labor market level: * = p$<$0.10. ** = p$<$0.05. *** = p$<$0.01. \textit{Sources:} Integrated Employment Biographies + Official Statistics of the German Federal Employment Agency + IAB Job Vacancy Survey, 2012-2022.
\end{tablenotes}
\end{center}
\end{table}

\clearpage

\begin{table}[ht!]
\begin{center}
\caption{IV Effects of Labor Market Tightness on Wages} \label{tab:baseline_iv_full}
\begin{center}
\begin{tabular}{lcccc}
\toprule
 & (1) & (2) & (3) & (4) \\
 & Log Wage & Log Wage & Log Wage & Log Wage \\
\midrule
\multirow{2}{*}{Log Tightness}                         & \hphantom{00}0.0124***\hphantom{0} & \hphantom{00}0.0114***\hphantom{0} & \hphantom{00}0.0114***\hphantom{0} & \hphantom{00}0.0113***\hphantom{0} \\
                                                                                       & \hphantom{0}(0.0018)\hphantom{000} & \hphantom{0}(0.0014)\hphantom{000} & \hphantom{0}(0.0015)\hphantom{000} & \hphantom{0}(0.0012)\hphantom{000} \\
\multirow{2}{*}{Hire}                                          & - & \hphantom{0}-0.0428***\hphantom{0} & - & \hphantom{0}-0.0384***\hphantom{0} \\
                                                                                       & - & \hphantom{0}(0.0007)\hphantom{000} & - & \hphantom{0}(0.0007)\hphantom{000} \\
\multirow{2}{*}{Eastern Germany}                       & - & \hphantom{0}-0.0328***\hphantom{0} & - & - \\
                                                                                       & - & \hphantom{0}(0.0026)\hphantom{000} & - & - \\
\multirow{2}{*}{Age}                                           & - & \hphantom{00}0.0648***\hphantom{0} & - & \hphantom{00}0.0525***\hphantom{0} \\
                                                                                       & - & \hphantom{0}(0.0033)\hphantom{000} & - & \hphantom{0}(0.0047)\hphantom{000} \\
\multirow{2}{*}{Age\textsuperscript{2}}        & - & \hphantom{0}-0.0006***\hphantom{0} & - & \hphantom{0}-0.0005***\hphantom{0} \\
                                                                                       & - & \hphantom{0}(0.0000)\hphantom{000} & - & \hphantom{0}(0.0000)\hphantom{000} \\
\multirow{2}{*}{Low Education}                         & - & base & - & base \\
                                                                                       & - & - & - & - \\
\multirow{2}{*}{Medium Education}                      & - & \hphantom{00}0.1614***\hphantom{0} & - & \hphantom{00}0.1412***\hphantom{0} \\
                                                                                       & - & \hphantom{0}(0.0042)\hphantom{000} & - & \hphantom{0}(0.0061)\hphantom{000} \\
\multirow{2}{*}{High Education}                        & - & \hphantom{00}0.3780***\hphantom{0} & - & \hphantom{00}0.3770***\hphantom{0} \\
                                                                                       & - & \hphantom{0}(0.0082)\hphantom{000} & - & \hphantom{0}(0.0115)\hphantom{000} \\
Year FE & yes & yes & yes & yes \\
Worker FE & yes & yes & yes & yes \\
Labor Market FE & yes & yes & yes & yes \\
Firm FE &  &  & yes & yes \\
Controls &  & yes &  & yes \\
\midrule
Observations \phantom{XXXXXX} &  8,584,317 &  8,584,317 &  8,454,543 &  8,454,543 \\
Clusters &     13,783 &     13,783 &     13,692 &     13,692 \\
\midrule
First-Stage Coefficient & \hphantom{00}0.8914***\hphantom{0} & \hphantom{00}0.8915***\hphantom{0} & \hphantom{00}0.8814***\hphantom{0} & \hphantom{00}0.8814***\hphantom{0} \\
\makecell[l]{First-Stage F-Statistic \\of Excluded Instrument} & \makecell[c]{     2,847} & \makecell[c]{     2,848} & \makecell[c]{     2,737} & \makecell[c]{     2,740} \\
\bottomrule
\end{tabular}
\end{center}
\begin{tablenotes}
\small \item \textit{Notes:} The table displays IV regressions of log real daily wages of regular full-time workers on log labor market tightness. The instrumental variable refers to leave-one-out averages of labor market tightness in all other commuting zones but for the same occupation and time period. Control variables include binary variables for new hires, workplace location in Eastern Germany, three levels of professional education, and continuous variables for age and squared age. Labor markets are combinations of detailed 3-digit occupations and commuting zones. Standard errors (in parentheses) are clustered at the labor market level: * = p$<$0.10. ** = p$<$0.05. *** = p$<$0.01. \textit{Sources:} Integrated Employment Biographies + Official Statistics of the German Federal Employment Agency + IAB Job Vacancy Survey, 2012-2022.
\end{tablenotes}
\end{center}
\end{table}

\clearpage

\begin{table}[ht!]
\centering
\begin{center}
\caption{Addressing Productivity Shocks at the National Level} \label{tab:productivity_shocks_full}
\begin{center}
\scalebox{0.85}{
\begin{tabular}{lcccc}
\toprule
 & (1) & (2) & (3) & (4) \\
 & Log Wage & Log Wage & Log Wage & Log Wage \\
\midrule
\multirow{2}{*}{Log Tightness}                                         & \hphantom{00}0.0060***\hphantom{0} & \hphantom{00}0.0101***\hphantom{0} & \hphantom{00}0.0057***\hphantom{0} & \hphantom{00}0.0044***\hphantom{0} \\
                                                                                                       & \hphantom{0}(0.0004)\hphantom{000} & \hphantom{0}(0.0028)\hphantom{000} & \hphantom{0}(0.0011)\hphantom{000} & \hphantom{0}(0.0017)\hphantom{000} \\
\multirow{2}{*}{Log Firm Productivity}                         & - & \hphantom{00}0.0214***\hphantom{0} & - & - \\
                                                                                                       & - & \hphantom{0}(0.0021)\hphantom{000} & - & - \\
\multirow{2}{*}{Log Leave-One-Out Sum V}                       & - & - & - & \hphantom{00}0.0114***\hphantom{0} \\
                                                                                                       & - & - & - & \hphantom{0}(0.0016)\hphantom{000} \\
\multirow{2}{*}{Hire}                                                          & \hphantom{0}-0.0378***\hphantom{0} & \hphantom{0}-0.0328***\hphantom{0} & \hphantom{0}-0.0375***\hphantom{0} & \hphantom{0}-0.0384***\hphantom{0} \\
                                                                                                       & \hphantom{0}(0.0007)\hphantom{000} & \hphantom{0}(0.0016)\hphantom{000} & \hphantom{0}(0.0007)\hphantom{000} & \hphantom{0}(0.0007)\hphantom{000} \\
\multirow{2}{*}{Age}                                                           & \hphantom{00}0.0531***\hphantom{0} & \hphantom{00}0.0746\hphantom{0000} & \hphantom{00}0.0525***\hphantom{0} & \hphantom{00}0.0526***\hphantom{0} \\
                                                                                                       & \hphantom{0}(0.0048)\hphantom{000} & \hphantom{0}(0.1112)\hphantom{000} & \hphantom{0}(0.0046)\hphantom{000} & \hphantom{0}(0.0047)\hphantom{000} \\
\multirow{2}{*}{Age\textsuperscript{2}}                        & \hphantom{0}-0.0005***\hphantom{0} & \hphantom{0}-0.0004***\hphantom{0} & \hphantom{0}-0.0005***\hphantom{0} & \hphantom{0}-0.0005***\hphantom{0} \\
                                                                                                       & \hphantom{0}(0.0000)\hphantom{000} & \hphantom{0}(0.0000)\hphantom{000} & \hphantom{0}(0.0000)\hphantom{000} & \hphantom{0}(0.0000)\hphantom{000} \\
\multirow{2}{*}{Low Education}                                         & base & base & base & base \\
                                                                                                       & - & - & - & - \\
\multirow{2}{*}{Medium Education}                                      & \hphantom{00}0.1418***\hphantom{0} & \hphantom{00}0.1756***\hphantom{0} & \hphantom{00}0.1396***\hphantom{0} & \hphantom{00}0.1411***\hphantom{0} \\
                                                                                                       & \hphantom{0}(0.0061)\hphantom{000} & \hphantom{0}(0.0271)\hphantom{000} & \hphantom{0}(0.0062)\hphantom{000} & \hphantom{0}(0.0061)\hphantom{000} \\
\multirow{2}{*}{High Education}                                        & \hphantom{00}0.3707***\hphantom{0} & \hphantom{00}0.3858***\hphantom{0} & \hphantom{00}0.3736***\hphantom{0} & \hphantom{00}0.3764***\hphantom{0} \\
                                                                                                       & \hphantom{0}(0.0114)\hphantom{000} & \hphantom{0}(0.0410)\hphantom{000} & \hphantom{0}(0.0116)\hphantom{000} & \hphantom{0}(0.0115)\hphantom{000} \\
Year FE & yes & yes & yes & yes \\
Worker FE & yes & yes & yes & yes \\
Labor Market FE & yes & yes & yes & yes \\
Firm FE & yes & yes & yes & yes \\
Occupation-by-Year FE & yes & & & \\
Industry-by-Year FE & & & yes & \\
Controls & yes & yes & yes & yes \\
\midrule
Source & 5\% IEB & LIAB & 5\% IEB & 5\% IEB \\
Survey Weights & & yes & & \\
\midrule
Observations \phantom{XXXXXX} &  8,454,869 &  3,117,429 &  8,454,541 &  8,454,541 \\
Clusters &     13,698 &      8,865 &     13,692 &     13,692 \\
\midrule
First-Stage Coefficient &  & \hphantom{00}0.8569***\hphantom{0} & \hphantom{00}0.8558***\hphantom{0} & \hphantom{00}0.7920***\hphantom{0} \\
\makecell[l]{First-Stage F-Statistic \\of Excluded Instrument} & \makecell[c]{} & \makecell[c]{     2,098} & \makecell[c]{     2,679} & \makecell[c]{     1,629} \\
\bottomrule
\end{tabular}
}
\end{center}
\begin{tablenotes}[para]
\small \item \textit{Notes:} The table displays OLS and IV regressions of log real daily wages of regular full-time workers on on log labor market tightness. The instrumental variable refers to leave-one-out averages of labor market tightness in all other commuting zones but for the same occupation and time period. Control variables include binary variables for new hires, workplace location in Eastern Germany, three levels of professional education and continuous variables for age and squared age. Labor markets are combinations of detailed 3-digit occupations and commuting zones. Column (1) shows OLS estimates when additionally controlling for detailed 3-digit-occupation-by-year fixed effects. Column (2) features IV estimates when controlling for log firm productivity in the LIAB worker sample. Column (3) displays IV estimates when further controlling for 1-digit industry-by-year fixed effects (based on the NACE 2.0 classification). Column (4) presents IV estimates when conditioning on the log leave-one-out sum of the number of vacancies in all other commuting zones but for the very same occupation and time period. Standard errors (in parentheses) are clustered at the labor market level: * = p$<$0.10. ** = p$<$0.05. *** = p$<$0.01. \textit{Sources:} Integrated Employment Biographies + Official Statistics of the German Federal Employment Agency + IAB Job Vacancy Survey + Linked Employer-Employee Dataset of the IAB, 2012-2022.
\end{tablenotes}
\end{center}
\end{table}

\clearpage

\begin{landscape}

\begin{table}[ht!]
\begin{center}
\caption{Robustness Checks I} \label{tab:robustness1}
\begin{center}
\scalebox{0.775}{
\begin{tabular}{lC{4cm}C{4cm}C{4cm}C{4cm}C{4cm}C{4cm}}
\toprule
 & (1) & (2) & (3) & (4) & (5) & (6) \\
 & \textbf{Detailed 2-Digit Occupations} & \textbf{Detailed 4-Digit Occupations} & \textbf{Flow-Adjusted V/U} & \textbf{Federal States} & \textbf{Government Regions} & \textbf{Districts} \\
 & Log Wage & Log Wage & Log Wage & Log Wage & Log Wage & Log Wage \\
\midrule
\multirow{2}{*}{Log Tightness}                         & \hphantom{00}0.0127***\hphantom{0} & \hphantom{00}0.0101***\hphantom{0} & \hphantom{00}0.0144***\hphantom{0} & \hphantom{00}0.0121***\hphantom{0} & \hphantom{00}0.0123***\hphantom{0} & \hphantom{00}0.0103***\hphantom{0} \\
                                                                                       & \hphantom{0}(0.0017)\hphantom{000} & \hphantom{0}(0.0011)\hphantom{000} & \hphantom{0}(0.0010)\hphantom{000} & \hphantom{0}(0.0016)\hphantom{000} & \hphantom{0}(0.0011)\hphantom{000} & \hphantom{0}(0.0008)\hphantom{000} \\
\multirow{2}{*}{Hire}                                          & \hphantom{0}-0.0384***\hphantom{0} & \hphantom{0}-0.0382***\hphantom{0} & \hphantom{0}-0.0379***\hphantom{0} & \hphantom{0}-0.0384***\hphantom{0} & \hphantom{0}-0.0383***\hphantom{0} & \hphantom{0}-0.0377***\hphantom{0} \\
                                                                                       & \hphantom{0}(0.0009)\hphantom{000} & \hphantom{0}(0.0006)\hphantom{000} & \hphantom{0}(0.0006)\hphantom{000} & \hphantom{0}(0.0009)\hphantom{000} & \hphantom{0}(0.0006)\hphantom{000} & \hphantom{0}(0.0005)\hphantom{000} \\
\multirow{2}{*}{Age}                                           & \hphantom{00}0.0518***\hphantom{0} & \hphantom{00}0.0501***\hphantom{0} & \hphantom{00}0.0505***\hphantom{0} & \hphantom{00}0.0518***\hphantom{0} & \hphantom{00}0.0514***\hphantom{0} & \hphantom{00}0.0488***\hphantom{0} \\
                                                                                       & \hphantom{0}(0.0045)\hphantom{000} & \hphantom{0}(0.0049)\hphantom{000} & \hphantom{0}(0.0046)\hphantom{000} & \hphantom{0}(0.0046)\hphantom{000} & \hphantom{0}(0.0046)\hphantom{000} & \hphantom{0}(0.0054)\hphantom{000} \\
\multirow{2}{*}{Age\textsuperscript{2}}        & \hphantom{0}-0.0005***\hphantom{0} & \hphantom{0}-0.0005***\hphantom{0} & \hphantom{0}-0.0005***\hphantom{0} & \hphantom{0}-0.0005***\hphantom{0} & \hphantom{0}-0.0005***\hphantom{0} & \hphantom{0}-0.0005***\hphantom{0} \\
                                                                                       & \hphantom{0}(0.0000)\hphantom{000} & \hphantom{0}(0.0000)\hphantom{000} & \hphantom{0}(0.0000)\hphantom{000} & \hphantom{0}(0.0000)\hphantom{000} & \hphantom{0}(0.0000)\hphantom{000} & \hphantom{0}(0.0000)\hphantom{000} \\
\multirow{2}{*}{Low Education}                         & base & base & base & base & base & base \\
                                                                                       & - & - & - & - & - & - \\
\multirow{2}{*}{Medium Education}                      & \hphantom{00}0.1428***\hphantom{0} & \hphantom{00}0.1386***\hphantom{0} & \hphantom{00}0.1416***\hphantom{0} & \hphantom{00}0.1424***\hphantom{0} & \hphantom{00}0.1407***\hphantom{0} & \hphantom{00}0.1361***\hphantom{0} \\
                                                                                       & \hphantom{0}(0.0068)\hphantom{000} & \hphantom{0}(0.0061)\hphantom{000} & \hphantom{0}(0.0059)\hphantom{000} & \hphantom{0}(0.0077)\hphantom{000} & \hphantom{0}(0.0051)\hphantom{000} & \hphantom{0}(0.0040)\hphantom{000} \\
\multirow{2}{*}{High Education}                        & \hphantom{00}0.3790***\hphantom{0} & \hphantom{00}0.3727***\hphantom{0} & \hphantom{00}0.3764***\hphantom{0} & \hphantom{00}0.3791***\hphantom{0} & \hphantom{00}0.3756***\hphantom{0} & \hphantom{00}0.3649***\hphantom{0} \\
                                                                                       & \hphantom{0}(0.0125)\hphantom{000} & \hphantom{0}(0.0118)\hphantom{000} & \hphantom{0}(0.0114)\hphantom{000} & \hphantom{0}(0.0135)\hphantom{000} & \hphantom{0}(0.0102)\hphantom{000} & \hphantom{0}(0.0087)\hphantom{000} \\
Year FE & yes & yes & yes & yes & yes & yes \\
Worker FE & yes & yes & yes & yes & yes & yes \\
Labor Market FE & yes & yes & yes & yes & yes & yes \\
Firm FE & yes & yes & yes & yes & yes & yes  \\
Controls & yes & yes & yes & yes & yes & yes \\
\midrule
Observations \phantom{XXXXXX} &  8,591,598 &  8,131,060 &  8,616,432 &  8,552,284 &  8,471,619 &  7,692,410 \\
Clusters &      6,112 &     23,919 &     33,966 &      5,498 &     12,204 &     63,707 \\
\midrule
First-Stage Coefficient & \hphantom{00}0.9084***\hphantom{0} & \hphantom{00}0.8434***\hphantom{0} & \hphantom{00}0.8360***\hphantom{0} & \hphantom{00}0.8591***\hphantom{0} & \hphantom{00}0.9381***\hphantom{0} & \hphantom{00}0.9434***\hphantom{0} \\
\makecell[l]{First-Stage F-Statistic \\of Excluded Instrument} & \makecell[c]{     1,863} & \makecell[c]{     3,576} & \makecell[c]{     6,225} & \makecell[c]{     2,197} & \makecell[c]{     3,452} & \makecell[c]{     2,558} \\
\bottomrule
\end{tabular}
}
\end{center}
\begin{tablenotes}
\footnotesize \item \textit{Notes:} The table displays IV regressions of log real daily wages of regular full-time workers on log labor market tightness. The instrumental variable refers to leave-one-out averages of labor market tightness in all other commuting zones but for the same occupation and time period. Control variables include binary variables for new hires, workplace location in Eastern Germany, three levels of professional education, and continuous variables for age and squared age. Labor markets are combinations of detailed 3-digit occupations and commuting zones. Standard errors (in parentheses) are clustered at the labor market level: * = p$<$0.10. ** = p$<$0.05. *** = p$<$0.01. \textit{Sources:} Integrated Employment Biographies + Official Statistics of the German Federal Employment Agency + IAB Job Vacancy Survey, 2012-2022.
\end{tablenotes}
\end{center}
\end{table}

\begin{table}[ht!]
\begin{center}
\caption{Robustness Checks II} \label{tab:robustness2}
\begin{center}
\scalebox{0.725}{
\begin{tabular}{lC{4cm}C{4cm}C{4cm}C{4cm}C{4cm}}
\toprule
 & (1) & (2) & (3) & (4) & (5) \\
 & \textbf{Time-Constant Notification Share} & \textbf{Only Registered Vacancies} & \textbf{Trim at 5\textsuperscript{th} and 95\textsuperscript{th} Percentile} & \textbf{Sum-Based Instrument} & \textbf{Squared Estimation} \\
 & Log Wage & Log Wage & Log Wage & Log Wage & Log Wage \\
\midrule
\multirow{2}{*}{Log Tightness}                         & \hphantom{00}0.0120***\hphantom{0} & \hphantom{00}0.0119***\hphantom{0} & \hphantom{00}0.0145***\hphantom{0} & \hphantom{00}0.0107***\hphantom{0} & \hphantom{00}0.0122***\hphantom{0} \\
                                                                                       & \hphantom{0}(0.0013)\hphantom{000} & \hphantom{0}(0.0013)\hphantom{000} & \hphantom{0}(0.0013)\hphantom{000} & \hphantom{0}(0.0012)\hphantom{000} & \hphantom{0}(0.0013)\hphantom{000} \\
\multirow{2}{*}{(Log Tightness)$^2$}           &  &  &  &  & \hphantom{00}0.0007\hphantom{0000} \\
                                                                                       &  &  &  &  & \hphantom{0}(0.0004)\hphantom{000} \\
\multirow{2}{*}{Hire}                                          & \hphantom{0}-0.0384***\hphantom{0} & \hphantom{0}-0.0384***\hphantom{0} & \hphantom{0}-0.0273***\hphantom{0} & \hphantom{0}-0.0384***\hphantom{0} & \hphantom{0}-0.0384***\hphantom{0} \\
                                                                                       & \hphantom{0}(0.0007)\hphantom{000} & \hphantom{0}(0.0007)\hphantom{000} & \hphantom{0}(0.0005)\hphantom{000} & \hphantom{0}(0.0007)\hphantom{000} & \hphantom{0}(0.0007)\hphantom{000} \\
\multirow{2}{*}{Age}                                           & \hphantom{00}0.0525***\hphantom{0} & \hphantom{00}0.0525***\hphantom{0} & \hphantom{00}0.0482***\hphantom{0} & \hphantom{00}0.0525***\hphantom{0} & \hphantom{00}0.0525***\hphantom{0} \\
                                                                                       & \hphantom{0}(0.0047)\hphantom{000} & \hphantom{0}(0.0047)\hphantom{000} & \hphantom{0}(0.0039)\hphantom{000} & \hphantom{0}(0.0047)\hphantom{000} & \hphantom{0}(0.0047)\hphantom{000} \\
\multirow{2}{*}{Age\textsuperscript{2}}        & \hphantom{0}-0.0005***\hphantom{0} & \hphantom{0}-0.0005***\hphantom{0} & \hphantom{0}-0.0005***\hphantom{0} & \hphantom{0}-0.0005***\hphantom{0} & \hphantom{0}-0.0005***\hphantom{0} \\
                                                                                       & \hphantom{0}(0.0000)\hphantom{000} & \hphantom{0}(0.0000)\hphantom{000} & \hphantom{0}(0.0000)\hphantom{000} & \hphantom{0}(0.0000)\hphantom{000} & \hphantom{0}(0.0000)\hphantom{000} \\
\multirow{2}{*}{Low Education}                         & base & base & base & base & base \\
                                                                                       & - & - & - & - & - \\
\multirow{2}{*}{Medium Education}                      & \hphantom{00}0.1411***\hphantom{0} & \hphantom{00}0.1411***\hphantom{0} & \hphantom{00}0.0362***\hphantom{0} & \hphantom{00}0.1412***\hphantom{0} & \hphantom{00}0.1413***\hphantom{0} \\
                                                                                       & \hphantom{0}(0.0061)\hphantom{000} & \hphantom{0}(0.0061)\hphantom{000} & \hphantom{0}(0.0021)\hphantom{000} & \hphantom{0}(0.0061)\hphantom{000} & \hphantom{0}(0.0061)\hphantom{000} \\
\multirow{2}{*}{High Education}                        & \hphantom{00}0.3770***\hphantom{0} & \hphantom{00}0.3770***\hphantom{0} & \hphantom{00}0.1253***\hphantom{0} & \hphantom{00}0.3769***\hphantom{0} & \hphantom{00}0.3771***\hphantom{0} \\
                                                                                       & \hphantom{0}(0.0115)\hphantom{000} & \hphantom{0}(0.0115)\hphantom{000} & \hphantom{0}(0.0030)\hphantom{000} & \hphantom{0}(0.0115)\hphantom{000} & \hphantom{0}(0.0115)\hphantom{000} \\
Year FE & yes & yes & yes & yes & yes \\
Worker FE & yes & yes & yes & yes & yes \\
Labor Market FE & yes & yes & yes & yes & yes \\
Firm FE & yes & yes & yes & yes & yes  \\
Controls & yes & yes & yes & yes & yes \\
\midrule
Observations \phantom{XXXXXX} &  8,454,549 &  8,454,549 &  6,862,884 &  8,454,700 &  8,454,543 \\
Clusters &     13,692 &     13,692 &     12,581 &     13,699 &     13,692 \\
\midrule
First-Stage Coefficient (Log Tightness) & \hphantom{00}0.8777***\hphantom{0} & \hphantom{00}0.8805***\hphantom{0} & \hphantom{00}0.8127***\hphantom{0} & \hphantom{00}0.8979***\hphantom{0} & \hphantom{00}0.8823***\hphantom{0} \\
\makecell[l]{First-Stage F-Statistic \\of Excluded Instrument (Log Tightness)} & \makecell[c]{     2,276} & \makecell[c]{     2,305} & \makecell[c]{     2,595} & \makecell[c]{     3,624} & \makecell[c]{     1,380} \\
First-Stage Coefficient (Log Tightness)$^2$ &   &   &   &   & \hphantom{00}0.8065***\hphantom{0} \\
\makecell[l]{First-Stage F-Statistic \\of Excluded Instrument (Log Tightness)$^2$} & \makecell[c]{ } & \makecell[c]{ } & \makecell[c]{ } & \makecell[c]{ } & \makecell[c]{     1,020} \\
\bottomrule
\end{tabular}
}
\end{center}
\begin{tablenotes}
\footnotesize \item \textit{Notes:} The table displays IV regressions of log real daily wages of regular full-time workers on log labor market tightness. The instrumental variable refers to leave-one-out averages of labor market tightness in all other commuting zones but for the same occupation and time period. Control variables include binary variables for new hires, workplace location in Eastern Germany, three levels of professional education, and continuous variables for age and squared age. Labor markets are combinations of detailed 3-digit occupations and commuting zones. Standard errors (in parentheses) are clustered at the labor market level: * = p$<$0.10. ** = p$<$0.05. *** = p$<$0.01. \textit{Sources:} Integrated Employment Biographies + Official Statistics of the German Federal Employment Agency + IAB Job Vacancy Survey, 2012-2022.
\end{tablenotes}
\end{center}
\end{table}

\end{landscape}

\clearpage
\begin{landscape}

\begin{table}[ht!]
\centering
\begin{center}
\caption{Heterogeneous Effects I} \label{tab:het_effects1}
\begin{center}
\scalebox{0.775}{
\begin{tabular}{lR{3.0cm}C{2.5cm}R{3.0cm}C{2.5cm}R{3.0cm}C{2.5cm}}
\toprule
 & \multicolumn{2}{c}{(1)} & \multicolumn{2}{c}{(2)} & \multicolumn{2}{c}{(3)} \\
 & \multicolumn{2}{c}{Log Wage} & \multicolumn{2}{c}{Log Wage} & \multicolumn{2}{c}{Log Wage} \\
\midrule
\multirow{2}{*}{Log Tightness * ...}   & \multirow{2}{*}{Hire}                 & \hphantom{00}0.0158***\hphantom{0} & \multirow{2}{*}{Helper}                      & \hphantom{00}0.0128***\hphantom{0} & \multirow{2}{*}{Low Education}               & \hphantom{00}0.0113***\hphantom{0} \\
                                                                               &                                                               & \hphantom{0}(0.0014)\hphantom{000} &                                                                      & \hphantom{0}(0.0018)\hphantom{000} &                                                                              & \hphantom{0}(0.0020)\hphantom{000}  \\
\multirow{2}{*}{Log Tightness * ...}   & \multirow{2}{*}{Incumbent}    & \hphantom{00}0.0106***\hphantom{0} & \multirow{2}{*}{Professional}        & \hphantom{00}0.0075***\hphantom{0} & \multirow{2}{*}{Med. Education}      & \hphantom{00}0.0068***\hphantom{0} \\
                                                                               &                                                               & \hphantom{0}(0.0012)\hphantom{000} &                                                                      & \hphantom{0}(0.0015)\hphantom{000} &                                                                              & \hphantom{0}(0.0014)\hphantom{000}  \\
\multirow{2}{*}{Log Tightness * ...}   & \multirow{2}{*}{}                     &  & \multirow{2}{*}{Specialist}          & \hphantom{00}0.0273***\hphantom{0} & \multirow{2}{*}{High Education}              & \hphantom{00}0.0262***\hphantom{0} \\
                                                                               &                                                               &  &                                                                      & \hphantom{0}(0.0020)\hphantom{000} &                                                                              & \hphantom{0}(0.0019)\hphantom{000}  \\
\multirow{2}{*}{Log Tightness * ...}   & \multirow{2}{*}{}                     &  & \multirow{2}{*}{Expert}                      & \hphantom{00}0.0156***\hphantom{0} & \multirow{2}{*}{}                                    &  \\
                                                                               &                                                               &  &                                                                      & \hphantom{0}(0.0020)\hphantom{000} &                                                                              &   \\
\multirow{2}{*}{Hire}                                          && \hphantom{0}-0.0349***\hphantom{0}       & & \hphantom{0}-0.0384***\hphantom{0} & & \hphantom{0}-0.0383***\hphantom{0} \\
                                                                                       && \hphantom{0}(0.0007)\hphantom{000}      & & \hphantom{0}(0.0007)\hphantom{000} & & \hphantom{0}(0.0007)\hphantom{000} \\
\multirow{2}{*}{Age}                                           && \hphantom{00}0.0524***\hphantom{0}      & & \hphantom{00}0.0529***\hphantom{0} & & \hphantom{00}0.0528***\hphantom{0} \\
                                                                                       && \hphantom{0}(0.0048)\hphantom{000}      & & \hphantom{0}(0.0048)\hphantom{000} & & \hphantom{0}(0.0048)\hphantom{000} \\
\multirow{2}{*}{Age\textsuperscript{2}}        && \hphantom{0}-0.0005***\hphantom{0}      & & \hphantom{0}-0.0005***\hphantom{0} & & \hphantom{0}-0.0005***\hphantom{0} \\
                                                                                       && \hphantom{0}(0.0000)\hphantom{000}      & & \hphantom{0}(0.0000)\hphantom{000} & & \hphantom{0}(0.0000)\hphantom{000} \\
\multirow{2}{*}{Low Education}                         && base      & & base & & base \\
                                                                                       && -      & & - & & - \\
\multirow{2}{*}{Medium Education}                      && \hphantom{00}0.1407***\hphantom{0}      & & \hphantom{00}0.1413***\hphantom{0} & & \hphantom{00}0.1381***\hphantom{0} \\
                                                                                       && \hphantom{0}(0.0061)\hphantom{000}      & & \hphantom{0}(0.0061)\hphantom{000} & & \hphantom{0}(0.0055)\hphantom{000} \\
\multirow{2}{*}{High Education}                        && \hphantom{00}0.3763***\hphantom{0}      & & \hphantom{00}0.3765***\hphantom{0} & & \hphantom{00}0.3832***\hphantom{0} \\
                                                                                       && \hphantom{0}(0.0114)\hphantom{000}      & & \hphantom{0}(0.0115)\hphantom{000} & & \hphantom{0}(0.0111)\hphantom{000} \\
Year FE && yes && yes && yes \\
Worker FE && yes && yes && yes \\
Labor Market FE && yes && yes && yes \\
Firm FE && yes && yes && yes  \\
Controls && yes && yes && yes \\
\midrule
Observations \phantom{XXXXXX} &&  8,454,541 &&  8,454,541 &&  8,454,541  \\
Clusters &&     13,692 &&     13,692 &&     13,692 \\
\bottomrule
\end{tabular}
}
\end{center}
\begin{tablenotes}[para]
\footnotesize \item \textit{Notes:} The table displays IV regressions of log real daily wages of regular full-time workers on log labor market tightness. The instrumental variable refers to leave-one-out averages of labor market tightness in all other commuting zones but for the same occupation and time period. Control variables include binary variables for new hires, workplace location in Eastern Germany, three levels of professional education, and continuous variables for age and squared age. Labor markets are combinations of detailed 3-digit occupations and commuting zones. Standard errors (in parentheses) are clustered at the labor market level: * = p$<$0.10. ** = p$<$0.05. *** = p$<$0.01. \textit{Sources:} Integrated Employment Biographies + Official Statistics of the German Federal Employment Agency + IAB Job Vacancy Survey, 2012-2022.
\end{tablenotes}
\end{center}
\end{table}

\begin{table}[ht!]
\centering
\begin{center}
\caption{Heterogeneous Effects II} \label{tab:het_effects2}
\begin{center}
\scalebox{0.775}{
\begin{tabular}{lR{3.0cm}C{2.5cm}R{3.0cm}C{2.5cm}R{3.0cm}C{2.5cm}}
\toprule
 & \multicolumn{2}{c}{(1)} & \multicolumn{2}{c}{(2)} & \multicolumn{2}{c}{(3)} \\
 & \multicolumn{2}{c}{Log Wage} & \multicolumn{2}{c}{Log Wage} & \multicolumn{2}{c}{Log Wage} \\
\midrule
\multirow{2}{*}{Log Tightness * ...}   & \multirow{2}{*}{Male}                 & \hphantom{00}0.0121***\hphantom{0} & \multirow{2}{*}{Native}                      & \hphantom{00}0.0109***\hphantom{0} & \multirow{2}{*}{Age $<$35}                   & \hphantom{00}0.0097***\hphantom{0} \\
                                                                               &                                                               & \hphantom{0}(0.0012)\hphantom{000} &                                                                      & \hphantom{0}(0.0012)\hphantom{000} &                                                                              & \hphantom{0}(0.0015)\hphantom{000}  \\
\multirow{2}{*}{Log Tightness * ...}   & \multirow{2}{*}{Female}               & \hphantom{00}0.0090***\hphantom{0} & \multirow{2}{*}{Foreign}             & \hphantom{00}0.0153***\hphantom{0} & \multirow{2}{*}{Age 35-49}                   & \hphantom{00}0.0095***\hphantom{0} \\
                                                                               &                                                               & \hphantom{0}(0.0017)\hphantom{000} &                                                                      & \hphantom{0}(0.0014)\hphantom{000} &                                                                              & \hphantom{0}(0.0012)\hphantom{000}  \\
\multirow{2}{*}{Log Tightness * ...}   & \multirow{2}{*}{}                     &  & \multirow{2}{*}{}                            &  & \multirow{2}{*}{Age $>=$50}                  & \hphantom{00}0.0146***\hphantom{0} \\
                                                                               &                                                               &  &                                                                      &  &                                                                              & \hphantom{0}(0.0012)\hphantom{000}  \\
\multirow{2}{*}{Hire}                                          && \hphantom{0}-0.0384***\hphantom{0}       & & \hphantom{0}-0.0384***\hphantom{0} & & \hphantom{0}-0.0384***\hphantom{0} \\
                                                                                       && \hphantom{0}(0.0007)\hphantom{000}      & & \hphantom{0}(0.0007)\hphantom{000} & & \hphantom{0}(0.0007)\hphantom{000} \\
\multirow{2}{*}{Age}                                           && \hphantom{00}0.0525***\hphantom{0}      & & \hphantom{00}0.0525***\hphantom{0} & & \hphantom{00}0.0532***\hphantom{0} \\
                                                                                       && \hphantom{0}(0.0047)\hphantom{000}      & & \hphantom{0}(0.0047)\hphantom{000} & & \hphantom{0}(0.0049)\hphantom{000} \\
\multirow{2}{*}{Age\textsuperscript{2}}        && \hphantom{0}-0.0005***\hphantom{0}      & & \hphantom{0}-0.0005***\hphantom{0} & & \hphantom{0}-0.0005***\hphantom{0} \\
                                                                                       && \hphantom{0}(0.0000)\hphantom{000}      & & \hphantom{0}(0.0000)\hphantom{000} & & \hphantom{0}(0.0000)\hphantom{000} \\
\multirow{2}{*}{Low Education}                         && base      & & base & & base \\
                                                                                       && -      & & - & & - \\
\multirow{2}{*}{Medium Education}                      && \hphantom{00}0.1412***\hphantom{0}      & & \hphantom{00}0.1410***\hphantom{0} & & \hphantom{00}0.1411***\hphantom{0} \\
                                                                                       && \hphantom{0}(0.0061)\hphantom{000}      & & \hphantom{0}(0.0061)\hphantom{000} & & \hphantom{0}(0.0061)\hphantom{000} \\
\multirow{2}{*}{High Education}                        && \hphantom{00}0.3770***\hphantom{0}      & & \hphantom{00}0.3768***\hphantom{0} & & \hphantom{00}0.3767***\hphantom{0} \\
                                                                                       && \hphantom{0}(0.0115)\hphantom{000}      & & \hphantom{0}(0.0115)\hphantom{000} & & \hphantom{0}(0.0115)\hphantom{000} \\
Year FE && yes && yes && yes \\
Worker FE && yes && yes && yes \\
Labor Market FE && yes && yes && yes \\
Firm FE && yes && yes && yes  \\
Controls && yes && yes && yes \\
\midrule
Observations \phantom{XXXXXX} &&  8,454,541 &&  8,454,541 &&  8,454,541  \\
Clusters &&     13,692 &&     13,692 &&     13,692 \\
\bottomrule
\end{tabular}
}
\end{center}
\begin{tablenotes}[para]
\small \item \textit{Notes:} The table displays IV regressions of log real daily wages of regular full-time workers on log labor market tightness. The instrumental variable refers to leave-one-out averages of labor market tightness in all other commuting zones but for the same occupation and time period. Control variables include binary variables for new hires, workplace location in Eastern Germany, three levels of professional education, and continuous variables for age and squared age. Labor markets are combinations of detailed 3-digit occupations and commuting zones. Standard errors (in parentheses) are clustered at the labor market level: * = p$<$0.10. ** = p$<$0.05. *** = p$<$0.01. \textit{Sources:} Integrated Employment Biographies + Official Statistics of the German Federal Employment Agency + IAB Job Vacancy Survey, 2012-2022.
\end{tablenotes}
\end{center}
\end{table}

\begin{table}[ht!]
\centering
\begin{center}
\caption{Heterogeneous Effects III} \label{tab:het_effects3}
\begin{center}
\scalebox{0.775}{
\begin{tabular}{lR{3.0cm}C{2.5cm}R{3.0cm}C{2.5cm}}
\toprule
 & \multicolumn{2}{c}{(1)} & \multicolumn{2}{c}{(2)} \\
 & \multicolumn{2}{c}{Log Wage} & \multicolumn{2}{c}{Log Wage} \\
\midrule
\multirow{2}{*}{Log Tightness * ...}   & \multirow{2}{*}{East}                 & \hphantom{00}0.0353***\hphantom{0} & \multirow{2}{*}{Manufacturing}       & \hphantom{00}0.0082***\hphantom{0}  \\
                                                                               &                                                               & \hphantom{0}(0.0014)\hphantom{000} &                                                                      & \hphantom{0}(0.0012)\hphantom{000} \\
\multirow{2}{*}{Log Tightness * ...}   & \multirow{2}{*}{West}                 & \hphantom{00}0.0072***\hphantom{0} & \multirow{2}{*}{Services}            & \hphantom{00}0.0145***\hphantom{0}  \\
                                                                               &                                                               & \hphantom{0}(0.0012)\hphantom{000} &                                                                      & \hphantom{0}(0.0015)\hphantom{000} \\
\multirow{2}{*}{Hire}                                          && \hphantom{0}-0.0383***\hphantom{0}       & & \hphantom{0}-0.0383***\hphantom{0} \\
                                                                                       && \hphantom{0}(0.0007)\hphantom{000}      & & \hphantom{0}(0.0007)\hphantom{000} \\
\multirow{2}{*}{Age}                                           && \hphantom{00}0.0525***\hphantom{0}      & & \hphantom{00}0.0525***\hphantom{0} \\
                                                                                       && \hphantom{0}(0.0046)\hphantom{000}      & & \hphantom{0}(0.0048)\hphantom{000} \\
\multirow{2}{*}{Age\textsuperscript{2}}        && \hphantom{0}-0.0005***\hphantom{0}      & & \hphantom{0}-0.0005***\hphantom{0} \\
                                                                                       && \hphantom{0}(0.0000)\hphantom{000}      & & \hphantom{0}(0.0000)\hphantom{000} \\
\multirow{2}{*}{Low Education}                         && base      & & base \\
                                                                                       && -      & & - \\
\multirow{2}{*}{Medium Education}                      && \hphantom{00}0.1419***\hphantom{0}      & & \hphantom{00}0.1411***\hphantom{0} \\
                                                                                       && \hphantom{0}(0.0061)\hphantom{000}      & & \hphantom{0}(0.0061)\hphantom{000} \\
\multirow{2}{*}{High Education}                        && \hphantom{00}0.3781***\hphantom{0}      & & \hphantom{00}0.3768***\hphantom{0} \\
                                                                                       && \hphantom{0}(0.0115)\hphantom{000}      & & \hphantom{0}(0.0115)\hphantom{000} \\
Year FE && yes && yes  \\
Worker FE && yes && yes  \\
Labor Market FE && yes && yes \\
Firm FE && yes && yes  \\
Controls && yes && yes  \\
\midrule
Observations \phantom{XXXXXX} &&  8,454,541 &&  8,454,543  \\
Clusters &&     13,692 &&     13,692 \\
\bottomrule
\end{tabular}
}
\end{center}
\begin{tablenotes}[para]
\footnotesize \item \textit{Notes:} The table displays IV regressions of log real daily wages of regular full-time workers on log labor market tightness. The instrumental variable refers to leave-one-out averages of labor market tightness in all other commuting zones but for the same occupation and time period. Control variables include binary variables for new hires, workplace location in Eastern Germany, three levels of professional education, and continuous variables for age and squared age. Labor markets are combinations of detailed 3-digit occupations and commuting zones. Standard errors (in parentheses) are clustered at the labor market level: * = p$<$0.10. ** = p$<$0.05. *** = p$<$0.01. \textit{Sources:} Integrated Employment Biographies + Official Statistics of the German Federal Employment Agency + IAB Job Vacancy Survey, 2012-2022.
\end{tablenotes}
\end{center}
\end{table}

\end{landscape}

\begin{table}[ht!]
\centering
\begin{center}
\caption{Heterogeneous Effects - First-Stage Regressions} \label{tab:het_effects_fs}
\begin{center}
\scalebox{0.85}{
\begin{tabular}{L{4.2cm}L{6.0cm}C{3.0cm}}
\toprule
 & & (1) \\
 & & Log Tightness  \\
\midrule
\multirow{2}{*}{Z\textsuperscript{1} * Hire}                           & Coefficient & \hphantom{00}0.9280***\hphantom{0} \\
                                                                                                                       & F-Statistic of Excluded Instrument &      3,332 \\
\multirow{2}{*}{Z\textsuperscript{1} * Incumbent}                      & Coefficient & \hphantom{00}0.9018***\hphantom{0} \\
                                                                                                                       & \makecell[l]{F-Statistic of Excluded Instrument} & \makecell[c]{     4,028} \\
\multirow{2}{*}{Z\textsuperscript{1} * Helper}                         & Coefficient & \hphantom{00}0.8637***\hphantom{0} \\
                                                                                                                       & \makecell[l]{F-Statistic of Excluded Instrument} & \makecell[c]{       347} \\
\multirow{2}{*}{Z\textsuperscript{1} * Professional}           & Coefficient & \hphantom{00}0.8350***\hphantom{0} \\
                                                                                                                       & \makecell[l]{F-Statistic of Excluded Instrument} & \makecell[c]{     2,449} \\
\multirow{2}{*}{Z\textsuperscript{1} * Specialist}                     & Coefficient & \hphantom{00}0.8146***\hphantom{0} \\
                                                                                                                       & \makecell[l]{F-Statistic of Excluded Instrument} & \makecell[c]{       653} \\
\multirow{2}{*}{Z\textsuperscript{1} * Expert}                         & Coefficient & \hphantom{00}0.9328***\hphantom{0} \\
                                                                                                                       & \makecell[l]{F-Statistic of Excluded Instrument} & \makecell[c]{       357} \\
\multirow{2}{*}{Z\textsuperscript{1} * Low Education}          & Coefficient & \hphantom{00}0.8696***\hphantom{0} \\
                                                                                                                       & \makecell[l]{F-Statistic of Excluded Instrument} & \makecell[c]{     5,353} \\
\multirow{2}{*}{Z\textsuperscript{1} * Medium Education}       & Coefficient & \hphantom{00}0.8614***\hphantom{0} \\
                                                                                                                       & \makecell[l]{F-Statistic of Excluded Instrument} & \makecell[c]{     8,569} \\
\multirow{2}{*}{Z\textsuperscript{1} * High Education}         & Coefficient & \hphantom{00}0.9138***\hphantom{0} \\
                                                                                                                       & \makecell[l]{F-Statistic of Excluded Instrument} & \makecell[c]{     2,586} \\
\multirow{2}{*}{Z\textsuperscript{1} * Male}                           & Coefficient & \hphantom{00}0.8925***\hphantom{0} \\
                                                                                                                       & \makecell[l]{F-Statistic of Excluded Instrument} & \makecell[c]{     8,270} \\
\multirow{2}{*}{Z\textsuperscript{1} * Female}                         & Coefficient & \hphantom{00}0.8622***\hphantom{0} \\
                                                                                                                       & \makecell[l]{F-Statistic of Excluded Instrument} & \makecell[c]{     7,391} \\
\multirow{2}{*}{Z\textsuperscript{1} * Native}                         & Coefficient & \hphantom{00}0.8862***\hphantom{0} \\
                                                                                                                       & \makecell[l]{F-Statistic of Excluded Instrument} & \makecell[c]{    10,447} \\
\multirow{2}{*}{Z\textsuperscript{1} * Foreign}                        & Coefficient & \hphantom{00}0.8844***\hphantom{0} \\
                                                                                                                       & \makecell[l]{F-Statistic of Excluded Instrument} & \makecell[c]{     8,677} \\
\multirow{2}{*}{Z\textsuperscript{1} * Age $<$35}                      & Coefficient & \hphantom{00}0.9390***\hphantom{0} \\
                                                                                                                       & \makecell[l]{F-Statistic of Excluded Instrument} & \makecell[c]{     5,977} \\
\multirow{2}{*}{Z\textsuperscript{1} * Age 35-49}                      & Coefficient & \hphantom{00}0.9297***\hphantom{0} \\
                                                                                                                       & \makecell[l]{F-Statistic of Excluded Instrument} & \makecell[c]{     4,191} \\
\multirow{2}{*}{Z\textsuperscript{1} * Age $>=$50}                     & Coefficient & \hphantom{00}0.9597***\hphantom{0} \\
                                                                                                                       & \makecell[l]{F-Statistic of Excluded Instrument} & \makecell[c]{     4,615} \\
\multirow{2}{*}{Z\textsuperscript{1} * East}                           & Coefficient & \hphantom{00}1.0428***\hphantom{0} \\
                                                                                                                       & \makecell[l]{F-Statistic of Excluded Instrument} & \makecell[c]{     3,318} \\
\multirow{2}{*}{Z\textsuperscript{1} * West}                           & Coefficient & \hphantom{00}0.8777***\hphantom{0} \\
                                                                                                                       & \makecell[l]{F-Statistic of Excluded Instrument} & \makecell[c]{     4,921} \\
\multirow{2}{*}{Z\textsuperscript{1} * Manufacturing}          & Coefficient & \hphantom{00}0.9256***\hphantom{0} \\
                                                                                                                       & \makecell[l]{F-Statistic of Excluded Instrument} & \makecell[c]{     7,989} \\
\multirow{2}{*}{Z\textsuperscript{1} * Services}                       & Coefficient & \hphantom{00}0.8477***\hphantom{0} \\
                                                                                                                       & \makecell[l]{F-Statistic of Excluded Instrument} & \makecell[c]{     8,695} \\
\bottomrule
\end{tabular}
}
\end{center}
\begin{tablenotes}[para]
\small \item \textit{Notes:} The table displays estimated coefficients and F-statistics of the excluded instruments of the first-stage regressions of log labor market tightness on the leave-one-out averages of log labor market tightness in all other commuting zones but for the same occupation and time period (Z$^1$). The first stages are estimated separately for each subgroup. Standard errors (in parentheses) are clustered at the labor market level: * = p$<$0.10. ** = p$<$0.05. *** = p$<$0.01. \textit{Sources:} Integrated Employment Biographies + Official Statistics of the German Federal Employment Agency + IAB Job Vacancy Survey, 2012-2022.
\end{tablenotes}
\end{center}
\end{table}

\clearpage

\begin{table}[ht!]
\centering
\begin{center}
\caption{Heterogeneous Effects along the Wage Distribution} \label{tab:het_effects_wage_deciles}
\begin{center}
\scalebox{0.85}{
\begin{tabular}{lcc}
\toprule
 & (1) & (2) \\
 & \textbf{Worker Distribution} & \textbf{Firm Distribution} \\
 & Log Wage & Log Wage  \\
\midrule
\multirow{2}{*}{Log Tightness * Wage Decile Group 1}   & \hphantom{00}0.1190***\hphantom{0} & \hphantom{00}0.0578***\hphantom{0} \\
                                                                                                               & \hphantom{0}(0.0041)\hphantom{000} & \hphantom{0}(0.0020)\hphantom{000} \\
\multirow{2}{*}{Log Tightness * Wage Decile Group 2}   & \hphantom{00}0.0476***\hphantom{0} & \hphantom{00}0.0264***\hphantom{0} \\
                                                                                                               & \hphantom{0}(0.0018)\hphantom{000} & \hphantom{0}(0.0017)\hphantom{000} \\
\multirow{2}{*}{Log Tightness * Wage Decile Group 3}   & \hphantom{00}0.0289***\hphantom{0} & \hphantom{00}0.0174***\hphantom{0} \\
                                                                                                               & \hphantom{0}(0.0015)\hphantom{000} & \hphantom{0}(0.0015)\hphantom{000} \\
\multirow{2}{*}{Log Tightness * Wage Decile Group 4}   & \hphantom{00}0.0165***\hphantom{0} & \hphantom{00}0.0082***\hphantom{0} \\
                                                                                                               & \hphantom{0}(0.0015)\hphantom{000} & \hphantom{0}(0.0016)\hphantom{000} \\
\multirow{2}{*}{Log Tightness * Wage Decile Group 5}   & \hphantom{00}0.0062***\hphantom{0} & \hphantom{00}0.0015\hphantom{0000} \\
                                                                                                               & \hphantom{0}(0.0016)\hphantom{000} & \hphantom{0}(0.0015)\hphantom{000} \\
\multirow{2}{*}{Log Tightness * Wage Decile Group 6}   & \hphantom{00}0.0019\hphantom{0000} & \hphantom{00}0.0017\hphantom{0000} \\
                                                                                                               & \hphantom{0}(0.0015)\hphantom{000} & \hphantom{0}(0.0015)\hphantom{000} \\
\multirow{2}{*}{Log Tightness * Wage Decile Group 7}   & \hphantom{00}0.0005\hphantom{0000} & \hphantom{00}0.0033**\hphantom{00} \\
                                                                                                               & \hphantom{0}(0.0014)\hphantom{000} & \hphantom{0}(0.0014)\hphantom{000} \\
\multirow{2}{*}{Log Tightness * Wage Decile Group 8}   & \hphantom{00}0.0006\hphantom{0000} & \hphantom{00}0.0058***\hphantom{0} \\
                                                                                                               & \hphantom{0}(0.0014)\hphantom{000} & \hphantom{0}(0.0014)\hphantom{000} \\
\multirow{2}{*}{Log Tightness * Wage Decile Group 9}   & \hphantom{00}0.0002\hphantom{0000} & \hphantom{00}0.0067***\hphantom{0} \\
                                                                                                               & \hphantom{0}(0.0014)\hphantom{000} & \hphantom{0}(0.0015)\hphantom{000} \\
\multirow{2}{*}{Log Tightness * Wage Decile Group 10}  & \hphantom{0}-0.0231***\hphantom{0} & \hphantom{00}0.0094***\hphantom{0} \\
                                                                                                               & \hphantom{0}(0.0023)\hphantom{000} & \hphantom{0}(0.0024)\hphantom{000} \\
\multirow{2}{*}{Hire}                                                                  & \hphantom{0}-0.0371***\hphantom{0} & \hphantom{0}-0.0379***\hphantom{0} \\
                                                                                                               & \hphantom{0}(0.0007)\hphantom{000} & \hphantom{0}(0.0007)\hphantom{000} \\
\multirow{2}{*}{Age}                                                                   & \hphantom{00}0.0505***\hphantom{0} & \hphantom{00}0.0523***\hphantom{0} \\
                                                                                                               & \hphantom{0}(0.0044)\hphantom{000} & \hphantom{0}(0.0045)\hphantom{000} \\
\multirow{2}{*}{Age\textsuperscript{2}}                                & \hphantom{0}-0.0005***\hphantom{0} & \hphantom{0}-0.0005***\hphantom{0} \\
                                                                                                               & \hphantom{0}(0.0000)\hphantom{000} & \hphantom{0}(0.0000)\hphantom{000} \\
\multirow{2}{*}{Low Education}                                                 & base & base \\
                                                                                                               & - & - \\
\multirow{2}{*}{Medium Education}                                              & \hphantom{00}0.1297***\hphantom{0} & \hphantom{00}0.1391***\hphantom{0} \\
                                                                                                               & \hphantom{0}(0.0059)\hphantom{000} & \hphantom{0}(0.0062)\hphantom{000} \\
\multirow{2}{*}{High Education}                                                & \hphantom{00}0.3641***\hphantom{0} & \hphantom{00}0.3760***\hphantom{0} \\
                                                                                                               & \hphantom{0}(0.0110)\hphantom{000} & \hphantom{0}(0.0115)\hphantom{000} \\
Year FE & yes & yes \\
Worker FE & yes & yes \\
Labor Market FE & yes & yes \\
Firm FE & yes & yes \\
Controls & yes & yes \\
\midrule
Observations \phantom{XXXXXX} &  8,454,541 &  8,454,541 \\
Clusters &     13,692 &     13,692 \\
\bottomrule
\end{tabular}
}
\end{center}
\begin{tablenotes}[para]
\small \item \textit{Notes:} The table displays IV regressions of log real daily wages of regular full-time workers on log labor market tightness for ten decile groups along workers' wage distribution (Column 1) and firms' wage distribution (Column 2). Workers are assigned into ten decile groups based on their real daily gross wage when they first appear in the IEB as full-time workers during 2012-2022. Firms are assigned into ten decile groups based on the average (real daily) wage of their (full-time) workers when the firms first appear in the IEB sample during 2012-2022. Both estimations refer to the baseline specification with fixed effects for years, workers, labor markets, and firms as well as control variables. Standard errors (in parentheses) are clustered at the labor market level: * = p$<$0.10. ** = p$<$0.05. *** = p$<$0.01. \textit{Sources:} Integrated Employment Biographies + Official Statistics of the German Federal Employment Agency + IAB Job Vacancy Survey, 2012-2022.
\end{tablenotes}
\end{center}
\end{table}

\begin{table}[ht!]
\centering
\begin{center}
\caption{Heterogeneous Effects along the Wage Distribution - First-Stage Regressions} \label{tab:het_effects_wage_deciles_fs}
\begin{center}
\scalebox{0.85}{
\begin{tabular}{llcc}
\toprule
 & & (1) & (2) \\
 & & \textbf{Worker Distribution} & \textbf{Firm Distribution} \\
 & & Log Tightness & Log Tightness  \\
\midrule
\multirow{3}{*}{Z\textsuperscript{1} * Decile Group 1}         & Coefficient & \hphantom{00}0.9163***\hphantom{0} & \hphantom{00}0.9287***\hphantom{0} \\
                                                                                                                       & F-Statistic of & \multirow{2}{*}{     1,975} & \multirow{2}{*}{     1,827} \\
                                                                                                                       & Excluded Instrument & & \\
\multirow{3}{*}{Z\textsuperscript{1} * Decile Group 2}         & Coefficient & \hphantom{00}0.9199***\hphantom{0} & \hphantom{00}0.9217***\hphantom{0} \\
                                                                                                                       & F-Statistic of & \multirow{2}{*}{     2,263} & \multirow{2}{*}{     2,350} \\
                                                                                                                       & Excluded Instrument & & \\
\multirow{3}{*}{Z\textsuperscript{1} * Decile Group 3}         & Coefficient & \hphantom{00}0.9210***\hphantom{0} & \hphantom{00}0.9184***\hphantom{0} \\
                                                                                                                       & F-Statistic of & \multirow{2}{*}{     2,300} & \multirow{2}{*}{     2,339} \\
                                                                                                                       & Excluded Instrument & & \\
\multirow{3}{*}{Z\textsuperscript{1} * Decile Group 4}         & Coefficient & \hphantom{00}0.9016***\hphantom{0} & \hphantom{00}0.8958***\hphantom{0} \\
                                                                                                                       & F-Statistic of & \multirow{2}{*}{     2,672} & \multirow{2}{*}{     2,548} \\
                                                                                                                       & Excluded Instrument & & \\
\multirow{3}{*}{Z\textsuperscript{1} * Decile Group 5}         & Coefficient & \hphantom{00}0.8811***\hphantom{0} & \hphantom{00}0.8794***\hphantom{0} \\
                                                                                                                       & F-Statistic of & \multirow{2}{*}{     2,942} & \multirow{2}{*}{     3,034} \\
                                                                                                                       & Excluded Instrument & & \\
\multirow{3}{*}{Z\textsuperscript{1} * Decile Group 6}         & Coefficient & \hphantom{00}0.8625***\hphantom{0} & \hphantom{00}0.8704***\hphantom{0} \\
                                                                                                                       & F-Statistic of & \multirow{2}{*}{     2,718} & \multirow{2}{*}{     2,933} \\
                                                                                                                       & Excluded Instrument & & \\
\multirow{3}{*}{Z\textsuperscript{1} * Decile Group 7}         & Coefficient & \hphantom{00}0.8517***\hphantom{0} & \hphantom{00}0.8590***\hphantom{0} \\
                                                                                                                       & F-Statistic of & \multirow{2}{*}{     2,253} & \multirow{2}{*}{     2,253} \\
                                                                                                                       & Excluded Instrument & & \\
\multirow{3}{*}{Z\textsuperscript{1} * Decile Group 8}         & Coefficient & \hphantom{00}0.8496***\hphantom{0} & \hphantom{00}0.8644***\hphantom{0} \\
                                                                                                                       & F-Statistic of & \multirow{2}{*}{     1,938} & \multirow{2}{*}{     1,742} \\
                                                                                                                       & Excluded Instrument & & \\
\multirow{3}{*}{Z\textsuperscript{1} * Decile Group 9}         & Coefficient & \hphantom{00}0.8526***\hphantom{0} & \hphantom{00}0.8619***\hphantom{0} \\
                                                                                                                       & F-Statistic of & \multirow{2}{*}{     1,333} & \multirow{2}{*}{     1,107} \\
                                                                                                                       & Excluded Instrument & & \\
\multirow{3}{*}{Z\textsuperscript{1} * Decile Group 10}        & Coefficient & \hphantom{00}0.8895***\hphantom{0} & \hphantom{00}0.8750***\hphantom{0} \\
                                                                                                                       & F-Statistic of & \multirow{2}{*}{       422} & \multirow{2}{*}{       414} \\
                                                                                                                       & Excluded Instrument & & \\
\bottomrule
\end{tabular}
}
\end{center}
\begin{tablenotes}[para]
\small \item \textit{Notes:} The table displays estimated coefficients and F-statistics of the excluded instruments of the first-stage regressions of log labor market tightness on the leave-one-out averages of log labor market tightness in all other commuting zones but for the same occupation and time period (Z$^1$). The first stages are estimated separately for each wage decile group. Column 1 shows the first stage estimations for ten decile groups along the worker wage distribution. Column 2 shows the first-stage estimations for ten decile groups along the firm wage distribution. Workers are assigned into ten decile groups based on their real daily gross wage when they first appear in the IEB as full-time workers during 2012-2022. Firms are assigned into ten decile groups based on the average (real daily) wage of their (full-time) workers when the firms first appear in the IEB sample during 2012-2022. Standard errors (in parentheses) are clustered at the labor market level: * = p$<$0.10. ** = p$<$0.05. *** = p$<$0.01. \textit{Sources:} Integrated Employment Biographies + Official Statistics of the German Federal Employment Agency + IAB Job Vacancy Survey, 2012-2022.
\end{tablenotes}
\end{center}
\end{table}

\begin{table}[ht!]
\begin{center}
\caption{IV Regressions of Wage Setting at the Firm Level} \label{tab:firm_wage_setting_full}
\begin{center}
\begin{tabular}{L{4.2cm}C{2.6cm}C{2.6cm}}
\toprule
 & (1) & (2) \\
 & Log Wage & Log $\overline{\text{Wage}}^{\text{Firm}}$ \\
\midrule
\multirow{2}{*}{Log Tightness}         & \hphantom{00}0.0024**\hphantom{00} & \hphantom{00}0.0088***\hphantom{0} \\
                                                                       & \hphantom{0}(0.0011)\hphantom{000} & \hphantom{0}(0.0012)\hphantom{000} \\
\multirow{2}{*}{Hire}                                                          & \hphantom{0}-0.0497***\hphantom{0} & \hphantom{0}-0.0120***\hphantom{0} \\
                                                                                                       & \hphantom{0}(0.0009)\hphantom{000} & \hphantom{0}(0.0003)\hphantom{000} \\
\multirow{2}{*}{Age}                                                           & \hphantom{00}0.0501***\hphantom{0} & \hphantom{00}0.0223***\hphantom{0} \\
                                                                                                       & \hphantom{0}(0.0041)\hphantom{000} & \hphantom{0}(0.0013)\hphantom{000} \\
\multirow{2}{*}{Age\textsuperscript{2}}                        & \hphantom{0}-0.0005***\hphantom{0} & \hphantom{0}-0.0002***\hphantom{0} \\
                                                                                                       & \hphantom{0}(0.0000)\hphantom{000} & \hphantom{0}(0.0000)\hphantom{000} \\
\multirow{2}{*}{Low Education}                                         & base & base \\
                                                                                                       & - & - \\
\multirow{2}{*}{Medium Education}                                      & \hphantom{00}0.1407***\hphantom{0} & \hphantom{00}0.0371***\hphantom{0} \\
                                                                                                       & \hphantom{0}(0.0079)\hphantom{000} & \hphantom{0}(0.0016)\hphantom{000} \\
\multirow{2}{*}{High Education}                                        & \hphantom{00}0.3843***\hphantom{0} & \hphantom{00}0.0705***\hphantom{0} \\
                                                                                                       & \hphantom{0}(0.0143)\hphantom{000} & \hphantom{0}(0.0020)\hphantom{000} \\
Year FE & & yes \\
Worker FE & yes & yes \\
Labor Market FE & yes & yes \\
Firm FE & & yes  \\
Firm-by-Year FE & yes & \\
Controls & yes & yes \\
\midrule
Observations \phantom{XXXXXX} &  6,424,250 &  8,454,541 \\
Clusters &     12,676 &     13,692 \\
\midrule
First-Stage Coefficient & \hphantom{00}0.8597***\hphantom{0} & \hphantom{00}0.8814***\hphantom{0} \\
\makecell[l]{First-Stage F-Statistic \\of Excluded Instrument} & \makecell[c]{     3,247} & \makecell[c]{     2,740} \\
\bottomrule
\end{tabular}
\end{center}
\begin{tablenotes}
\small \item \textit{Notes:} The table displays IV regressions of log (average firm-level) real daily wages of regular full-time workers on log labor market tightness. The instrumental variable refers to leave-one-out averages of labor market tightness in all other commuting zones but for the same occupation and time period. Control variables include binary variables for new hires, workplace location in Eastern Germany, three levels of professional education, and continuous variables for age and squared age. Labor markets are combinations of detailed 3-digit occupations and commuting zones. Column (1) shows IV estimates when additionally controlling for firm-by-year fixed effects. Column (2) displays IV estimates when regressing the firm-level average of real daily wages on log labor market tightness. Standard errors (in parentheses) are clustered at the labor market level: * = p$<$0.10. ** = p$<$0.05. *** = p$<$0.01. \textit{Sources:} Integrated Employment Biographies + Official Statistics of the German Federal Employment Agency + IAB Job Vacancy Survey, 2012-2022.
\end{tablenotes}
\end{center}
\end{table}

\clearpage

\clearpage
\setcounter{figure}{0}
\setcounter{table}{0}
\renewcommand*{\thefigure}{\thesection\arabic{figure}}
\renewcommand*{\thetable}{\thesection\arabic{table}}
\section{Heterogeneous Effects along the Wage Distribution: Further Details}
\label{app:ddistribution}

\vspace*{2.5cm}

\begin{figure}[ht!]
\centering
\begin{center}
\caption{Heterogeneous Effects along the Wage Distribution by Requirement Level}
\label{fig:level_by_wage_decile_level_interaction}
\begin{center}
		\includegraphics[scale=1.0]{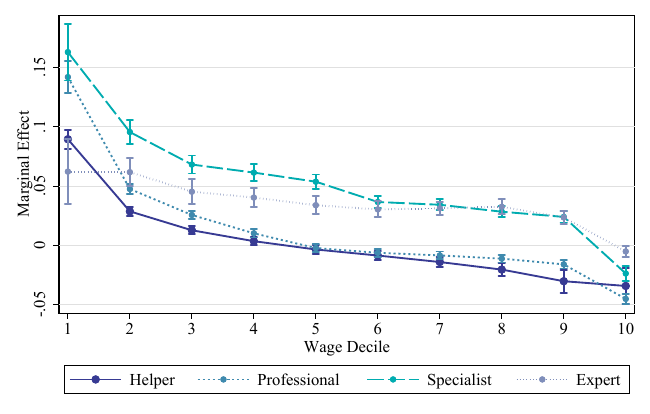}
\end{center}
\begin{tablenotes}
\small \item \textit{Notes:} The figure displays estimated elasticities and 95\% confidence intervals from IV regressions of log real daily wages of regular full-time workers on log labor market tightness for ten decile groups along the worker wage distribution and for each of the four job requirement levels as stated in the legend. Workers are assigned into ten decile groups based on their real daily gross wage when they first appear in the IEB as full-time workers during 2012-2022. The curves refer to the baseline specification with fixed effects for years, workers, labor markets, and firms as well as control variables. \textit{Sources:} Integrated Employment Biographies + Official Statistics of the German Federal Employment Agency + IAB Job Vacancy Survey, 2012-2022.
\end{tablenotes}
\end{center}
\end{figure}

\clearpage

\begin{figure}[ht!]
\centering
\begin{center}
\caption{Requirement Levels by Decile Groups}
\label{fig:level_by_wage_decile}
\begin{center}
\includegraphics[scale=1.0]{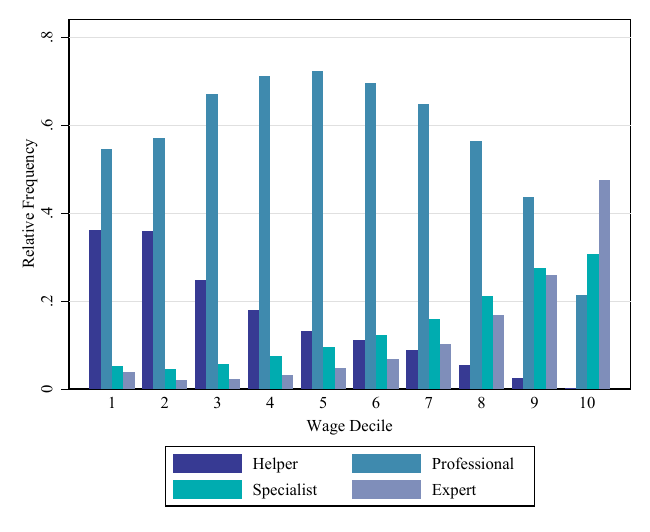}
\end{center}
\begin{tablenotes}
\small \item \textit{Notes:} The figure illustrates the share of the four different requirement levels in employment, separately by decile groups of workers' wage distribution. Requirement levels refer to the fifth digit of the KldB-2010 occupation variable. \textit{Sources:} Integrated Employment Biographies, 2012-2022
\end{tablenotes}
\end{center}
\end{figure}

\begin{figure}[ht!]
\centering
\begin{center}
\caption{Real Wage Growth By Decile Group}
\label{fig:wage_growth_by_decile}
\begin{center}
\includegraphics[scale=1.0]{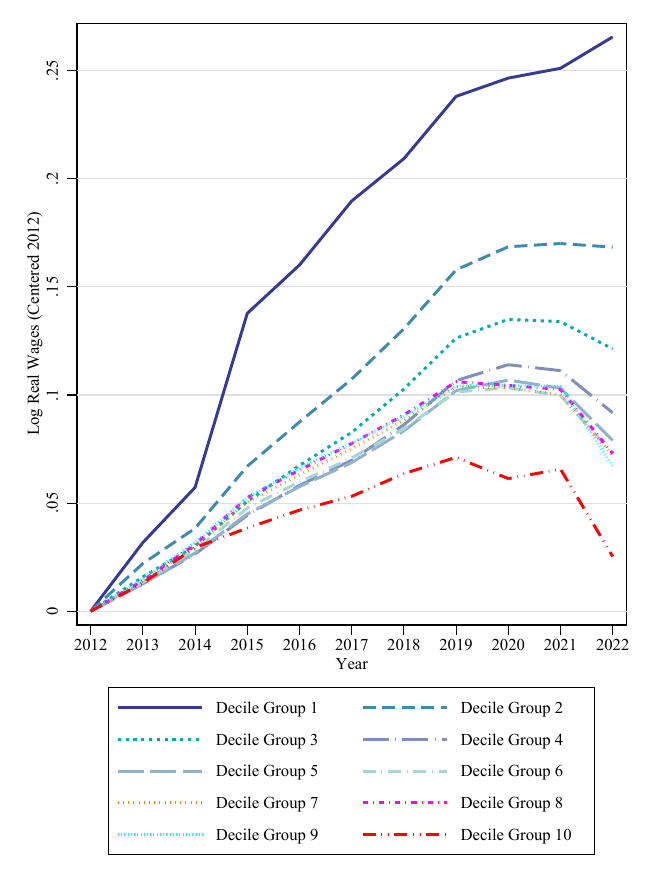}
\end{center}
\begin{tablenotes}
\small \item \textit{Notes:} The figure illustrates the average growth of real daily wages for ten decile groups of workers' wage distribution since 2012 (in percent). The respective time series of log real wages are centered at 2012 by subtracting the 2012 log real wage of the respective decile group \textit{Sources:} Integrated Employment Biographies, 2012-2022.
\end{tablenotes}
\end{center}
\end{figure}

\clearpage

\begin{figure}[ht!]
\centering
\begin{center}
\caption{Counterfactual Real Wage Growth in Absence of Tightness Increase}
\label{fig:wage_growth_by_decile_counterfactual}
\begin{center}
\includegraphics[scale=1.0]{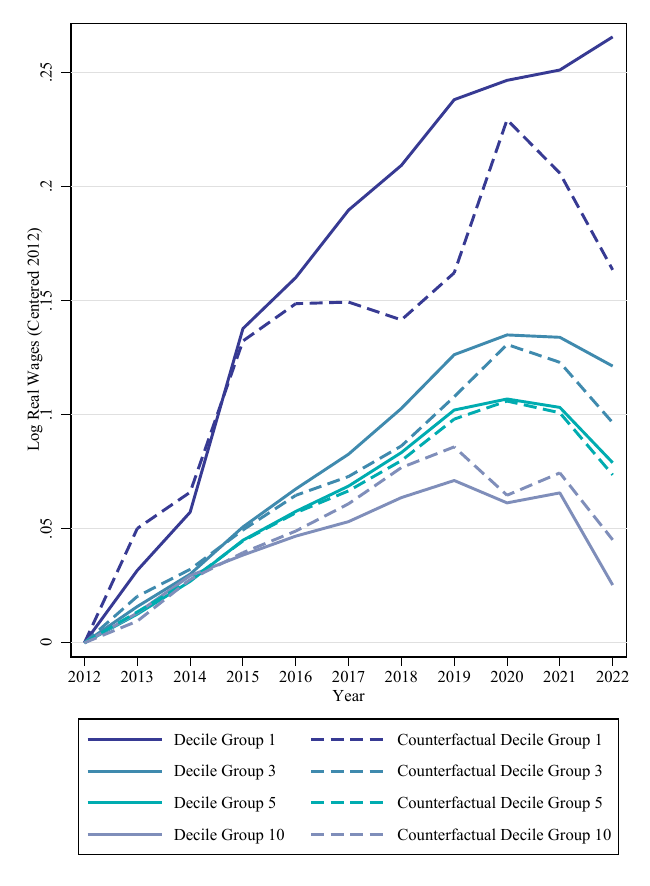}
\end{center}
\begin{tablenotes}
\small \item \textit{Notes:} The figure compares the wage growth in the first, second, third, and tenth decile group since 2012 with a counterfactual scenario in which labor market tightness is held constant. The solid lines illustrate average growth of real daily gross wages since 2012 (in percent). The dashed lines illustrate the counterfactual wage growth in the absence of a tightness increase. For each decile group, the counterfactual is calculated by subtracting the product of the wage elasticity with respect to tightness and the observed tightness increase (see Appendix Figure~\ref{fig:tightness_growth_by_decile}) from observed values. \textit{Sources:} Integrated Employment Biographies + Official Statistics of the German Federal Employment Agency + IAB Job Vacancy Survey, 2012-2022.
\end{tablenotes}
\end{center}
\end{figure}

\clearpage

\begin{figure}[ht!]
\centering
\begin{center}
\caption{Labor Market Tightness Growth by Decile Group}
\label{fig:tightness_growth_by_decile}
\begin{center}
\includegraphics[scale=1.0]{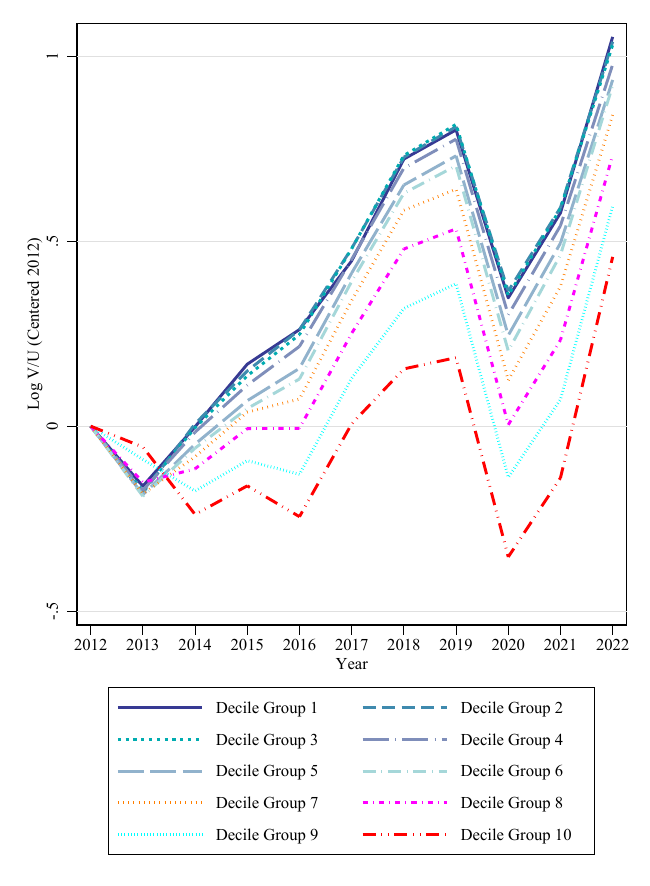}
\end{center}
\begin{tablenotes}
\small \item \textit{Notes:} The figure illustrates the average growth of labor market tightness for ten decile groups of workers' wage distribution since 2012 (in percent). The respective time series of log labor market tightness are centered at 2012 by subtracting the 2012 log tightness of the respective decile group \textit{Sources:} Official Statistics of the German Federal Employment Agency + IAB Job Vacancy Survey, 2012-2022.
\end{tablenotes}
\end{center}
\end{figure}

\clearpage
\addcontentsline{toc}{section}{References} 
\printbibliography[title=References] 

\end{refsection}
\end{appendix}


\end{document}